%% file: main.tex
\documentclass{egpubl}

\usepackage{booktabs} 
\usepackage{multirow}
\usepackage{array}
\usepackage{amssymb}
\usepackage{arydshln}
\usepackage{subcaption}
\usepackage{algorithm}
\usepackage{algorithmic}
\usepackage{amsmath}
\usepackage{soul}
\usepackage[normalem]{ulem}
\usepackage[bb=boondox]{mathalfa}

\usepackage[T1]{fontenc}
\usepackage{dfadobe}

\usepackage{cite}  
\bibtexVersion
\BibtexOrBiblatex

\electronicVersion
\PrintedOrElectronic
\ifpdf \usepackage[pdftex]{graphicx} \pdfcompresslevel=9
\else \usepackage[dvips]{graphicx} \fi

\usepackage{egweblnk} 

\title[Scheduled Inpainting]%
      {Interactive Generative Motion Editing via Scheduled Inpainting}

\author[Agrawal et. al.]
{\parbox{\textwidth}{\centering Dhruv Agrawal$^{1,2}$\orcid{0009-0005-0442-9781}, Dominik Borer $^{2}$\orcid{0009-0006-5343-3450}, Luca Vo\"geli$^{2}$\orcid{0009-0000-1175-7680}, Robert W. Sumner$^{1,2}$\orcid{0000-0002-1909-8082}, Martin Guay$^{2}$\orcid{0009-0002-7496-6185}, Jakob Buhmann$^{2}$\orcid{0009-0008-3038-4881}}%
\\
\centering $^1$ETH Z\"urich, Switzerland \quad
         $^2$DisneyResearch|Studios, Switzerland
}

\begin{document}

\input{custom_definitions}
\input{content/figure_definitions}
\teaser{
 \captionsetup{hypcap=false}
 \centering
    \includegraphics[width=0.95\textwidth]{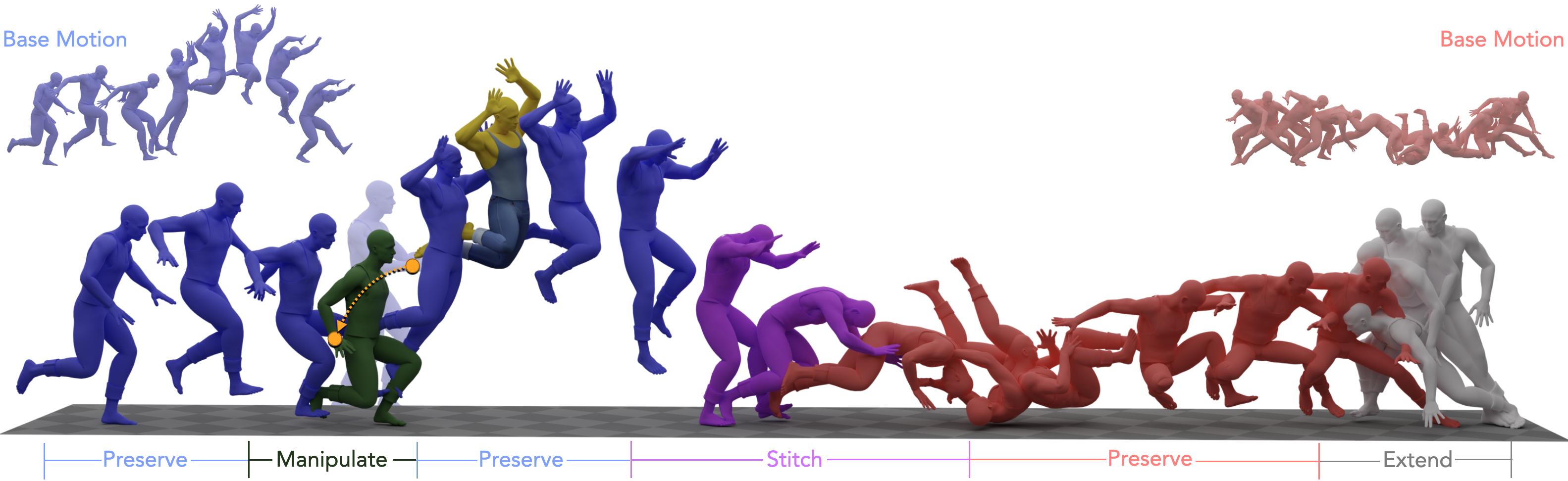}
    \caption{Scheduled inpainting allows for multiple generative motion editing tasks within the same real-time framework, while resulting in natural motion. We see here two clips in blue and red being stitched together (purple), extended (grey), as well as directly modified by dragging points (yellow dots).}
\label{fig:teaser}
}

\maketitle
\thispagestyle{empty}
\makeatletter
\renewcommand{\@oddhead}{}
\renewcommand{\@evenhead}{}
\renewcommand{\@oddfoot}{\hfil\thepage\hfil}
\renewcommand{\@evenfoot}{\hfil\thepage\hfil}
\makeatother
\begin{abstract}
Motion editing is central to VFX and game development, where it is used extensively to modify and augment existing movements to conform to new environments or changes in artistic direction. While traditional motion editing can do small modifications, it cannot accommodate larger structural edits, resulting in visual warping artifacts that require authoring new motion. Conversely, recent advances in large-scale generative modeling have unlocked newfound capabilities for authoring entire movements by directly manipulating sparse spatial constraints. While impressive at creating new movements, these methods lack the capability to preserve and edit existing motion interactively. In this work, we introduce scheduled inpainting, a method that enables interactive generative motion editing, a novel paradigm unifying motion synthesis and editing by leveraging generative models. Scheduled inpainting is a simple yet powerful inference-based technique that enables fine-grained spatiotemporal control over the balance between preserving the original motion and generating new content. By building atop generative models that support direct manipulation, our system allows artists to interactively refine existing animations while ensuring results remain natural and consistent with the learned motion distribution. Scheduled inpainting is versatile and supports many editing applications, such as extending, stitching, and compositing different clips. Finally, we extensively validate our approach by comparing with four baselines, conducting ablations of our design, and reporting user feedback.

\ccsdesc[500]{Computing methodologies~Artificial intelligence}
\ccsdesc[300]{Human-centered computing~User centered design}
\ccsdesc[500]{Computing methodologies~Animation}

\printccsdesc   
\end{abstract}  
\section{Introduction}
\input{content/01_Introduction}

\section{Related Work}
\input{content/02_RelatedWork}

\section{Scheduled Inpainting}
\input{content/03_Method}

\section{Applications}\label{sec:applications}
\input{content/application}

\input{content/04_Evaluation}

\section{Discussion \& Limitations}

\input{content/05_Discussion}

\section{Conclusion}
\input{content/06_Conclusion}
\bibliographystyle{eg-alpha-doi}
\bibliography{bibliography}


\newpage
\appendix
\input{content/07_Appendix}

\end{document}

%% file: custom_definitions.tex
\newcommand{\figref}[1]{Figure~\ref{#1}}
\newcommand{\tabref}[1]{Table~\ref{#1}}
\newcommand{\secref}[1]{Section~\ref{#1}}
\newcommand{\algoref}[1]{Algorithm~\ref{#1}} %
\newcommand{\myeqref}[1]{Eq.~\eqref{#1}}
\newcommand{\appref}[1]{Appendix~\ref{#1}}

\newcommand{\remove}[1]{}
\newcommand{\add}[1]{#1}
\newcommand{\replace}[2]{\remove{#1}\add{#2}}

\newcommand{\baseMotion}[0]{\mathcal{M}_{base}}
\newcommand{\constraints}[0]{\mathcal{C}}
\newcommand{\motionModel}[0]{\varphi} 
\newcommand{\genMotion}[1]{\mathcal{M}_{gen}^{#1}}
\newcommand{\appMask}[0]{\alpha_{mask}}
\newcommand{\timeMask}[1]{\alpha_{time}^{#1}}

\newcommand{\before}[1]{\color{red} \sout{#1} \color{black}}
\newcommand{\after}[1]{\color{blue}#1\color{black}}

%% file: content/figure_definitions.tex
\newcommand{\InferenceMethodComparison}[0]{

\begin{figure*}
    \centering
    \begin{subfigure}{\linewidth}
        \begin{subfigure}{0.25\linewidth}
            \centering
            \includegraphics[width=\linewidth, trim={1cm 5cm 1cm 5cm}, clip]{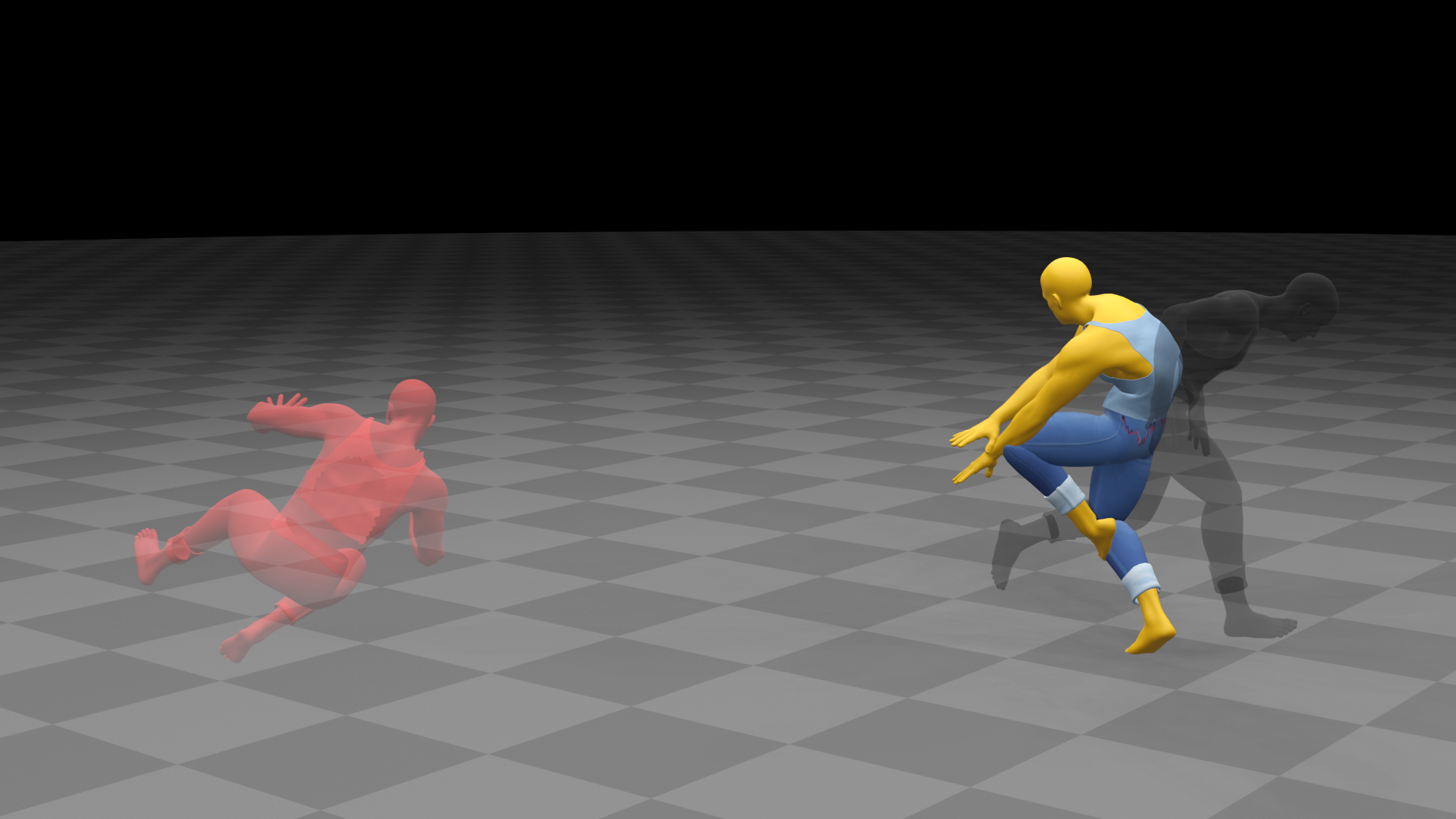}
        \end{subfigure}%
        \begin{subfigure}{0.25\linewidth}
            \centering
            \includegraphics[width=\linewidth, trim={1cm 5cm 1cm 5cm}, clip]{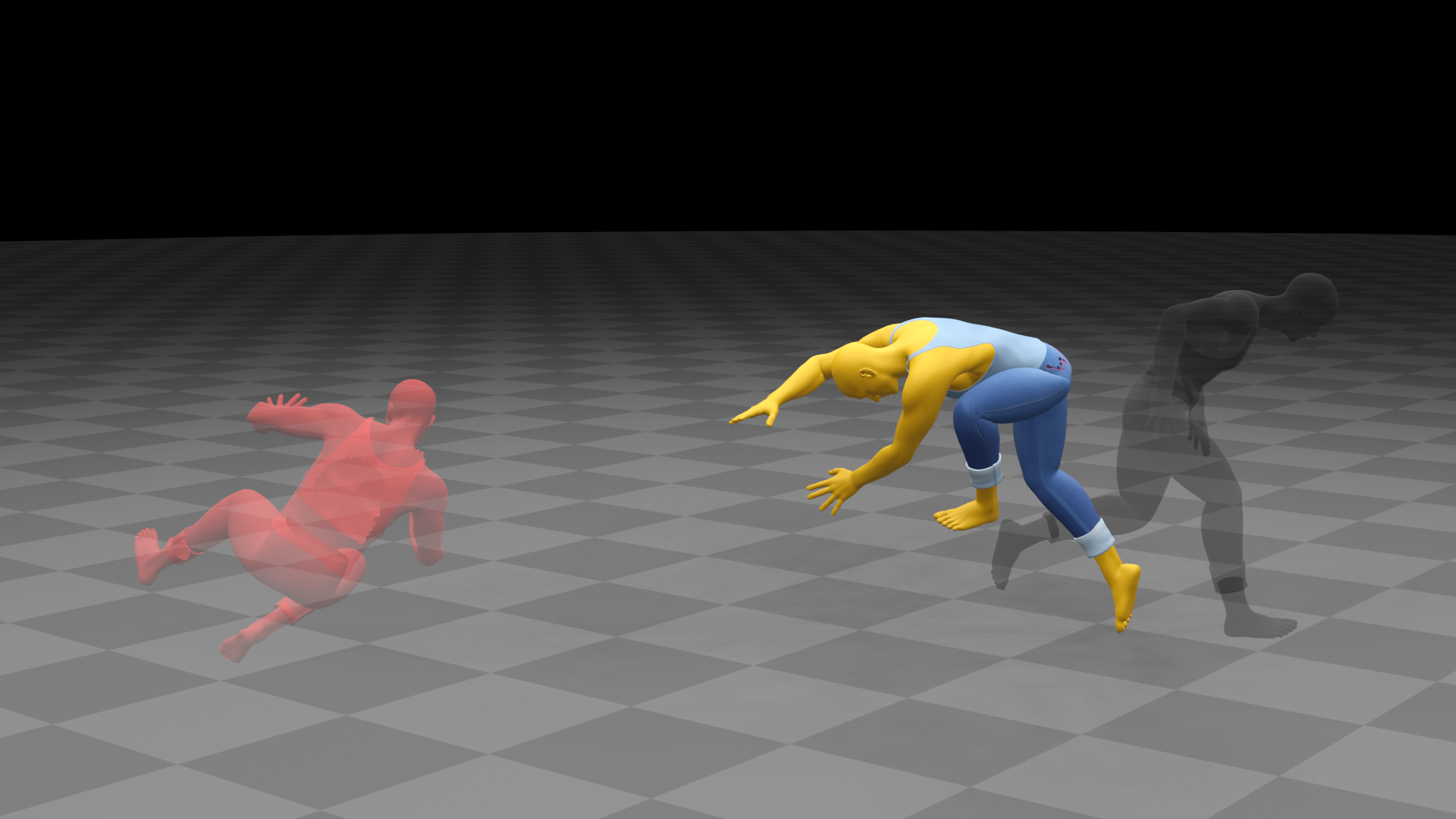}
        \end{subfigure}%
        \begin{subfigure}{0.25\linewidth}
            \centering
            \includegraphics[width=\linewidth, trim={1cm 5cm 1cm 5cm}, clip]{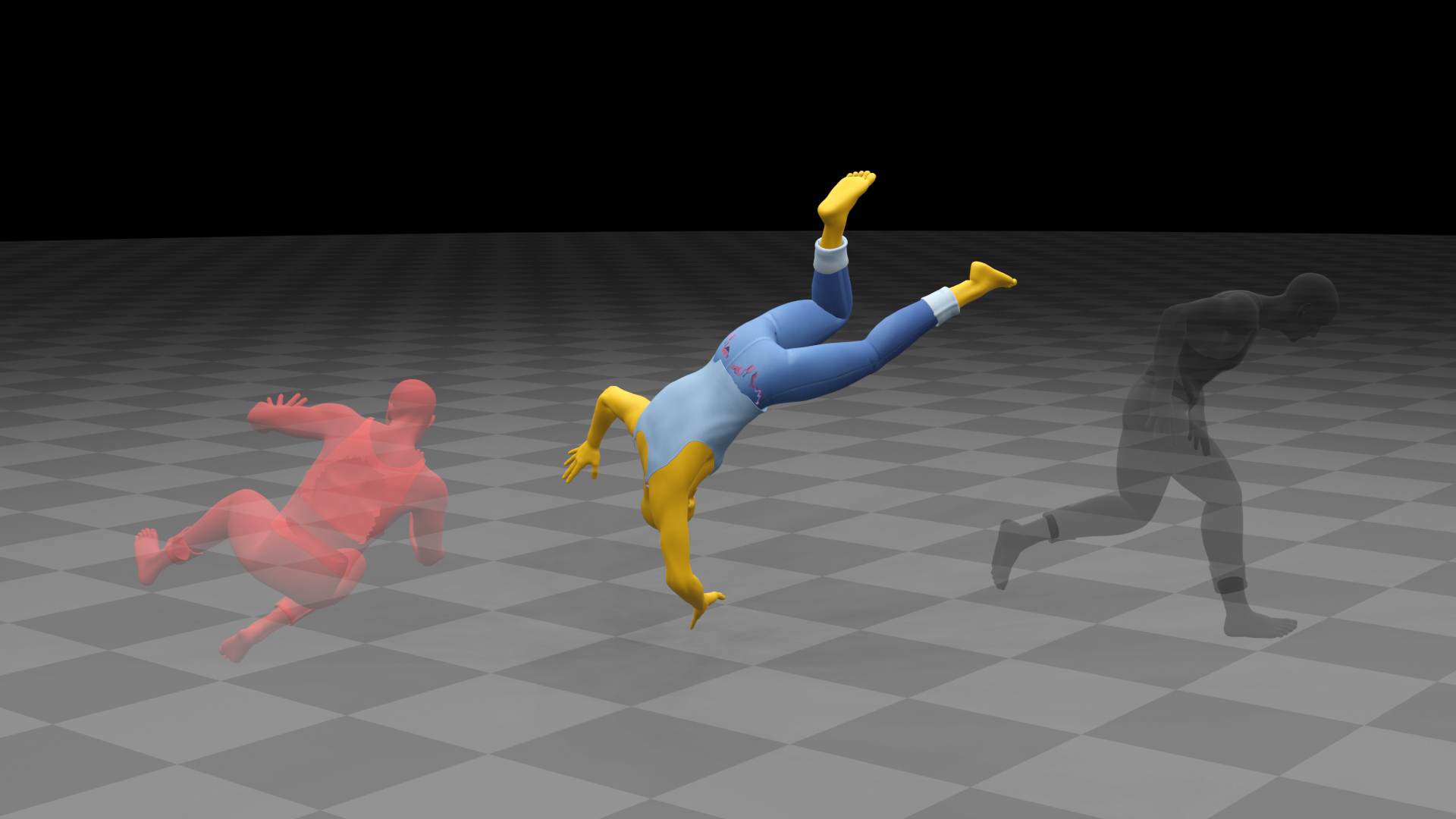}
        \end{subfigure}%
        \begin{subfigure}{0.25\linewidth}
            \centering
            \includegraphics[width=\linewidth, trim={1cm 5cm 1cm 5cm}, clip]{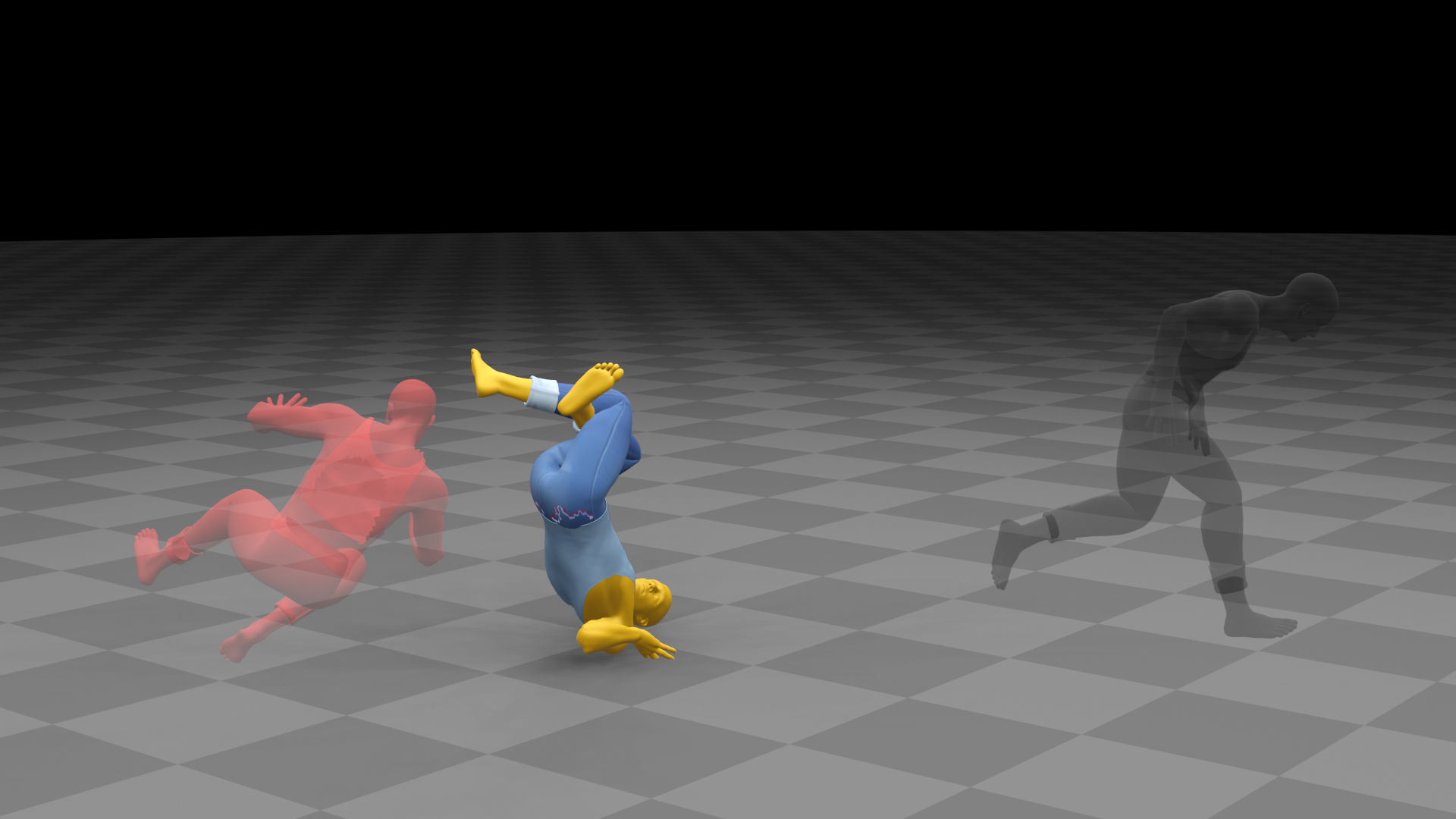}
        \end{subfigure}
    \end{subfigure}

    \begin{subfigure}{\linewidth}
        \begin{subfigure}{0.25\linewidth}
            \centering
            \includegraphics[width=\linewidth, trim={1cm 5cm 1cm 5cm}, clip]{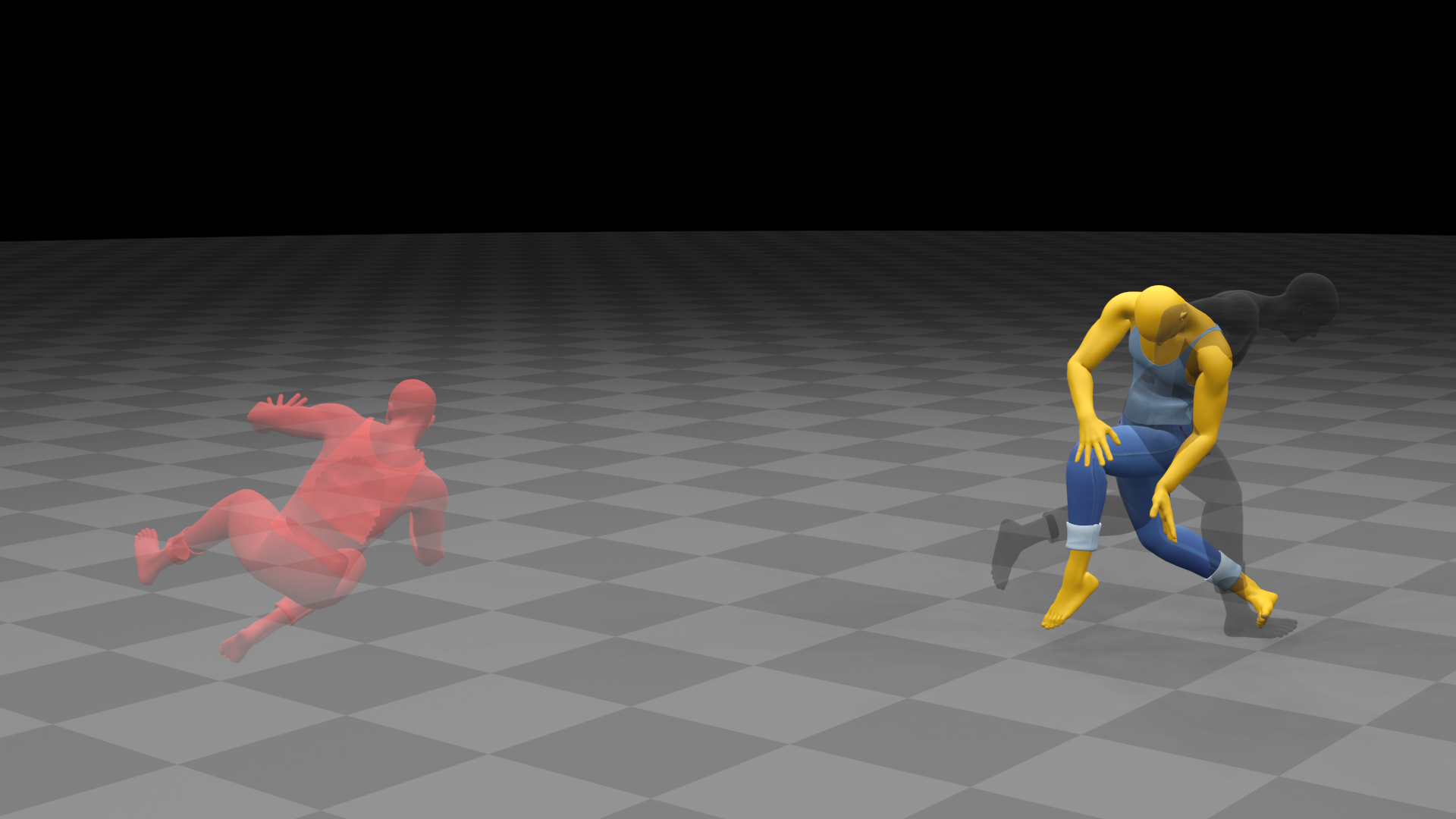}
        \end{subfigure}%
        \begin{subfigure}{0.25\linewidth}
            \centering
            \includegraphics[width=\linewidth, trim={1cm 5cm 1cm 5cm}, clip]{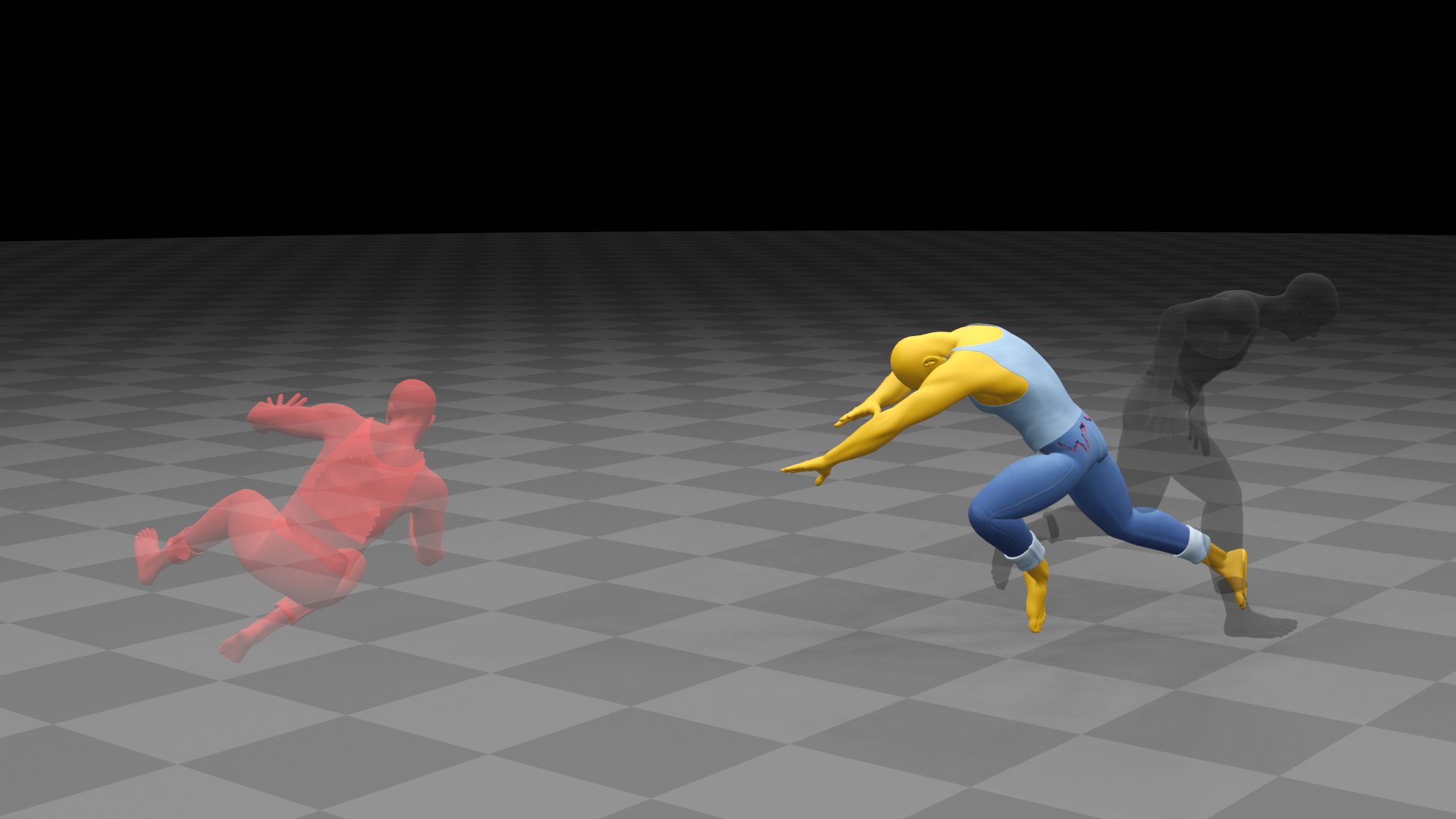}
        \end{subfigure}%
        \begin{subfigure}{0.25\linewidth}
            \centering
            \includegraphics[width=\linewidth, trim={1cm 5cm 1cm 5cm}, clip]{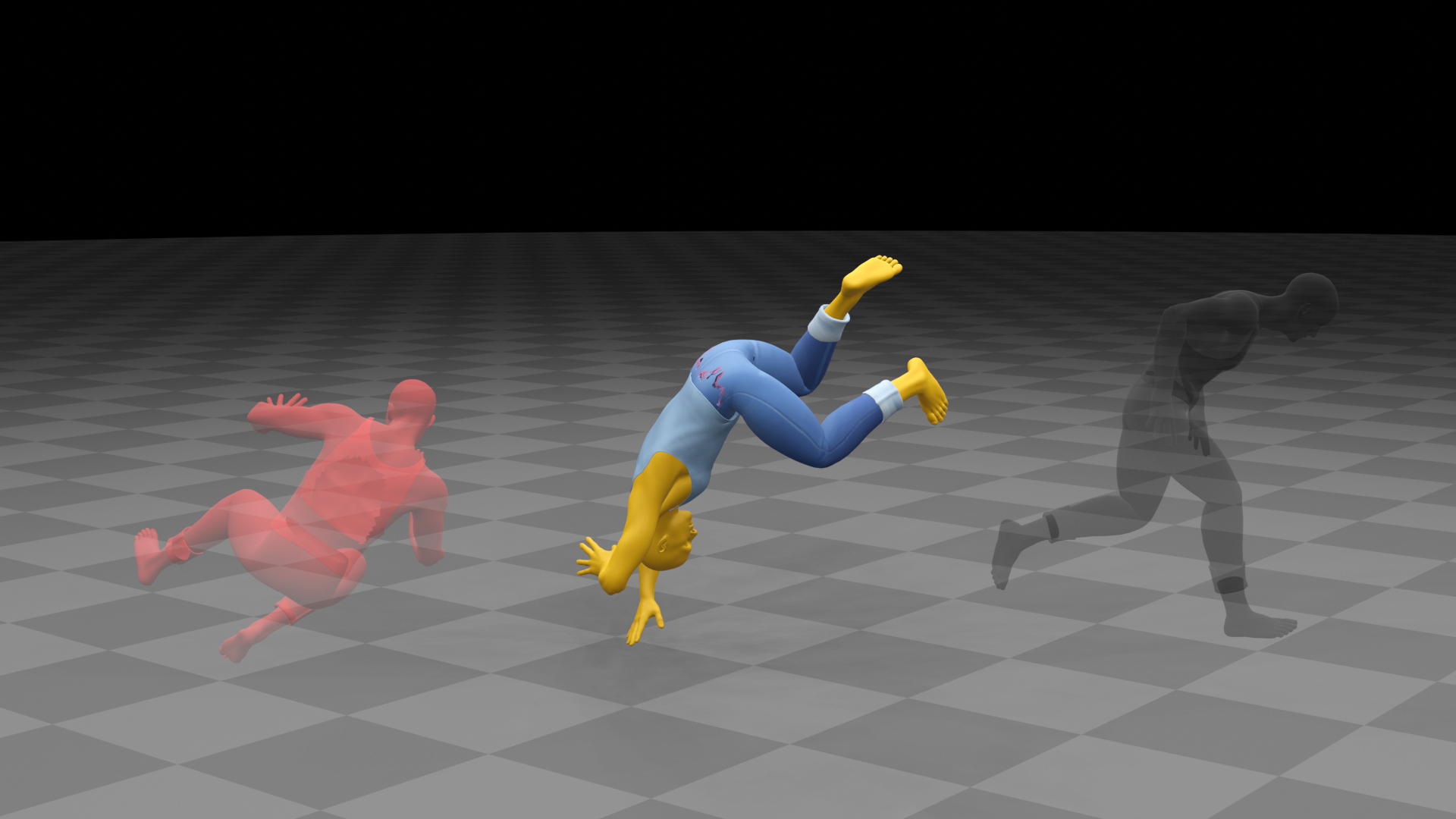}
        \end{subfigure}%
        \begin{subfigure}{0.25\linewidth}
            \centering
            \includegraphics[width=\linewidth, trim={1cm 5cm 1cm 5cm}, clip]{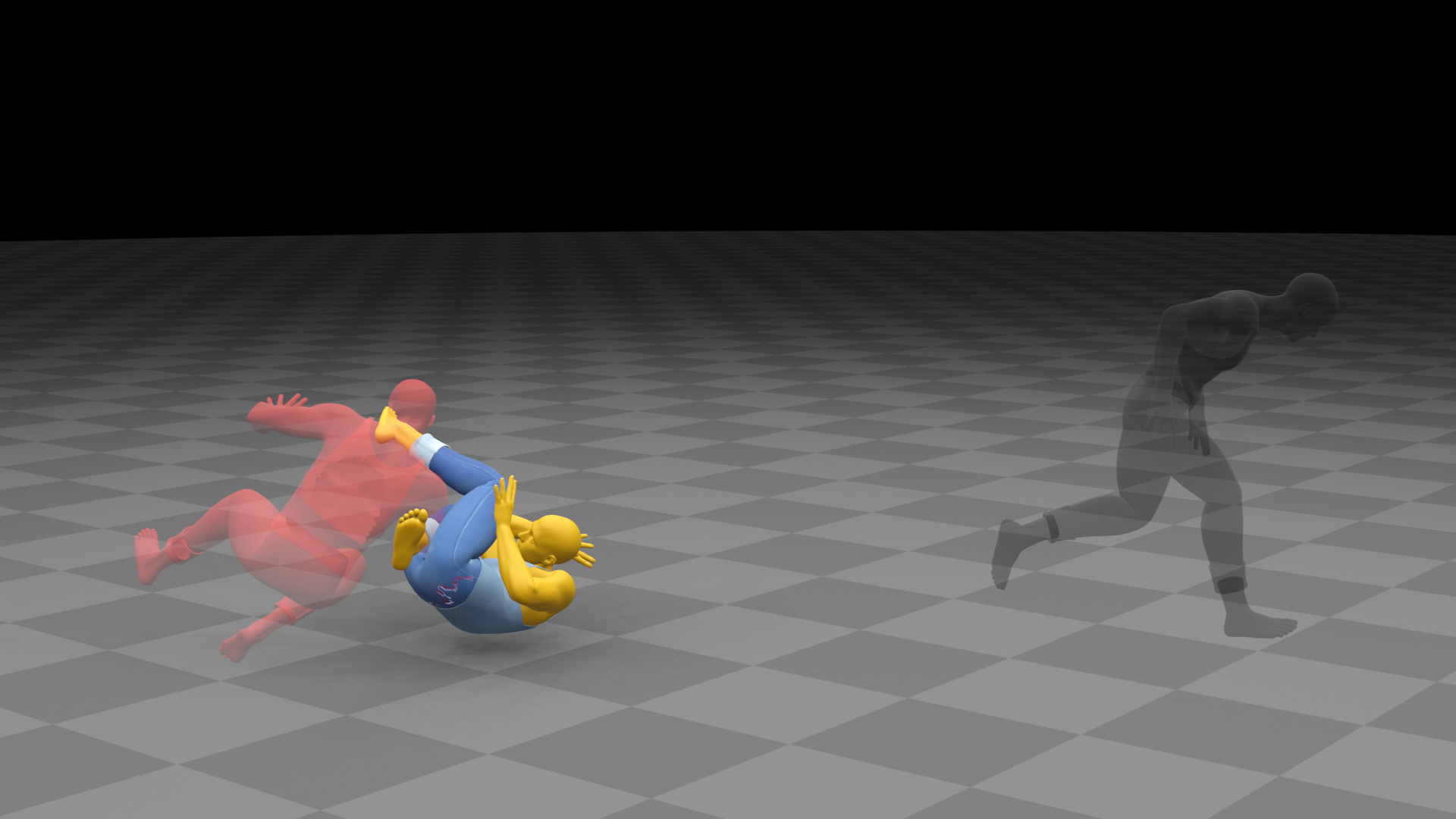}
        \end{subfigure}
    \end{subfigure}

    \begin{subfigure}{\linewidth}
        \begin{subfigure}{0.25\linewidth}
            \centering
            \includegraphics[width=\linewidth, trim={1cm 5cm 1cm 5cm}, clip]{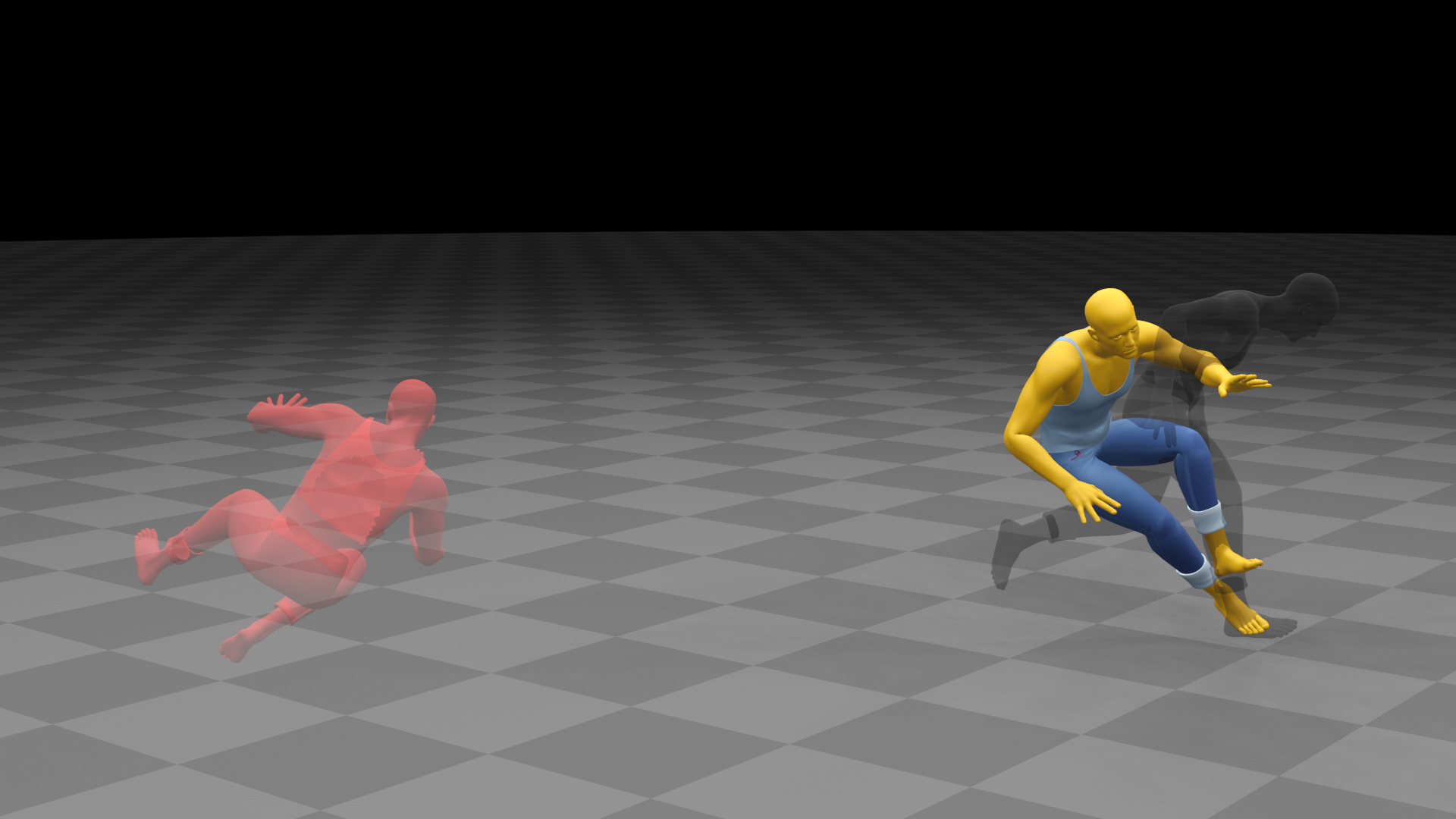}
        \end{subfigure}%
        \begin{subfigure}{0.25\linewidth}
            \centering
            \includegraphics[width=\linewidth, trim={1cm 5cm 1cm 5cm}, clip]{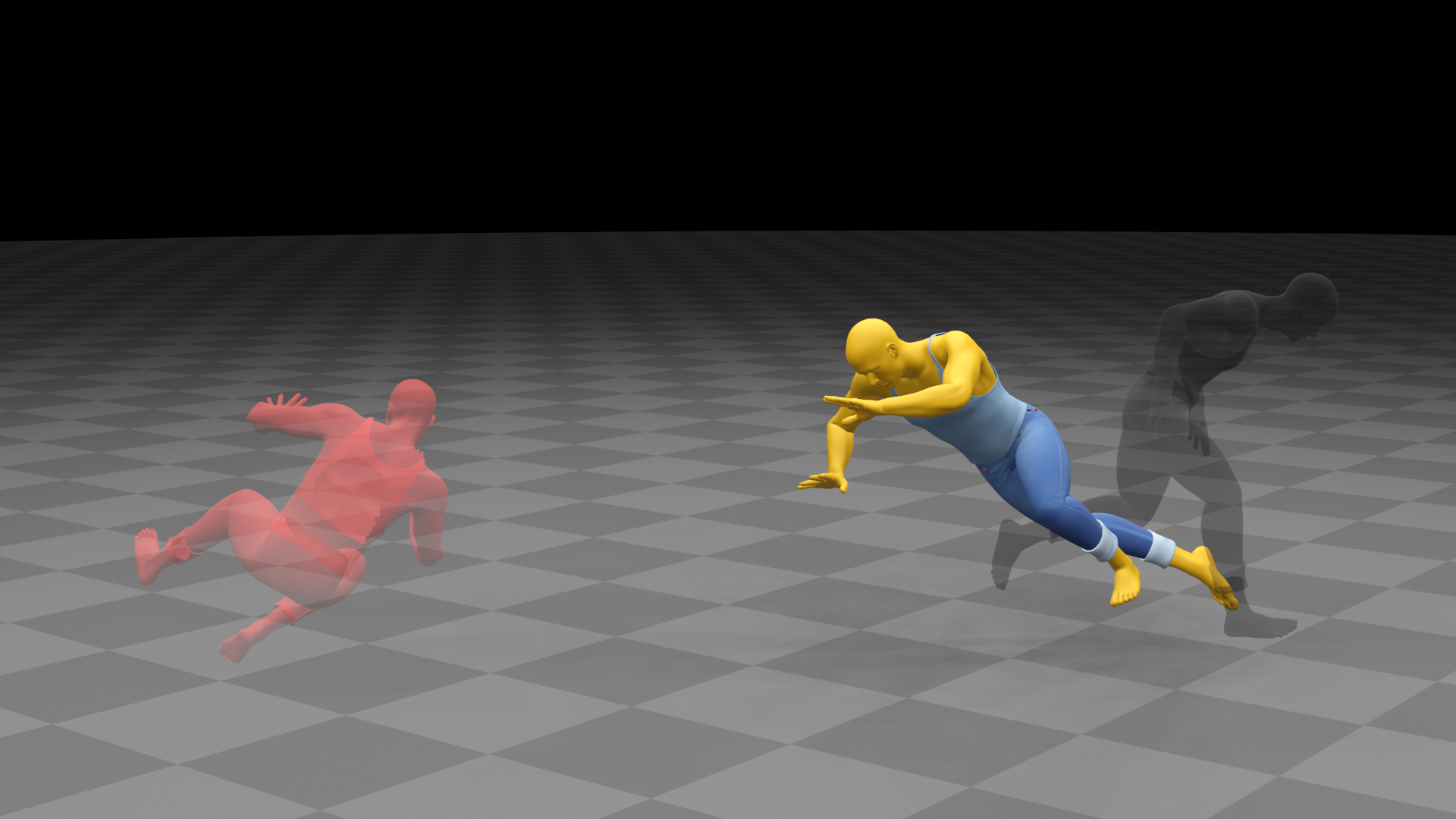}
        \end{subfigure}%
        \begin{subfigure}{0.25\linewidth}
            \centering
            \includegraphics[width=\linewidth, trim={1cm 5cm 1cm 5cm}, clip]{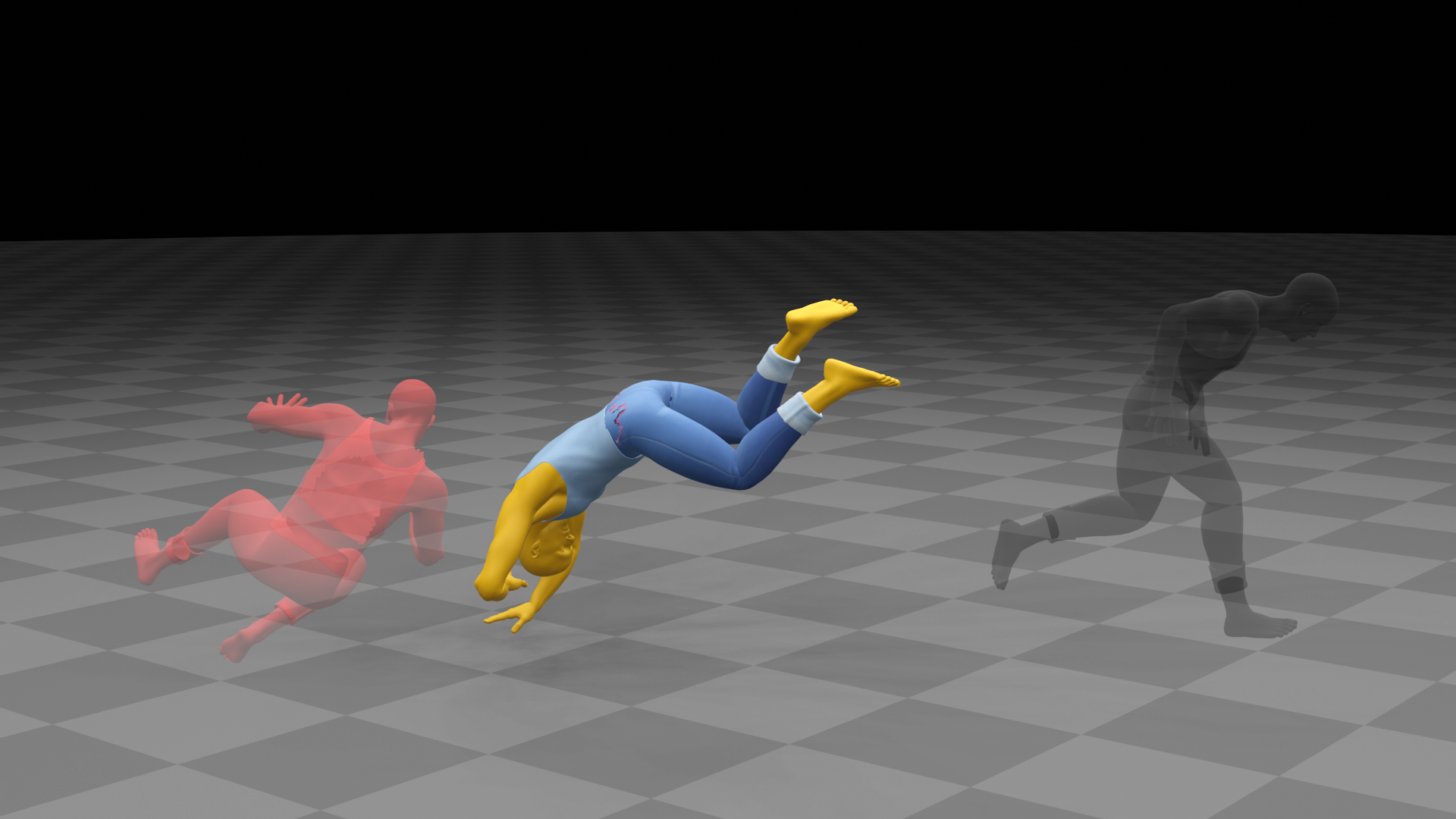}
        \end{subfigure}%
        \begin{subfigure}{0.25\linewidth}
            \centering
            \includegraphics[width=\linewidth, trim={1cm 5cm 1cm 5cm}, clip]{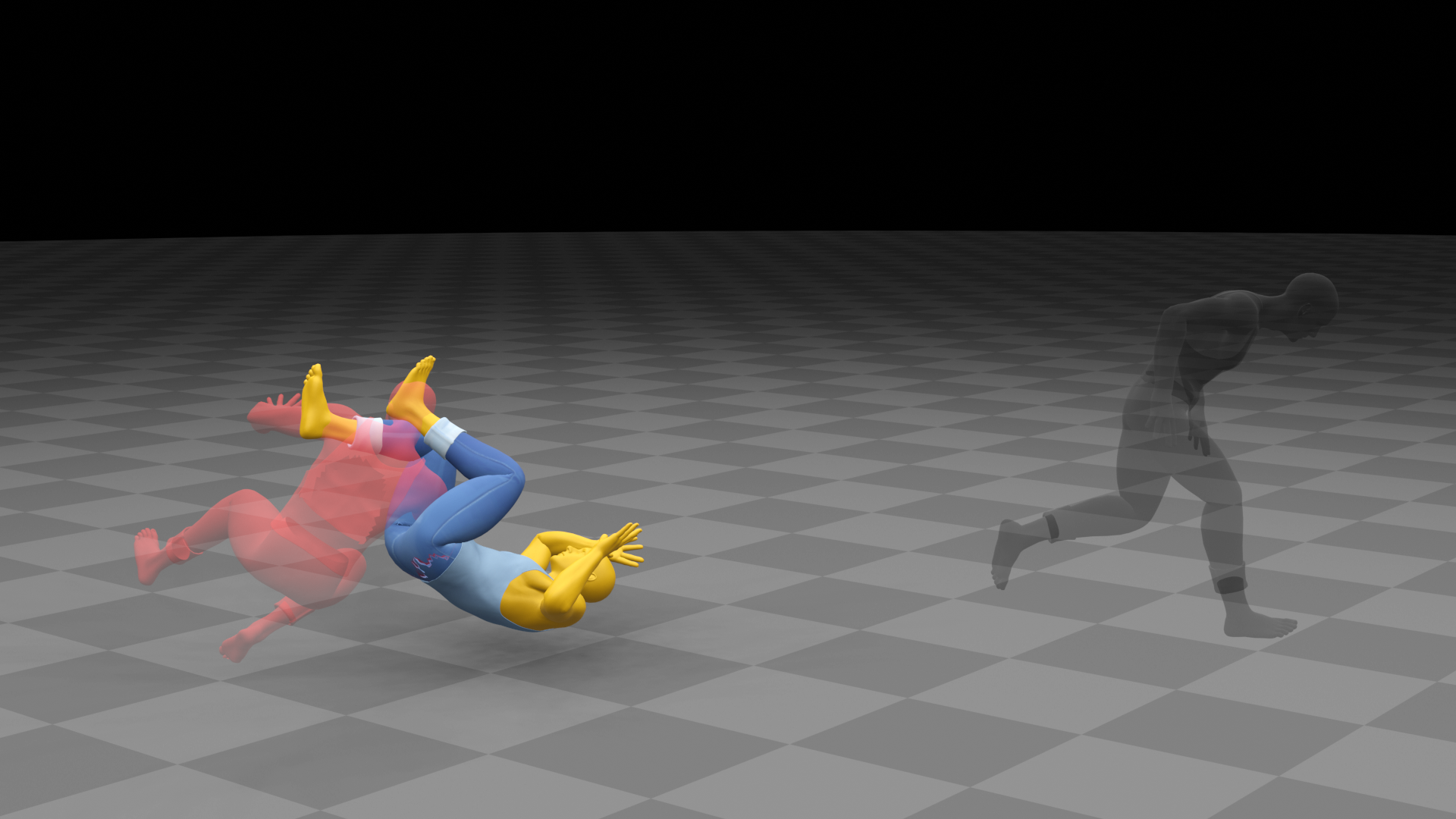}
        \end{subfigure}
    \end{subfigure}
    
    \caption{We edit the end frame (in red) of a rolling forwards clip to be behind the starting frame (in black). The three figures show the output from noise-inversion (top), DNO (middle), and our Scheduled Inpainting method (bottom).}
    \label{fig:motion_roll_baseline_comparison}
\end{figure*}
}

%% file: content/01_Introduction.tex
High-quality motion is both time-consuming and challenging to author and acquire. Motion editing is often used in game and movie productions to fix, alter, or augment movements, but is limited to small changes and often results in unnatural motion due to warping when larger structural changes are required.
For example, substantial manual effort is required to maintain physical plausibility when changes in the height of a platform cause the character to jump down instead of stepping down, as shown in our accompanying video.

Recent advances in generative motion authoring allow users to create entire motions either via textual \cite{hmdm, tseng2023edge, shafir2024human} or spatial \cite{studer2024factorized, hwang2025motion, vogeli2025implicit} conditioning. Among these models, only spatially conditioned models can interactively modify their own output via direct manipulation of sparse spatial constraints.
Crucially, no prior model has demonstrated interactively editing existing motion clips; that is, motion sequences from another source and not generated by the model itself.
The closest existing works for generative motion editing include MotionFix \cite{athanasiou2024motionfix} and MotionLab~\cite{guo2025motionlab}, which edit using text commands. However, text is a coarse condition modality and does not allow interactive spatial control, which is essential for artists who require precision when manipulating movements.

In this paper, we introduce a novel workflow that we call \textit{interactive} generative motion editing. In contrast to the limited and coarse control offered by prior motion editing systems, interactive generative motion editing enables large-scale structural changes to existing animations---such as extension, stitching, and composition---while simultaneously allowing users to manipulate these motion sequences both spatially and temporally.
Moreover, it combines synthesis and editing into a common framework, allowing artists to seamlessly switch between creating and editing animations.

To solve this task, we introduce \textit{scheduled inpainting}, a novel inference-based technique that extends pretrained generative motion models with the ability to edit existing movements---without requiring any additional training and at virtually no additional computational cost. By dynamically blending source motion at inference time, we leverage the generative prior of these models in a controllable way, enabling both large and small edits, while consistently resulting in natural motion.

While simple inpainting~\cite{hmdm, tseng2023edge} and blending~\cite{shafir2024human} have been explored for motion generation, these methods render the inpainted parts of motion completely uneditable and uncontrollable, making them problematic for generative motion editing. Beyond extending naive inpainting, our primary contribution lies in building the first comprehensive framework that solves this novel task.
More specifically, our scheduled inpainting framework comprises three components: a user-controlled schedule that modulates the amount of preservation, a spatiotemporal strength mask controlling the location of the preservation, and a robust representation space optimized through extensive ablations to enable compositing movements from very different global orientations and scales. Although simple, our framework is versatile and supports many applications. By configuring the spatiotemporal mask, artists can alter movements directly and procedurally, as well as stitch, extend, and compose clips, as shown in \figref{fig:teaser} and our accompanying video.

\noindent In summary, our main contributions include:

\begin{itemize}
    \item \textbf{Interactive Generative Motion Editing}: A novel generative workflow that simultaneously allows structurally editing and interactively manipulating existing animations.
    \item \textbf{Scheduled Inpainting}: An inference technique that extends binary inpainting with a user-controllable schedule and a user-defined spatiotemporal mask to enable the first framework capable of interactive generative motion editing.
    \item \textbf{Unified Motion Framework}: A versatile system with animation software integration that solves a wide range of animation tasks, including stitching, extension, and composition, within a single interactive framework.
\end{itemize}

To our knowledge, scheduled inpainting is the first interactive system for generative motion editing. To evaluate its performance, we compare with several adapted baselines, including noise inversion~\cite{song2022denoisingdiffusionimplicitmodels} and noise optimization~\cite{dno}, and ablate key design choices.
We further demonstrate our system's capabilities through a series of results in our supporting video and report qualitative feedback from professional artists.

%% file: content/02_RelatedWork.tex
Seminal work in motion editing introduced layered cubic spline interpolation  \cite{witkin1995motionWarping} and remains the standard in many authoring and game engine software \cite{motionbuilder, kinefx}. While both optimization \cite{gleicher1997motion, lee1999hierarchical} and statistical models \cite{chai2007constraintBasedMotionOptimization, min2009interactiveGenerationHumanAnimation} have been explored to go beyond warping movements, they could not generalize to large datasets, and could not make big changes due to local minima issues.

Recently, advances in deep learning, in particular diffusion models, have enabled training on large datasets that incorporate a vast range of motion into a single model. The first wave of motion diffusion focused on text conditioning \cite{hmdm}, while subsequent efforts led to a vast expansion into different capabilities such as contact-aware generation \cite{ma2024contact}, multi-character interaction \cite{xu2025multi, wang2025leader, ruiz2025mixermdm}, keyframe-centric generation \cite{bae2025less, goel2025generative}, sketch-based conditioning \cite{zhong2025sketch2anim}, and multi-modal conditioning \cite{zhang2025motion, li2025genmo, li2025motion}. \add{Binary spatiotemporal masks over frames and joints have been used as training-time objectives or conditioning for generating new motion~\cite{pinyoanuntapong2025maskcontrol, chen2024mmdm, tessler2024maskedmimic, cohan2024flexible}.} While these models are powerful generators, they are not designed for preserving and editing existing motion.

\subsection{Direct Manipulation Models}
Most related to our work are models focused on interaction such as CondMDI~\cite{cohan2024flexible}, which enable interactive authoring by conditioning on full-body keyframes or joint trajectories. Studer et al.~\cite{studer2024factorized} and V\"ogeli et al.~\cite{vogeli2025implicit} showed that compressing the motion representation spatially and temporally improves constraint satisfaction, allowing for both efficient and precise manipulation. Similarly, Hwang et al.~\cite{hwang2025motion} use a two-stage approach but adopt a slower generative model instead of the efficient neural IK model \cite{protores} used in previous methods. While these models can interactively modify their generated motion, they cannot edit an exemplar motion clip in a structure-preserving manner. \add{A related line of work targets real-time interactive control of motion diffusion models~\cite{shi2024interactive, gou2025control}, but likewise steers online generation rather than editing an existing clip.}

\subsection{Motion Editing} 

\begin{figure*}
    \centering
    \includegraphics[width=\linewidth]{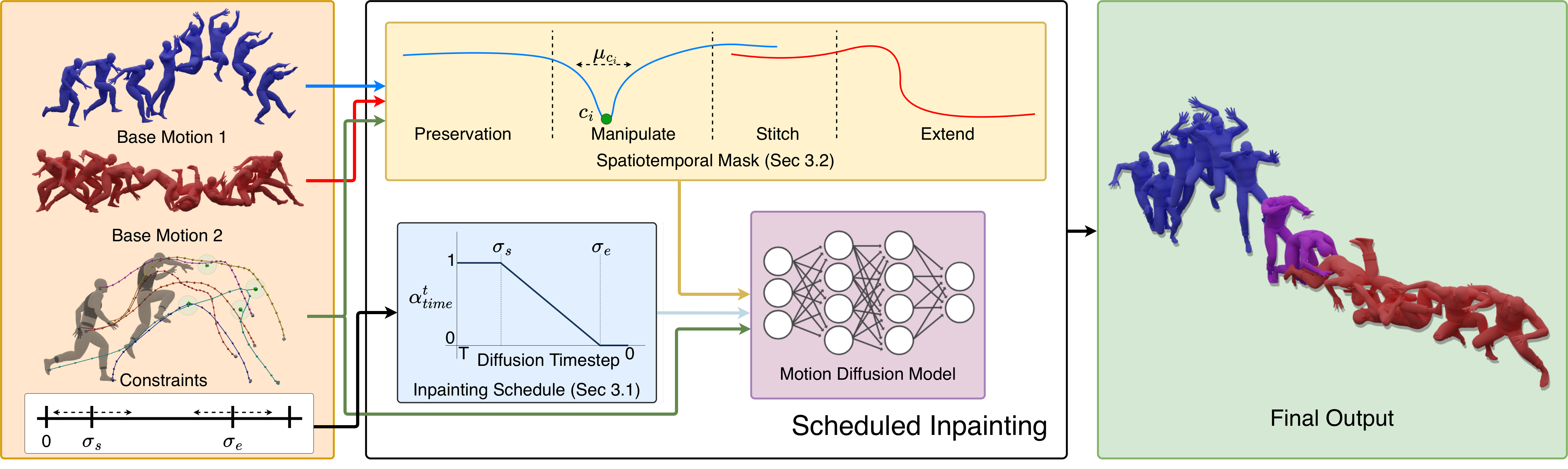}
    \caption{Overview of our scheduled inpainting for interactive generative motion editing. A user can specify which motion should be used (red and blue), as well as control where and how much the motion should be preserved via controls impacting the \remove{spacetime}\add{spatiotemporal} mask, which in turn dictates the schedule. }
    \label{fig:overview}
\end{figure*}

To our knowledge, the only model explicitly designed for motion editing in a direct manipulation setting is the work of Agrawal et al.~\cite{agrawal2024skel}, which trains a deterministic motion editing model (which is limited to smaller datasets). Their technique also requires training a specific model, in contrast to our approach that works at inference with different pretrained diffusion models.

Text-based motion editing has been explored using LLM-extracted keywords to guide motion generation \cite{zhang2023finemogen,huang2024controllable}. Chen et al.~\cite{chen2024pay} then showed that modifying the attention between motion and word tokens could replace, erase, and re-sequence parts of movements. Recently, MotionFix \cite{athanasiou2024motionfix}, STANCE \cite{jiang2025dynamic}, and FineMotion \cite{wu2025finemotion} released motion editing datasets with triplets of original motion, text, and final motion. To improve quality, Jiang et al.~\cite{jiang2025dynamic} explored leveraging unannotated data, Guo et al.~\cite{guo2025motionlab} jointly training for editing and synthesis, while Li et al.~\cite{li2025simmotionedit} explored predicting the similarity between edited and original motion. Although text-based motion editing is useful to make broad changes, it does not provide the fine-grained and direct spatial control artists need in order to achieve their full vision.

\subsection{Inference-based techniques} These were first developed in the image domain working alongside StableDiffusion~\cite{stable_diffusion}. Shi et al.~\cite{shi2024dragdiffusion} use DDIM noise inversion~\cite{song2022denoisingdiffusionimplicitmodels} to recover the latent code for a given image, to then optimize in latent space to match displaced points on the image. Similarly, Liu et al.~\cite{liu2024drag} optimized the added noise to perform edits. In the motion domain, optimizing the initial noise for better control was explored by Karunratanakul et al.~\cite{dno}. While it is possible to edit movements once inverted, the optimization is too costly to enable iterative workflows. Each time an artist needs to edit a different motion, a new offline optimization is required, breaking real-time workflows and preventing fluid interactions. In contrast, our inpainting approach operates at a constant cost and allows for larger changes in real-time.

%% file: content/03_Method.tex
\label{sec:inpainting}

To achieve the full potential of what we call interactive generative motion editing, we need spatial control over the amount of preservation and generation.
While inpainting an example clip during inference is an obvious choice for its efficiency, we observed it completely replaced the generated motion, resulting in unnatural transitions and leaving no flexibility to edit. 

We introduce a user-controllable schedule that modulates the inpainting strength and then augment it with a spatiotemporal mask that provides control over regions of generation and preservation. Note that in order to support direct manipulation, we focus our exposition and results on models such as IBMM~\cite{vogeli2025implicit} and SF-control~\cite{hwang2025motion}.
To achieve consistent movement alignment, we carefully construct a normalization state as a naive approach of inpainting clips from disparate spaces, like a front roll with a back roll, would introduce spurious motion artifacts.

\figref{fig:overview} shows an overview of our method. Formally, \textbf{scheduled inpainting} works with an example motion, which we call base motion $\baseMotion$, with $T$ frames and $J$ joints, together with a set of user-defined constraints $\constraints$. Our inference-based method operates with a diffusion model $\motionModel$, and we refer to the generated motion as $\genMotion{0} = \motionModel (\constraints, \genMotion{t})$ where $\genMotion{t}$ is the noisy motion according to the diffusion timestep $t$. The inpainted motion at timestep $t$ is then given by:

\begin{align}
    \widehat{\genMotion{0}} &= \alpha^t \times \baseMotion + (1 - \alpha^t) \times \genMotion{0} \label{eq:scheduled_inpainting} \\ 
    \text{with } \alpha^t &= \timeMask{t} \times \appMask, \label{eq:mask_construction}
\end{align} 
where a timestep-regulated scalar $\timeMask{t}$ and application mask $\appMask$ modulate the inpainting strength. The input to $\motionModel$ at timestep $t$ is then $\widehat{\genMotion{t}}$ instead of $\genMotion{t}$.

\subsection{Inpainting Schedule}
The temporal scalar $\timeMask{t}$ controls the strength of inpainting through the diffusion process. It is parameterized by two user-controlled variables $\sigma_s$ and $\sigma_e$  that define the start and the end of the schedule respectively. It is defined as:
\begin{equation}
    \timeMask{t} = \begin{cases}
        1 & \sigma_s < t \\
        \frac{t - \sigma_e}{\sigma_s - \sigma_e} & \sigma_e \leq t \leq \sigma_s \\
        0 & t < \sigma_e
    \end{cases}.
\end{equation}
Hence, for noise levels larger than $\sigma_s$, the generated motion is completely overwritten by the base motion $\baseMotion$ and conversely, for noise levels lower than $\sigma_e$, no inpainting is applied.
Between the two noise levels, we linearly interpolate between the base and generated motions.
By manipulating $\sigma_{s/e}$, a user can set the strength of preservation versus generation during inference. Note that we show the effect of different schedules $\sigma_{s/e}$ in \secref{sec:ablation_inpainting_schedule}.

\subsection{Spatiotemporal Mask}
\label{sec:mask}

\begin{figure}
    \centering
    \includegraphics[width=\linewidth]{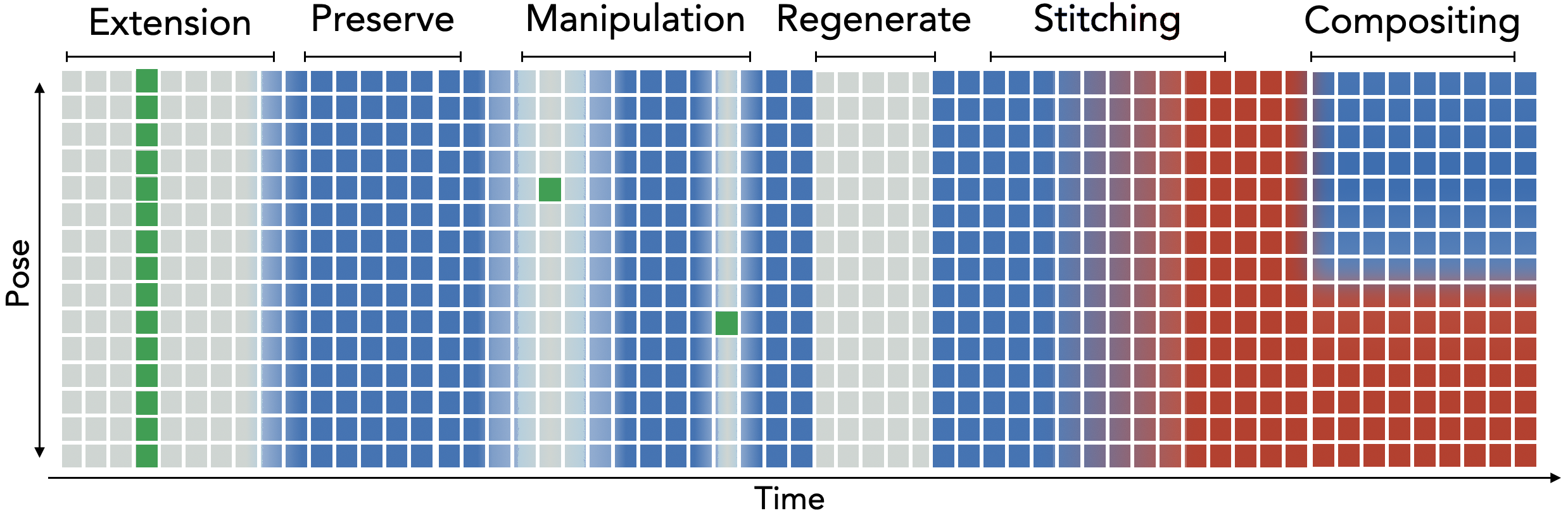}
    \caption{Conceptual overview of our spacetime mask $\appMask$ for the different motion editing applications \add{using two different motions (red and blue)}: direct manipulation with full-body and sparse constraints (\remove{pink}\add{green}), extending (grey), as well as stitching\remove{ (blue and red)}, and compositing\remove{ (red and green)} clips.}
    \label{fig:motion_editing_mask_combined}
\end{figure}

The spatiotemporal mask $\appMask \in \mathbb{R}^{T \times J}$ in \myeqref{eq:mask_construction} weighs how much the base motion should be preserved, both per joint $j$ and per frame $t$. It can be constructed differently to support various applications as shown in \figref{fig:motion_editing_mask_combined}. For simplicity of exposition, we focus here on constructing a direct manipulation mask and defer to \secref{sec:applications} for remaining applications.

\begin{figure}
    \centering
    \includegraphics[width=\linewidth, trim={ 0 0.5cm 0 0.5cm}, clip]{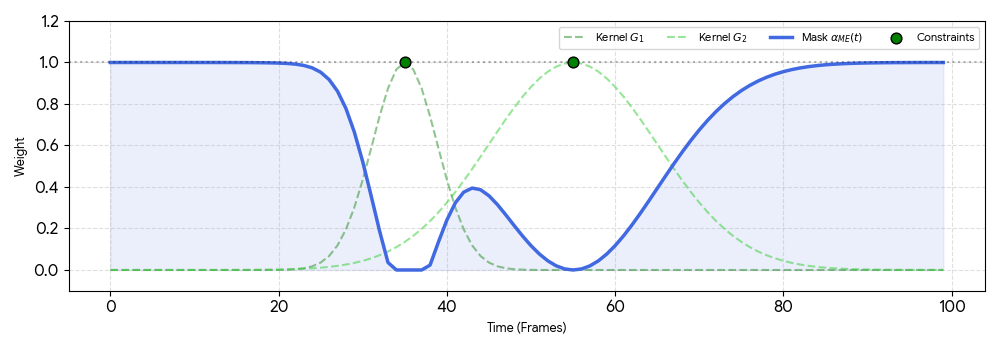}
    \caption{An example of the spatiotemporal mask $\alpha_{mask}$ \add{for a single joint} \remove{being constructed} when a user manipulates two constraints at frames $35$ and $55$. The motion is allowed to be generated in these areas, and is steered towards preservation in other parts of the timeline. }
    \label{fig:motion_editing_mask}
\end{figure}

By default, scheduled inpainting completely preserves the base motion and our spatiotemporal mask is set to 1. To manipulate the motion at a user-specified joint location called constraint $c_i \in \constraints$, we modify the mask to reduce the inpainting weight in the neighborhood of the constraint, with user-controllable kernel functions, as shown in \figref{fig:motion_editing_mask}. Given constraints $\constraints$, we construct the mask with the following equation:

\begin{align}
\label{eq:appDirectManip}
    G(j, t; c_i) &= \text{exp}\left(-\frac{(t - \text{FrameIdx}(c_i))^2}{\mu_{c_i}}\right)\\
    \alpha_{mask}(j,t) &= \text{max}\left(1 - \sum_{{c_i} \in \constraints} G(t; c_i), 0\right),
\end{align}
where the user can control the influence width via the scalar $\mu_{c_i}$, and the $\text{max}$ function ensures the mask is never negative.

\subsection{Inpainting Space}\label{sec:inpainting_space}
Naively mixing two unrelated motions can lead to unwanted artifacts.
For example, mixing two clips that start at the same position but walk away in opposite directions results in an average motion that is mostly static near the starting position. Or mixing two clips that are of very different speeds introduces discontinuities in accelerations.

Therefore, when inpainting $\baseMotion$ onto $\genMotion{0}$, it is important to align and normalize both sequences to avoid such artifacts. First, we align each motion such that the first frame is at the origin, and the direction from the first frame to the last is along the positive x-axis.
Additionally, we individually normalize both sequences to $0$ mean and $1$ variance to ensure that the two motions are on the same scale during blending.
To avoid discontinuities when extending or mixing animations, we represent the root using differential coordinates, together with root-relative positions and orientations for the remaining joints.

Note that our representation space is independent of the pretrained model's representation space and normalization. As $\motionModel$ generates clean motion at each diffusion timestep, we transform the generation into our representation space for blending before inverting to the model's native representation. 
Hence, the pretrained model is completely ignorant of our alignment and normalization, which are applied in addition to any of its own data processing. 

We show ablations of the different components of our inpainting space in \secref{sec:ablation_motion_alignment} and Appendix B.

%% file: content/application.tex
By constructing $\appMask$ in \myeqref{eq:mask_construction} differently, our scheduled inpainting technique can enable a number of generative motion editing tasks, as shown in \figref{fig:motion_editing_mask_combined}---each time enabling sampling variations due to the underlying generative motion. In \secref{sec:mask}, we described forming a direct manipulation mask and now turn to describing generative extension, stitching, compositing, and retiming.

\subsection{Motion Extension}
A given motion sequence $\baseMotion$ of length $T$ can be expanded by $T_e$ frames at any position by setting the strength mask $\alpha_{extend} \in \mathbb{R}^{(T + T_e) \times J}$ to $0$ for the interval $[t_s,t_e)$, which defines the region where the new motion is generated:
\begin{equation}
    \label{eq:appMaskExtension}
    \alpha_{extend}[t, :] = \begin{cases}
        0 &  t_s \leq t < t_e \\
        1 & \text{otherwise} \\ 
    \end{cases}.
\end{equation}
Note that $\baseMotion$ must be padded with additional frames to align with $\alpha_{extend}$ in \myeqref{eq:scheduled_inpainting}.

\begin{figure}
    \centering
    \begin{subfigure}{0.5\linewidth}
        \includegraphics[width=0.99\linewidth, trim={4cm 4cm 4cm 4cm}, clip]{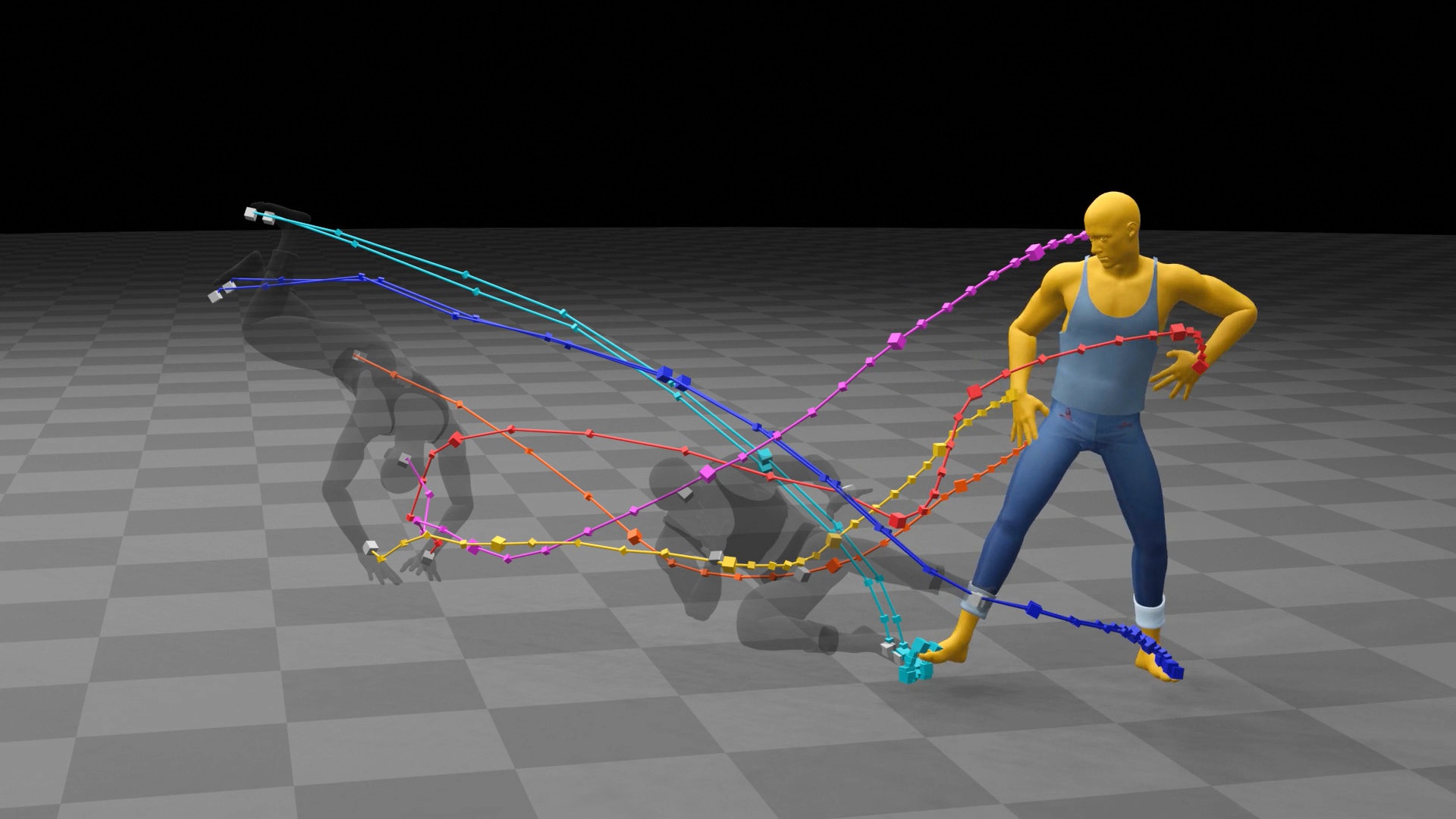}
    \end{subfigure}%
    \begin{subfigure}{0.5\linewidth}
        \includegraphics[width=0.99\linewidth, trim={4cm 4cm 4cm 4cm}, clip]{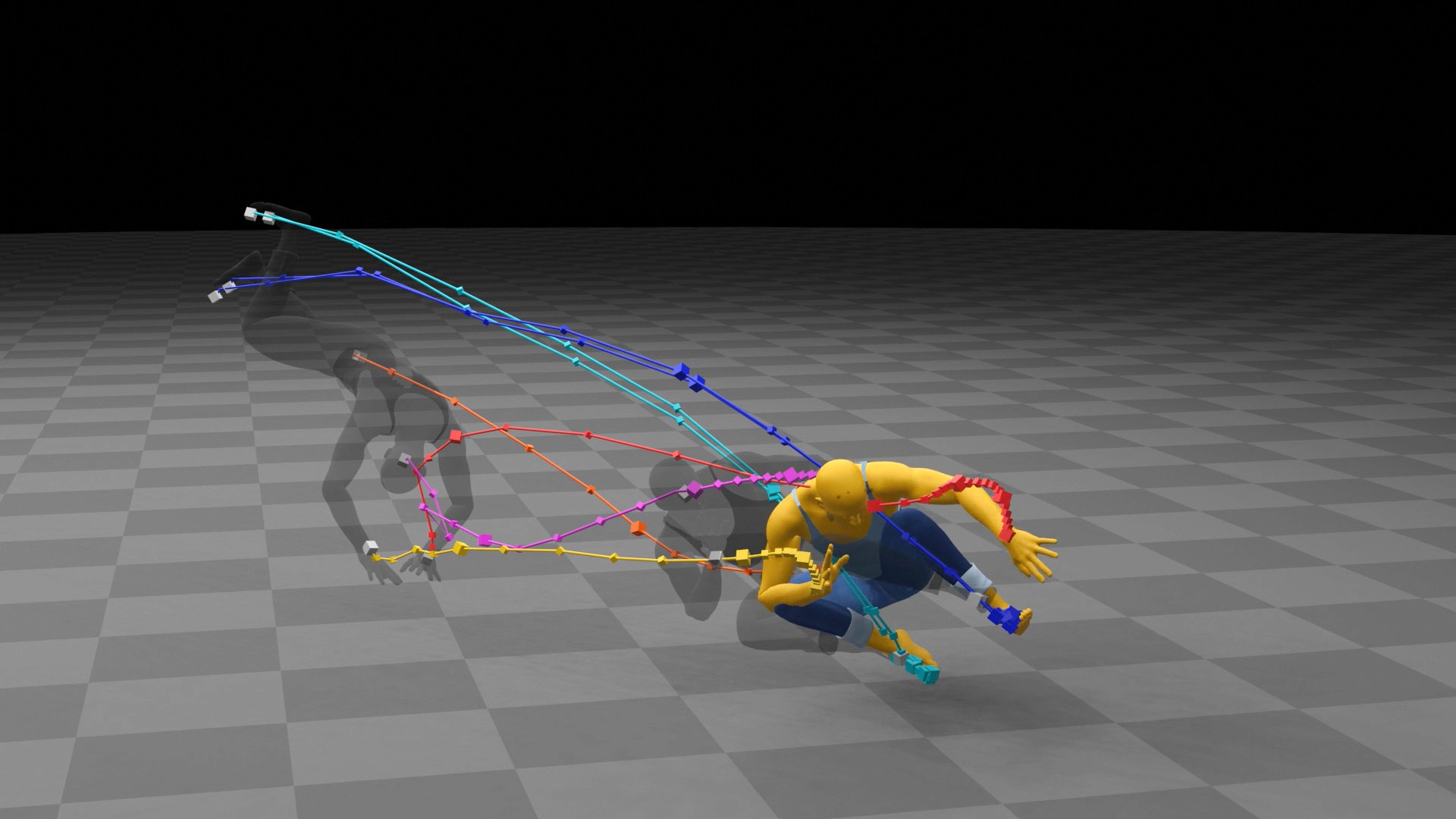}
    \end{subfigure}
    \begin{subfigure}{0.5\linewidth}
        \includegraphics[width=0.99\linewidth, trim={4cm 4cm 4cm 4cm}, clip]{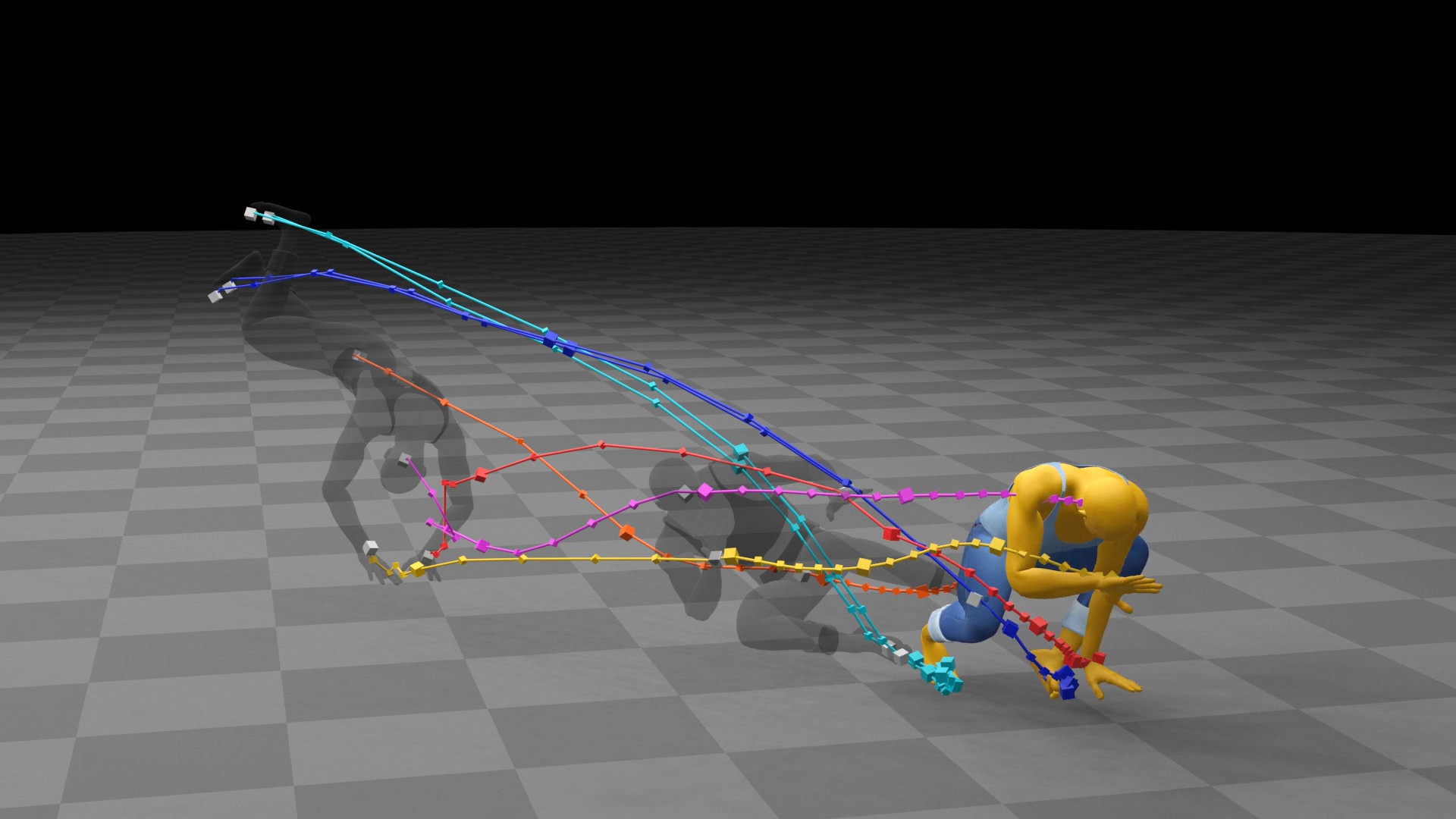}
    \end{subfigure}%
    \begin{subfigure}{0.5\linewidth}
        \includegraphics[width=0.99\linewidth, trim={4cm 4cm 4cm 4cm}, clip]{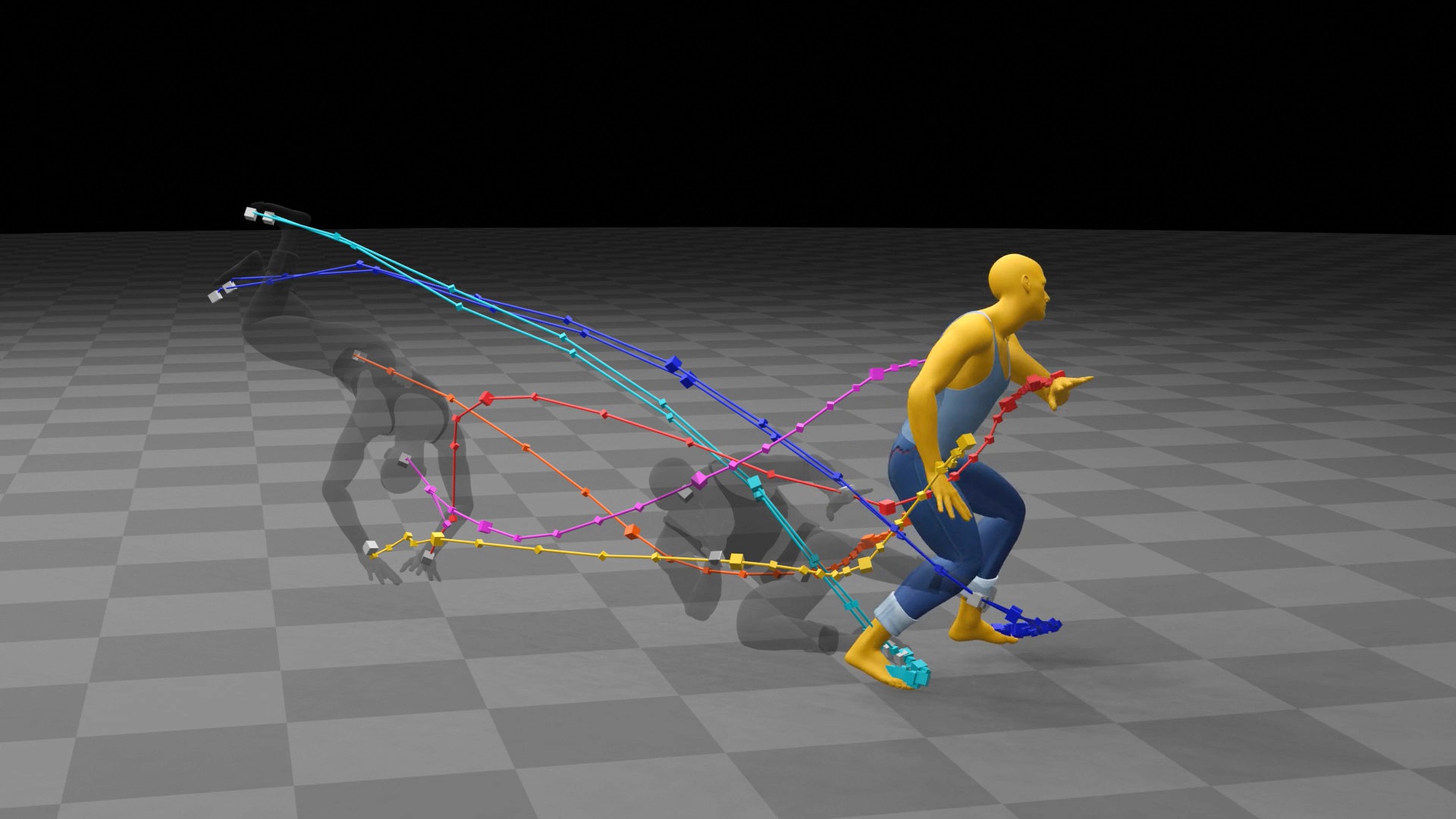}
    \end{subfigure}
    \caption{Generative motion editing allows sampling different motions, such as different extensions of a rolling motion. Each curve represents the motion path for an end-effector joint and the two black silhouettes represent the start and end of the base motion.}
    \label{fig:roll_extending_end}
\end{figure}

In \figref{fig:roll_extending_end}, we extend a roll beyond the second silhouette, which is where the original motion ends ($t_s = T$ and $t_e = T+T_e$).
Because we use a generative model, we can sample various extensions for the same base clip, as demonstrated by the different end poses. Furthermore, since the preserved $\baseMotion$ is inpainted as opposed to input as constraints, the entire motion remains fully editable via direct manipulation described in \myeqref{eq:appDirectManip}.

In the case where the generated motion is in the middle of the base motion, our inpainting space of root velocities and root-relative transforms for the remaining joints allows the second half of the motion to continue from the end of the generated sequence without any discontinuity in root trajectory.
Please refer to our video for these results.

\subsection{Motion Stitching}
Another common animation task involves creating transitions between independent motion clips. Unfortunately, naively stitching clips creates discontinuity or sliding artifacts, requiring manual clean-up. Given two motion clips $\baseMotion^0$ and $\baseMotion^1$ of lengths $T_0$ and $T_1$, respectively, we can set the spatiotemporal mask to $\alpha_{mask} \in \mathbb{R}^{(T_0 + T_{gen} + T_1) \times J}$, such that we inpaint the first frames $T_0$ in the sequence with the first base motion $\baseMotion^0$, smoothly transitioning to generating frames $T_{gen}$ in the middle, and finally smoothly transitioning to inpainting the last frames $T_1$ with the second base motion $\baseMotion^1$.

\begin{figure}[t]
    \centering
    \begin{subfigure}{\linewidth}
        \begin{subfigure}{0.5\linewidth}
            \includegraphics[width=0.99\linewidth]{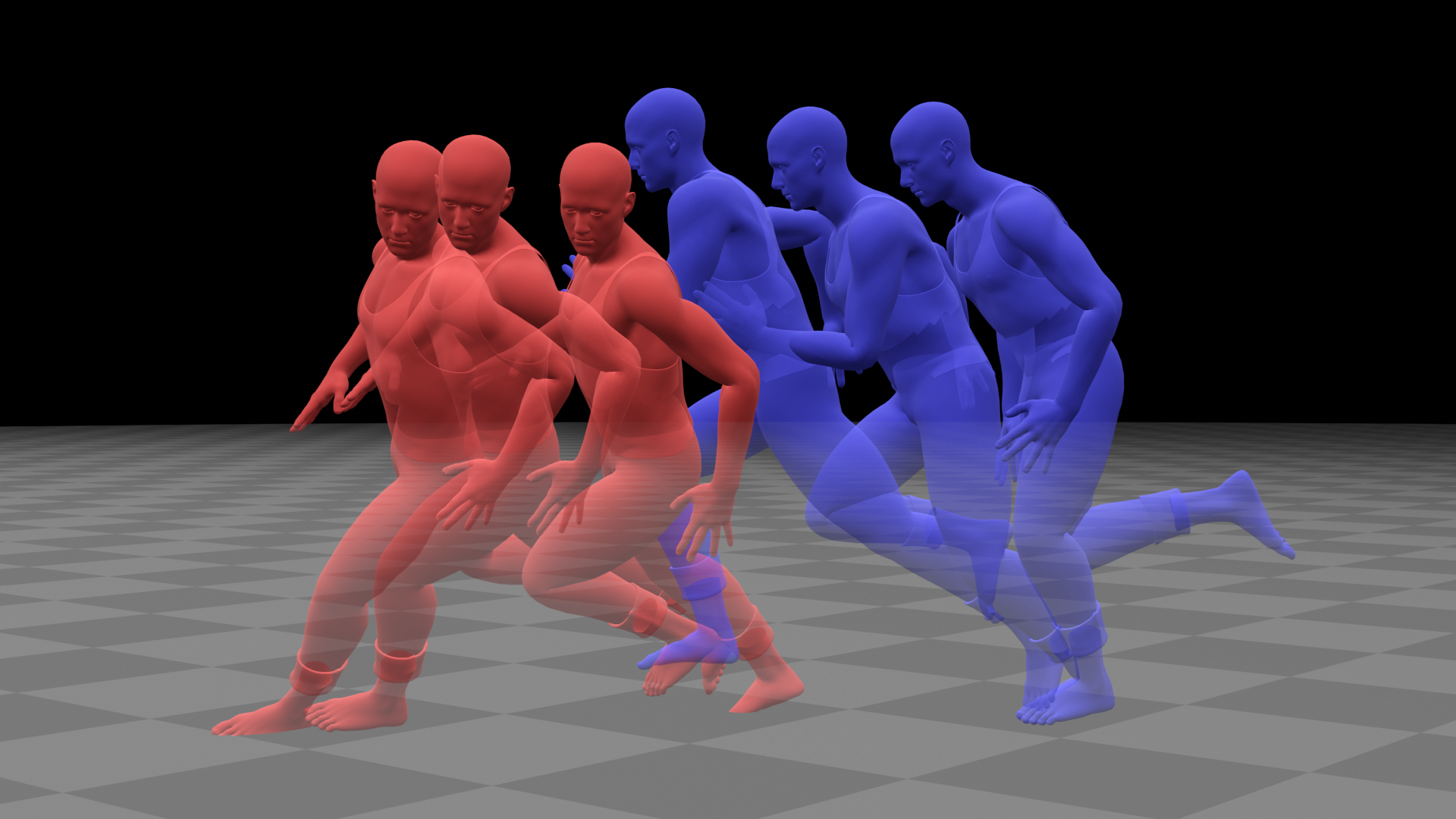}
        \end{subfigure}%
        \begin{subfigure}{0.5\linewidth}
            \includegraphics[width=0.99\linewidth]{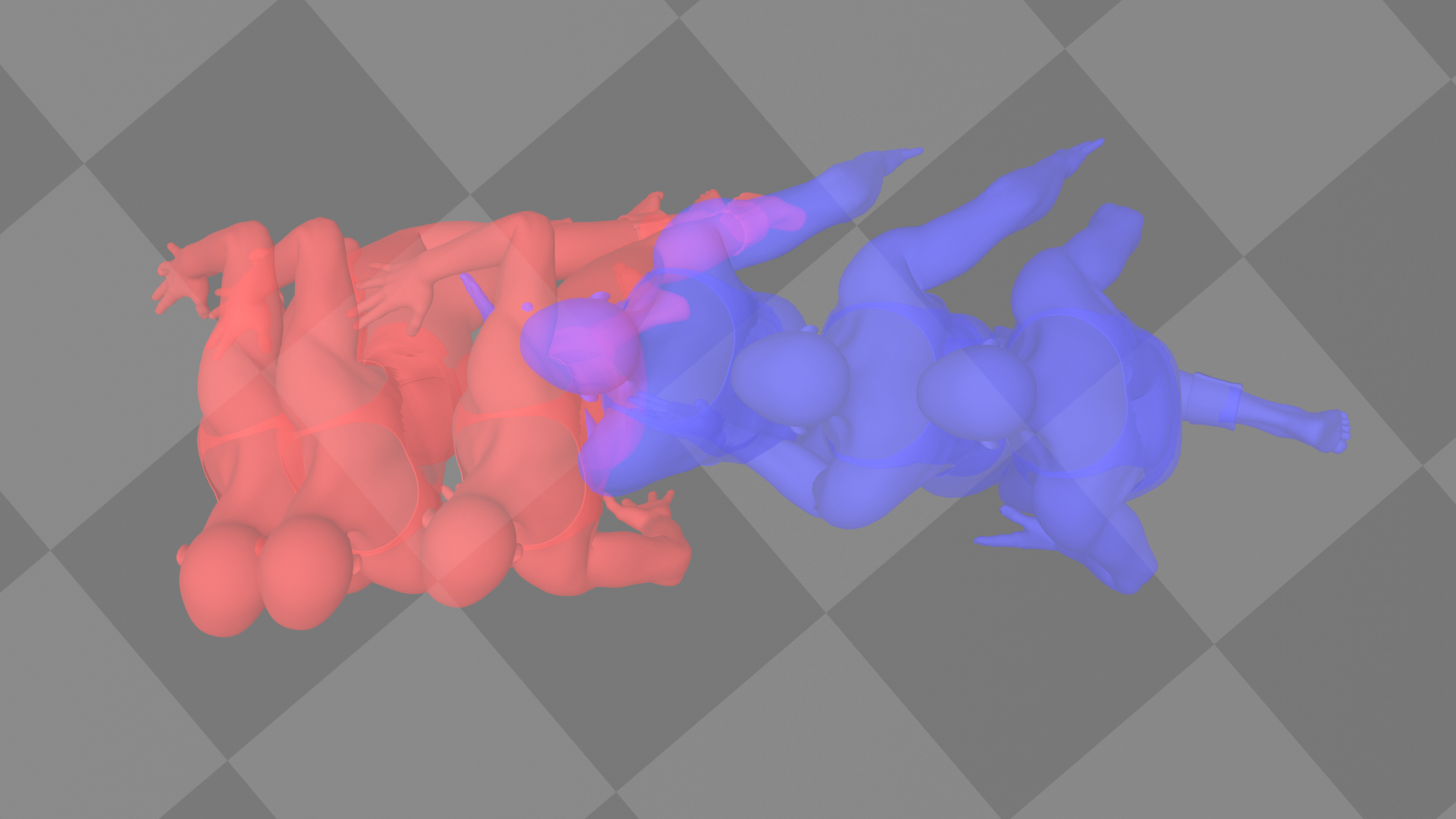}
        \end{subfigure}    
        \caption{Naive stitch}\label{fig:motion_stitching_compare_naive-naive}
    \end{subfigure}
    \begin{subfigure}{\linewidth}
        \begin{subfigure}{0.5\linewidth}
            \includegraphics[width=0.99\linewidth]{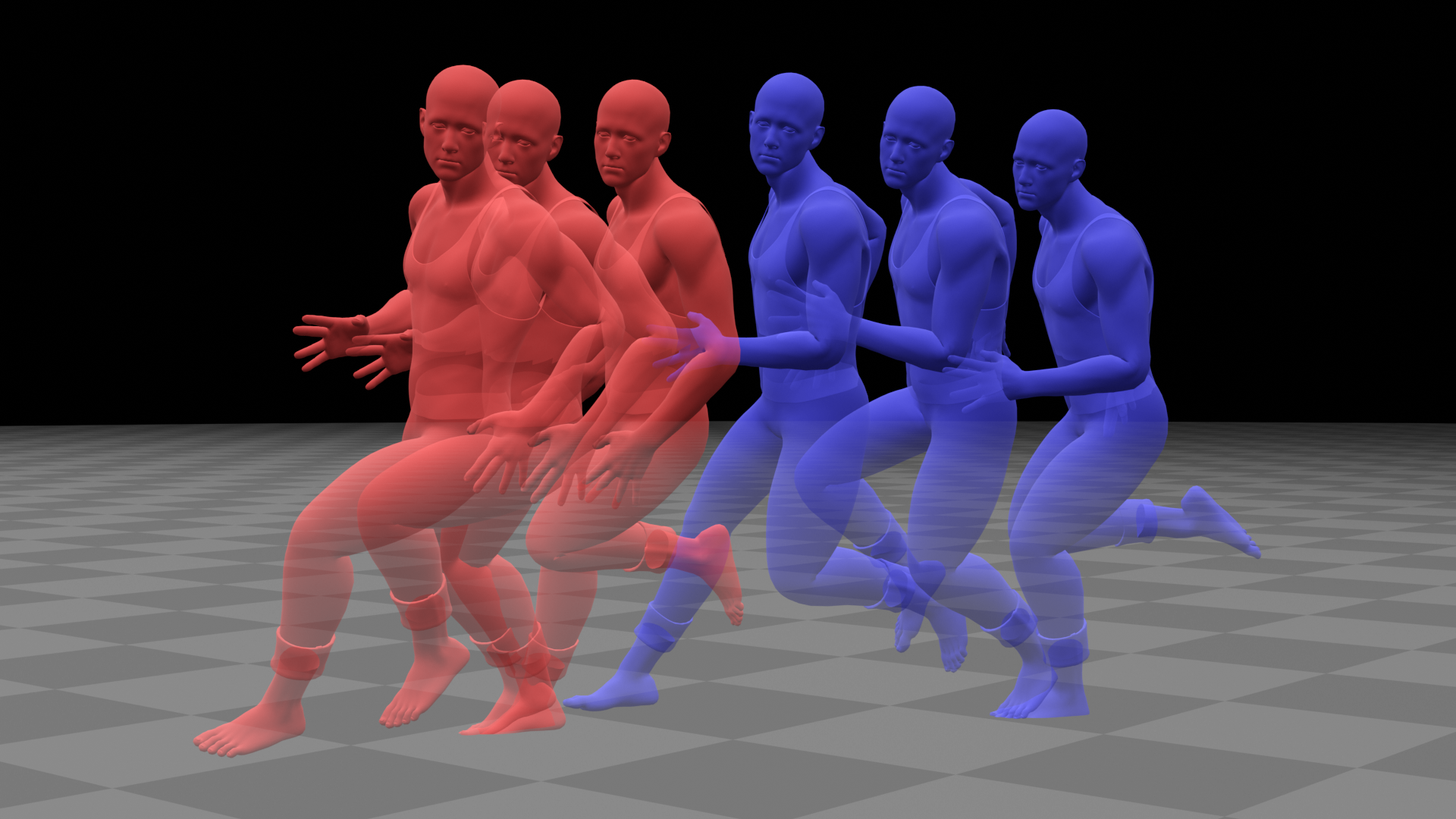}
        \end{subfigure}%
        \begin{subfigure}{0.5\linewidth}
            \includegraphics[width=0.99\linewidth]{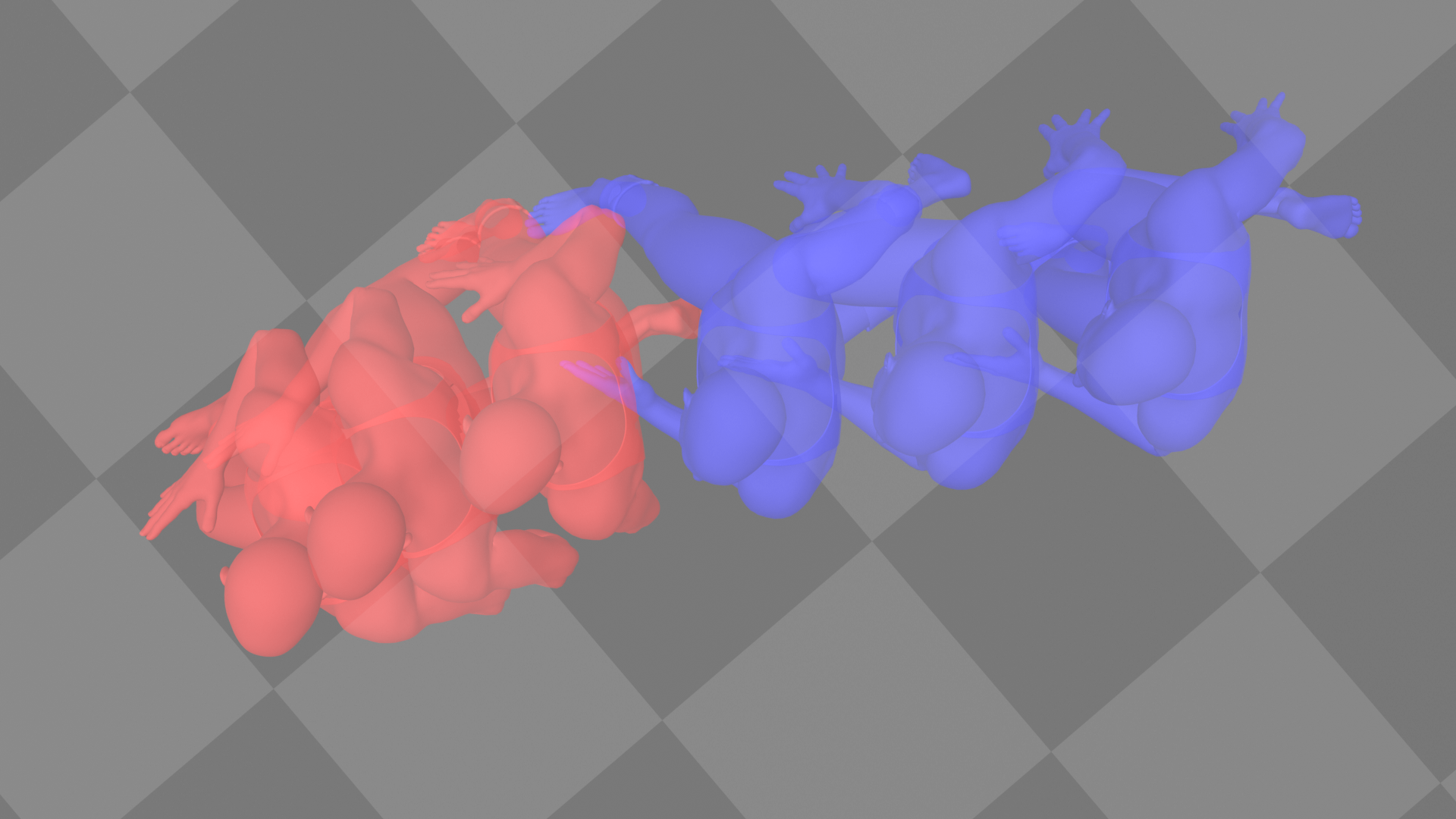}
        \end{subfigure} 
        \caption{Scheduled Inpainting}\label{fig:motion_stitching_compare_naive-inpainting}
    \end{subfigure}
    \caption{Comparison of stitching clips naively to our scheduled inpainting method. Top down view shows that scheduled inpainting resolves discontinuity artifacts seen in naive stitching.}
    \label{fig:motion_stitching_compare_naive}
\end{figure}

Note that stitching clips is a good example of where our normalization is essential for quality results. We show in \figref{fig:motion_stitching_compare_naive-naive} that even for manually aligned clips there is a sharp change in poses and root orientations at the boundary. In contrast, scheduled inpainting modifies the boundary between the clips to generate a realistic transition, as shown in \figref{fig:motion_stitching_compare_naive-inpainting} and our accompanying video. 

\subsection{Motion Compositing}
We define motion composition as applying different animation clips to different body parts within the same sequence. Motion is highly correlated, and combining different movements easily results in unrealistic movements. In contrast, composing clips with our method results in natural motion due to the underlying generative prior.

Note that this task is similar to motion stitching, but instead we blend different motions along the joint dimension instead of time. Consequently, the compositing mask can be constructed similarly to a stitching clip, but by varying the mask across joints rather than frames, where a set of joints such as the upper body get inpainted with a first clip, while the second set of joints are inpainted from a second clip.

\subsection{Adapting animation cycles}
Much production motion data consists of animation cycles that are short sequences of repetitive actions such as walking and running that can be looped seamlessly to create arbitrarily long animations.
Such cycles are often used for characters in games or for crowds in movies.
However, motion controllers typically override the root motion to adapt in real-time to different speeds and environments, introducing foot sliding.

Using scheduled inpainting, we can adapt an animation cycle for arbitrary trajectories by either fully generating the root joint or constraining it to a user-drawn curve while inpainting the remaining joints with the animation cycle.
The resulting motion naturally follows the desired path, but without foot sliding artifacts.
As shown in our accompanying video, the artist can interactively update the path with resulting motion updated in real-time. The generated motion follows the specified path but can maintain highly stylized animation cycles such as \textit{Zombie}, \textit{Chicken}, and \textit{Monkey}.

\subsection{Retiming Motion}

\begin{figure*}[t]
    \centering
    \begin{subfigure}{0.33\linewidth}
        \includegraphics[width=0.99\linewidth, trim={7cm 3cm 0cm 1cm}, clip]{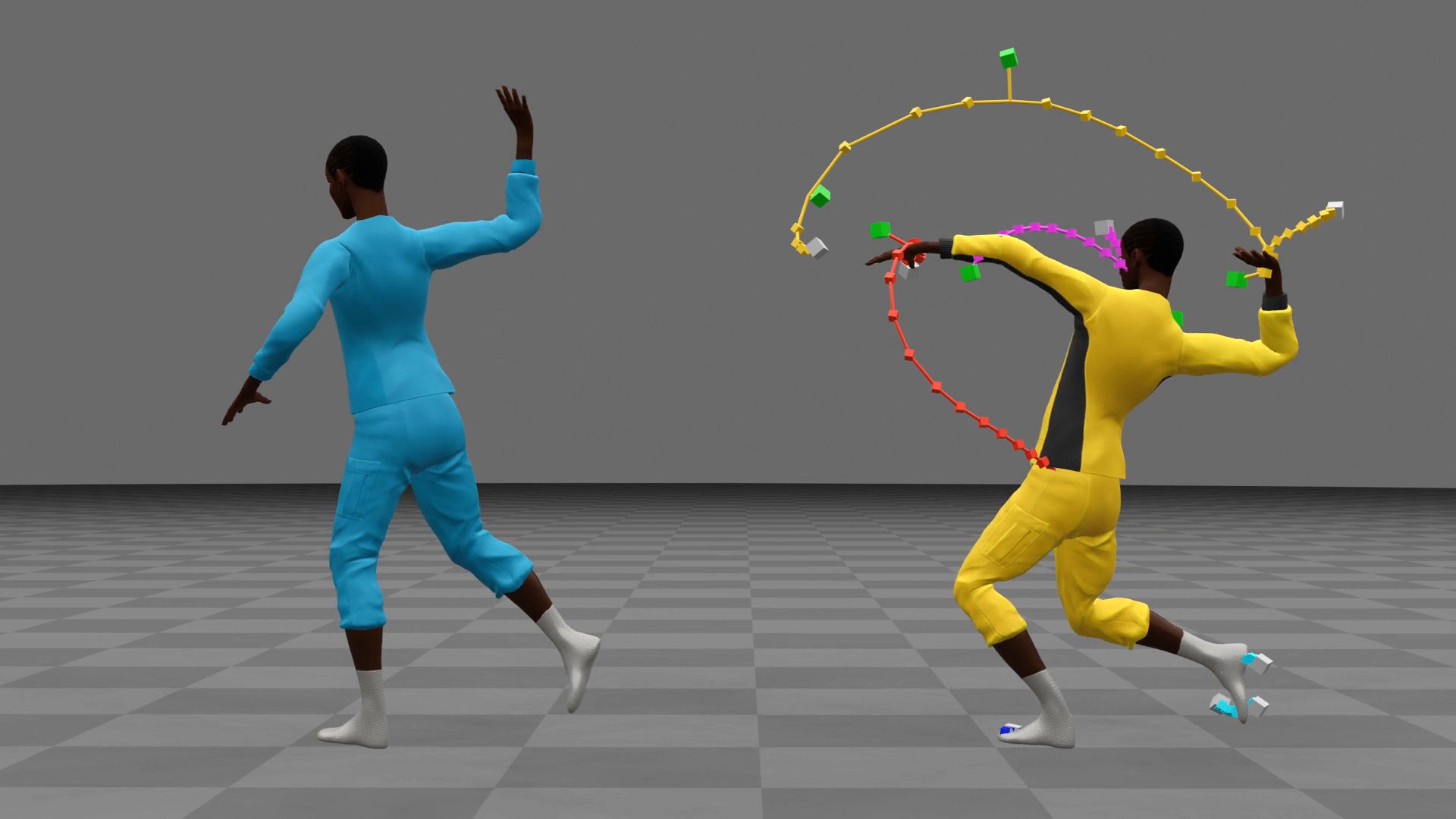}
        \caption{AMASS dataset \cite{AMASS:ICCV:2019}}\label{fig:motion_editing_multiple_datasets-amass}
    \end{subfigure}%
    \begin{subfigure}{0.33\linewidth}
        \includegraphics[width=0.99\linewidth, trim={7cm 3cm 0cm 1cm}, clip]{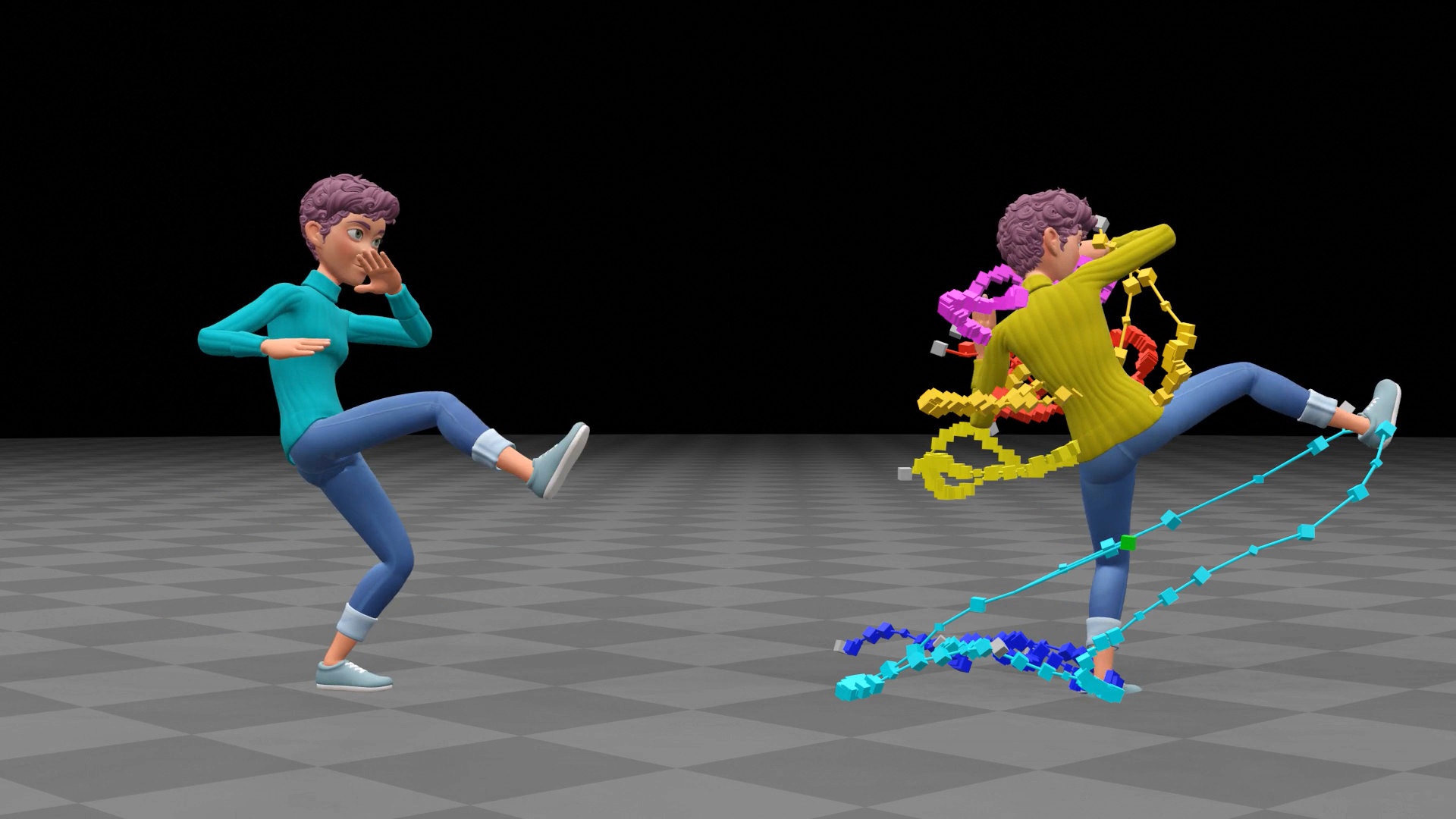}
        \caption{LaFan1 dataset \cite{robustInbetweening}}\label{fig:motion_editing_multiple_datasets-lafan}
    \end{subfigure}%
    \begin{subfigure}{0.33\linewidth}
        \includegraphics[width=0.99\linewidth, trim={7cm 3cm 0cm 1cm}, clip]{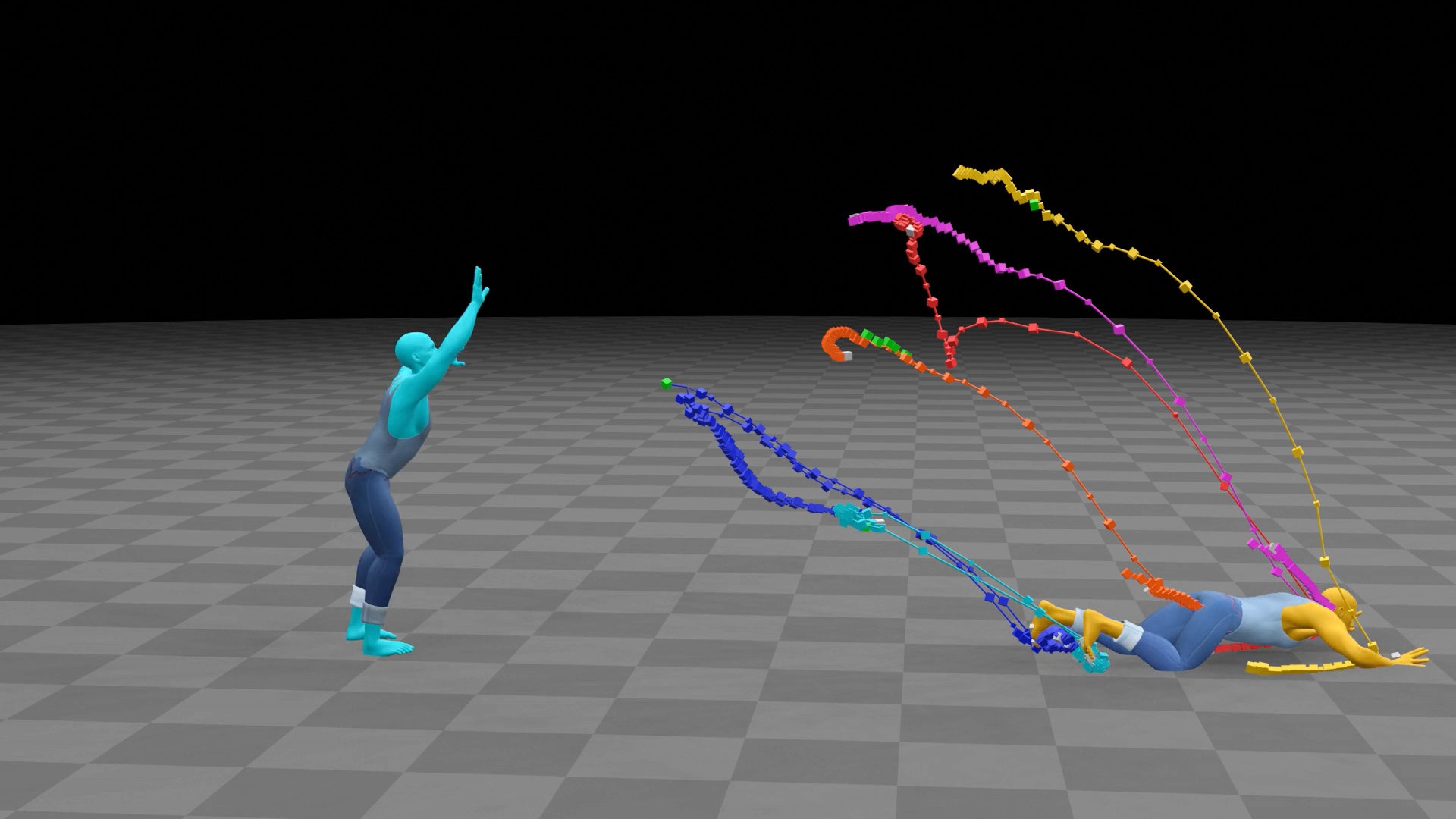}
        \caption{Internal dataset}\label{fig:motion_editing_multiple_datasets-ilm}
    \end{subfigure}
    \caption{Motion editing examples using point-based constraints (green) for AMASS (left), LaFan1 (middle) and internal (right) datasets. Original motion is shown in cyan and edited motion in yellow.}
    \label{fig:motion_editing_multiple_datasets}
\end{figure*}

In practice, motion is often adapted to new pacing, dialogue, or music.
A simple approach is to fit a parametric curve to the original animation and resample it with the new timing.
However, this only warps the motion and cannot accommodate structural changes when timings change significantly. By setting $\alpha_{mask} = 0.8\cdot\mathbb{1}_{J \times T}$, our scheduled inpainting method uses the warped motion as the base motion to recover natural movement through generation. We refer to our supporting video for retiming results.

%% file: content/04_Evaluation.tex
\section{Evaluation \& Results}
\label{sec:evaluation}

In this section, we validate the benefits of scheduled inpainting, as well as each of its components. After introducing alternative generative motion editing baselines---which did not demonstrate direct manipulation before---we show multiple comparisons, as well as showcase the universality of our method by applying it to different models and datasets. 

\subsection{Baselines}
Our first baseline is MotionLab~\cite{guo2025motionlab}, trained on the MotionFix~\cite{athanasiou2024motionfix} dataset. Note that although it supports spatial constraints, it requires dense trajectory conditioning and is not interactive, requiring $0.88$ secs for a single sample.

As interactive generative motion editing has not been demonstrated before, we developed three additional baselines for holistic evaluation.
First, we introduce \textit{CondEditor}, which adapts CondMDI~\cite{cohan2024flexible} using the motion editing training scheme proposed by \cite{agrawal2024skel} and trains on synthetic triplets of original, edited, and final motion.

Our last two baselines, DNO~\cite{dno} and noise-inversion are inference techniques that recover noise for an example motion via optimization and DDIM-inversion~\cite{song2022denoisingdiffusionimplicitmodels}, respectively.
Note that in our experiments, noise-inversion could not run in real-time, as it requires at least $400$ denoising steps for high-fidelity motion reconstruction in comparison to only $25$ steps with our method. See Appendix D for further details.

\subsection{Results \& Comparisons}
\label{sec:evaluation_motion_editing}

We now report using scheduled inpainting with different models and datasets and compare with the previously described baselines. Note that our general workflow for these results consisted in setting full-body constraints to make large edits, followed by individual position and orientation constraints to refine the motion curves. 

\subsubsection{Results}
\begin{figure}
    \centering
    \begin{subfigure}{0.5\linewidth}
        \centering
        \includegraphics[width=0.99\linewidth]{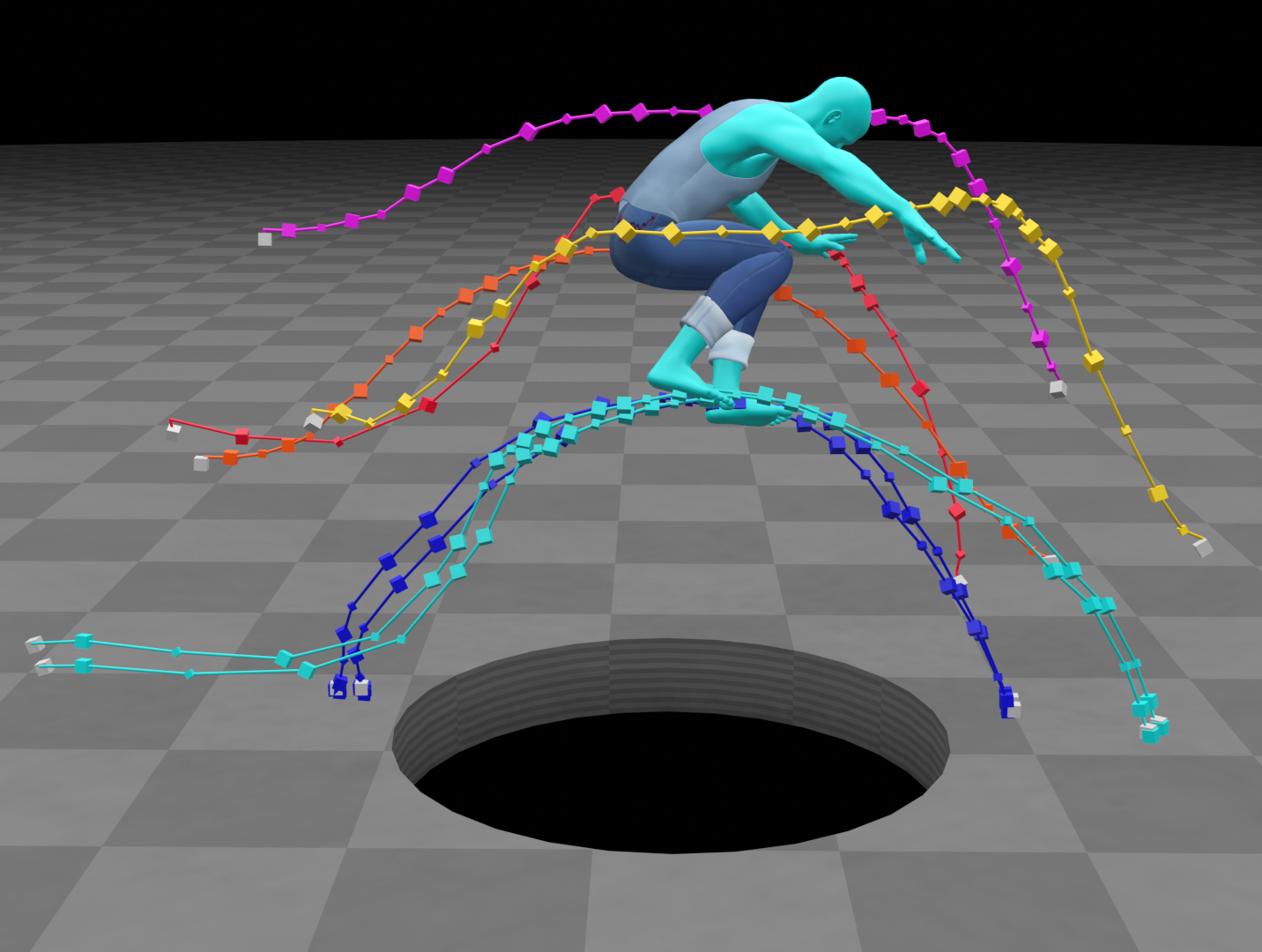}
    \end{subfigure}%
    \begin{subfigure}{0.5\linewidth}
        \centering
        \includegraphics[width=0.99\linewidth]{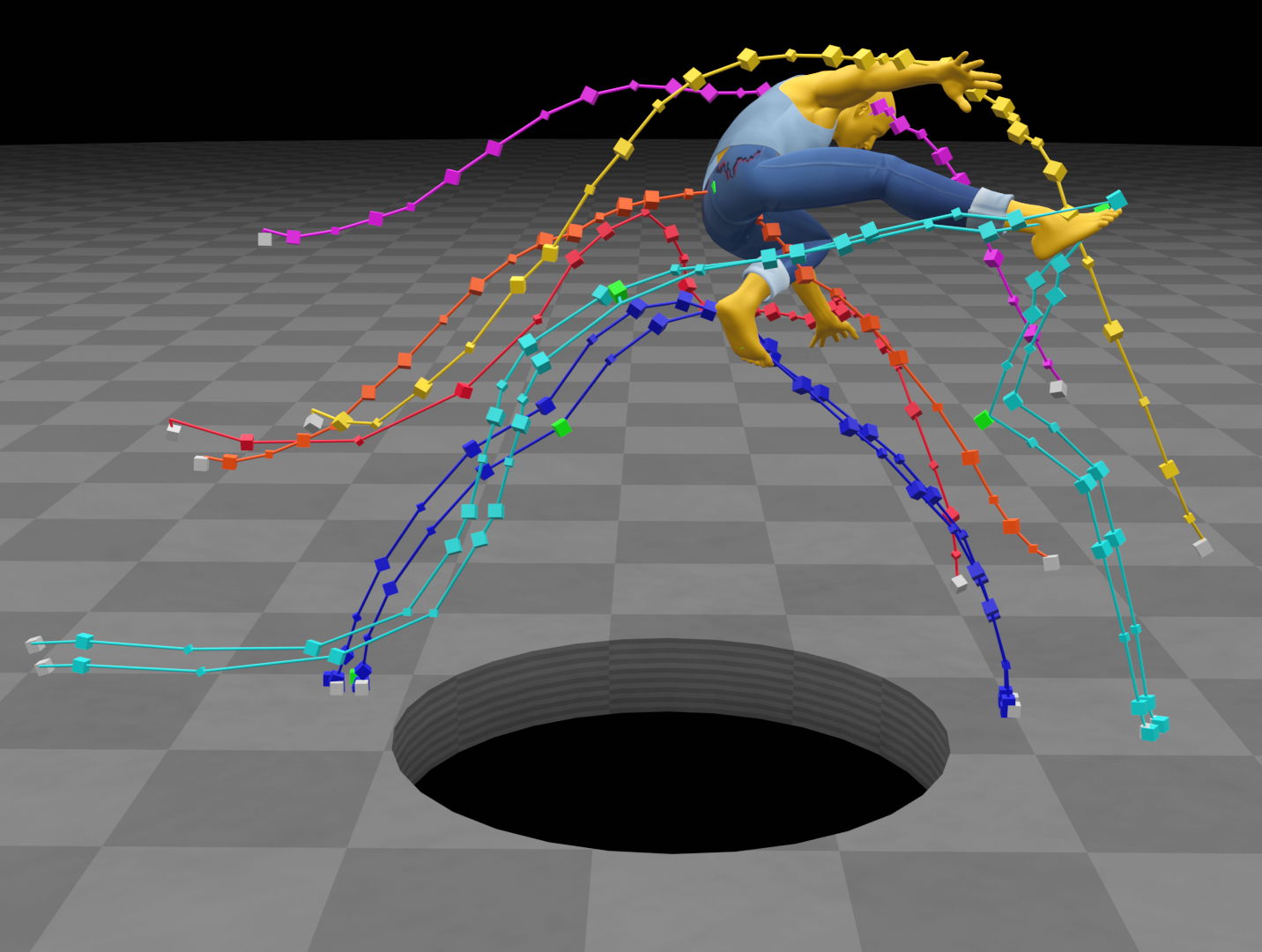}
    \end{subfigure}
    \caption{With scheduled inpainting, we can directly manipulate a few joints to modify a jumping motion (left) to include a natural kick in midair (right). The manipulated constraints are highlighted in green and also shown in our video. }\label{fig:motion_editing_with_sparse_constraints}
\end{figure}

First, we use the IBMM~\cite{vogeli2025implicit} motion model to edit a jumping motion to include a kick in midair, shown in \figref{fig:motion_editing_with_sparse_constraints}. Notice in our accompanying video how without our scheduled inpainting, the generated motion simply runs between the keyframes. As IBMM is a synthesis model, it is unaware of the jump without introducing additional constraints. In contrast, with our method the user only needs to manipulate a single kicking pose to obtain a coherent and natural midair kicking motion.

Additionally, we show results of using scheduled inpainting with two interactive models and three different datasets. In \figref{fig:motion_editing_multiple_datasets-amass}, we use the SF-control model \cite{hwang2025motion} to exaggerate a throwing motion clip from AMASS~\cite{AMASS:ICCV:2019}, by moving the hand constraint backwards. Similarly, \figref{fig:motion_editing_multiple_datasets-lafan} illustrates a kick from the LaFan1 dataset \cite{robustInbetweening}, edited using IBMM to add body twist and height. Finally, we use IBMM to edit a 'falling when stretching' motion from our internal dataset, in order to change the timing of the fall and the final pose.

\subsubsection{Trained baselines}

\begin{figure*}[ht]
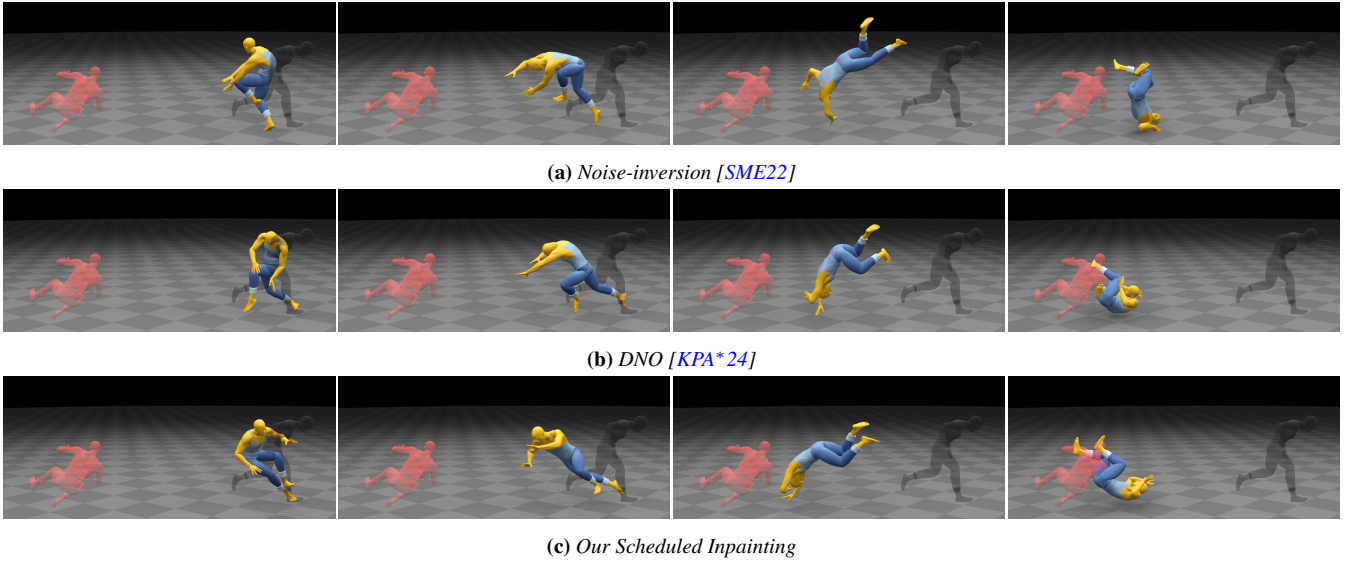

    \centering
    \begin{subfigure}{\linewidth}
        \begin{subfigure}{0.25\linewidth}
            \centering
            \includegraphics[width=0.99\linewidth, trim={1cm 5cm 1cm 5cm}, clip]{figures_brighter/motion_roll/motion_roll_compare_baseline/inversion_0.png}
        \end{subfigure}%
        \begin{subfigure}{0.25\linewidth}
            \centering
            \includegraphics[width=0.99\linewidth, trim={1cm 5cm 1cm 5cm}, clip]{figures_brighter/motion_roll/motion_roll_compare_baseline/inversion_1.png}
        \end{subfigure}%
        \begin{subfigure}{0.25\linewidth}
            \centering
            \includegraphics[width=0.99\linewidth, trim={1cm 5cm 1cm 5cm}, clip]{figures_brighter/motion_roll/motion_roll_compare_baseline/inversion_2.png}
        \end{subfigure}%
        \begin{subfigure}{0.25\linewidth}
            \centering
            \includegraphics[width=0.99\linewidth, trim={1cm 5cm 1cm 5cm}, clip]{figures_brighter/motion_roll/motion_roll_compare_baseline/inversion_3.png}
        \end{subfigure}
        \subcaption{Noise-inversion \cite{song2022denoisingdiffusionimplicitmodels}}
    \end{subfigure}

    \begin{subfigure}{\linewidth}
        \begin{subfigure}{0.25\linewidth}
            \centering
            \includegraphics[width=0.99\linewidth, trim={1cm 5cm 1cm 5cm}, clip]{figures_brighter/motion_roll/motion_roll_compare_baseline/optimization_0.png}
        \end{subfigure}%
        \begin{subfigure}{0.25\linewidth}
            \centering
            \includegraphics[width=0.99\linewidth, trim={1cm 5cm 1cm 5cm}, clip]{figures_brighter/motion_roll/motion_roll_compare_baseline/optimization_1.png}
        \end{subfigure}%
        \begin{subfigure}{0.25\linewidth}
            \centering
            \includegraphics[width=0.99\linewidth, trim={1cm 5cm 1cm 5cm}, clip]{figures_brighter/motion_roll/motion_roll_compare_baseline/optimization_2.png}
        \end{subfigure}%
        \begin{subfigure}{0.25\linewidth}
            \centering
            \includegraphics[width=0.99\linewidth, trim={1cm 5cm 1cm 5cm}, clip]{figures_brighter/motion_roll/motion_roll_compare_baseline/optimization_3.png}
        \end{subfigure}
        \subcaption{DNO \cite{dno}}
    \end{subfigure}

    \begin{subfigure}{\linewidth}
        \begin{subfigure}{0.25\linewidth}
            \centering
            \includegraphics[width=0.99\linewidth, trim={1cm 5cm 1cm 5cm}, clip]{figures_brighter/motion_roll/motion_roll_compare_baseline/inpainting_0.png}
        \end{subfigure}%
        \begin{subfigure}{0.25\linewidth} 
            \centering
            \includegraphics[width=0.99\linewidth, trim={1cm 5cm 1cm 5cm}, clip]{figures_brighter/motion_roll/motion_roll_compare_baseline/inpainting_1.png}
        \end{subfigure}%
        \begin{subfigure}{0.25\linewidth}
            \centering
            \includegraphics[width=0.99\linewidth, trim={1cm 5cm 1cm 5cm}, clip]{figures_brighter/motion_roll/motion_roll_compare_baseline/inpainting_2.png}
        \end{subfigure}%
        \begin{subfigure}{0.25\linewidth}
            \centering
            \includegraphics[width=0.99\linewidth, trim={1cm 5cm 1cm 5cm}, clip]{figures_brighter/motion_roll/motion_roll_compare_baseline/inpainting_3.png}
        \end{subfigure}
        \subcaption{Our Scheduled Inpainting}
    \end{subfigure}
    
    \caption{We edit the end frame (shown in red) of a rolling forward clip to go behind the starting frame (shown in black). Each row shows the output from noise-inversion \cite{song2022denoisingdiffusionimplicitmodels} (top), DNO \cite{dno} (middle), and our scheduled inpainting method (bottom) at three different moments or frames of the motion. While noise-inversion and DNO give an unnatural turn that is too fast in the first two frames, our method generates a realistic turn before the roll.}
    \label{fig:motion_roll_baseline_comparison}
\end{figure*}

\begin{figure}
    \centering
    \begin{subfigure}{\linewidth}
        \begin{subfigure}{0.5\linewidth}
        \includegraphics[width=0.99\linewidth]{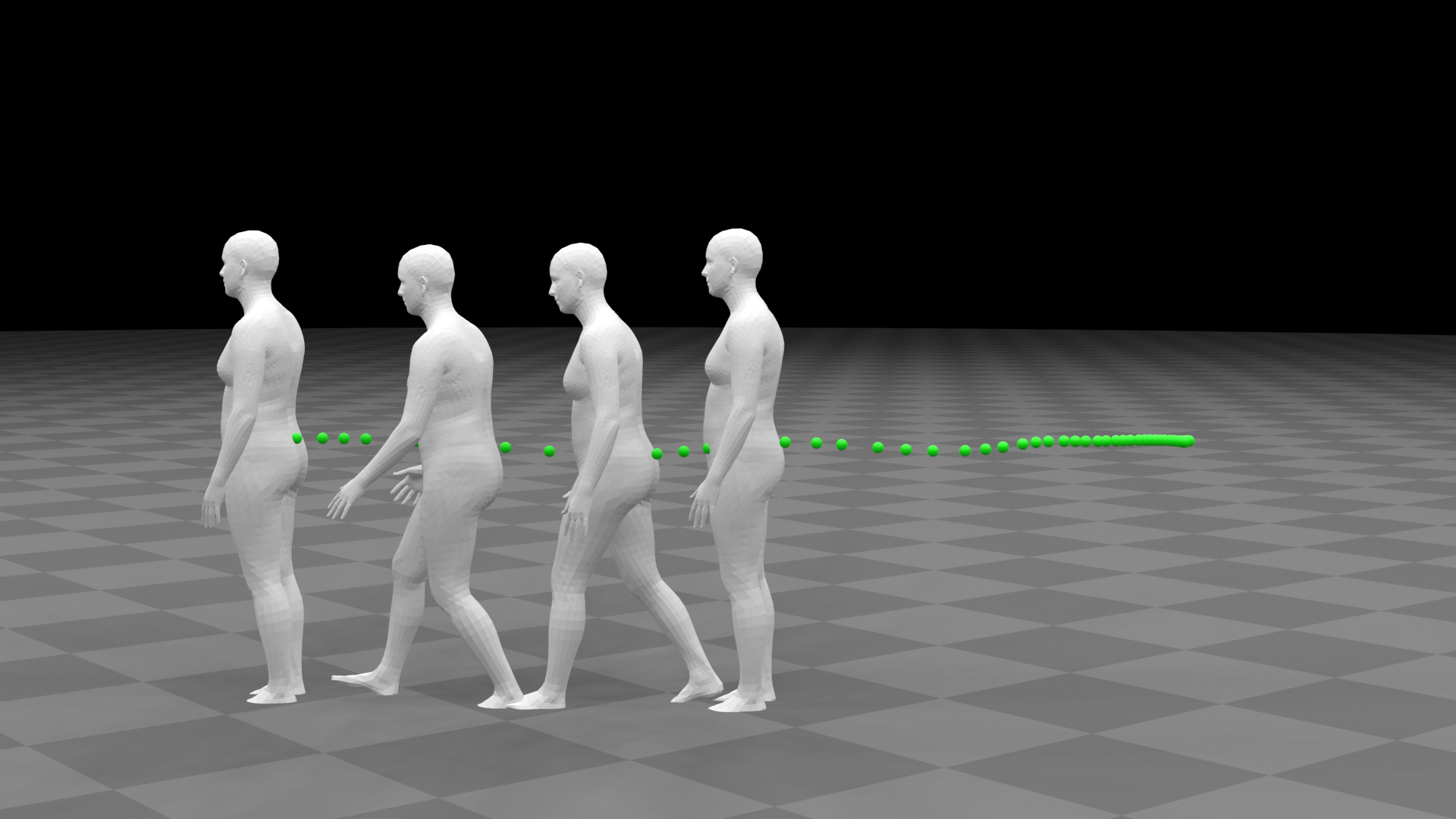}
        \end{subfigure}%
        \begin{subfigure}{0.5\linewidth}
        \includegraphics[width=0.99\linewidth]{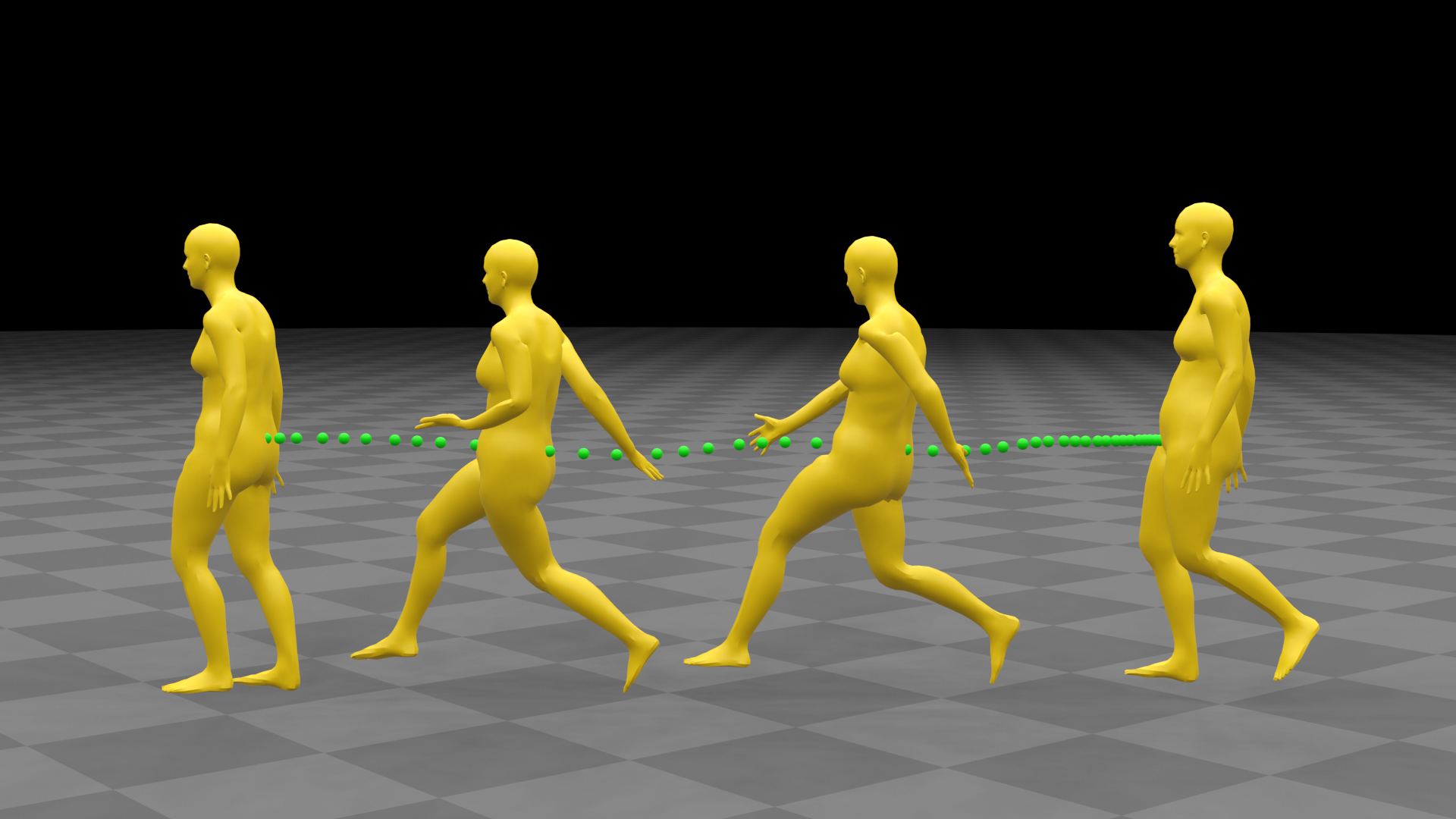}
        \end{subfigure}
        \caption{Editing a forward walk to walk backward}\label{fig:comparison_motion_lab_move_backwards}
    \end{subfigure}        
    \begin{subfigure}{\linewidth}
        \begin{subfigure}{0.5\linewidth}
        \includegraphics[width=0.99\linewidth]{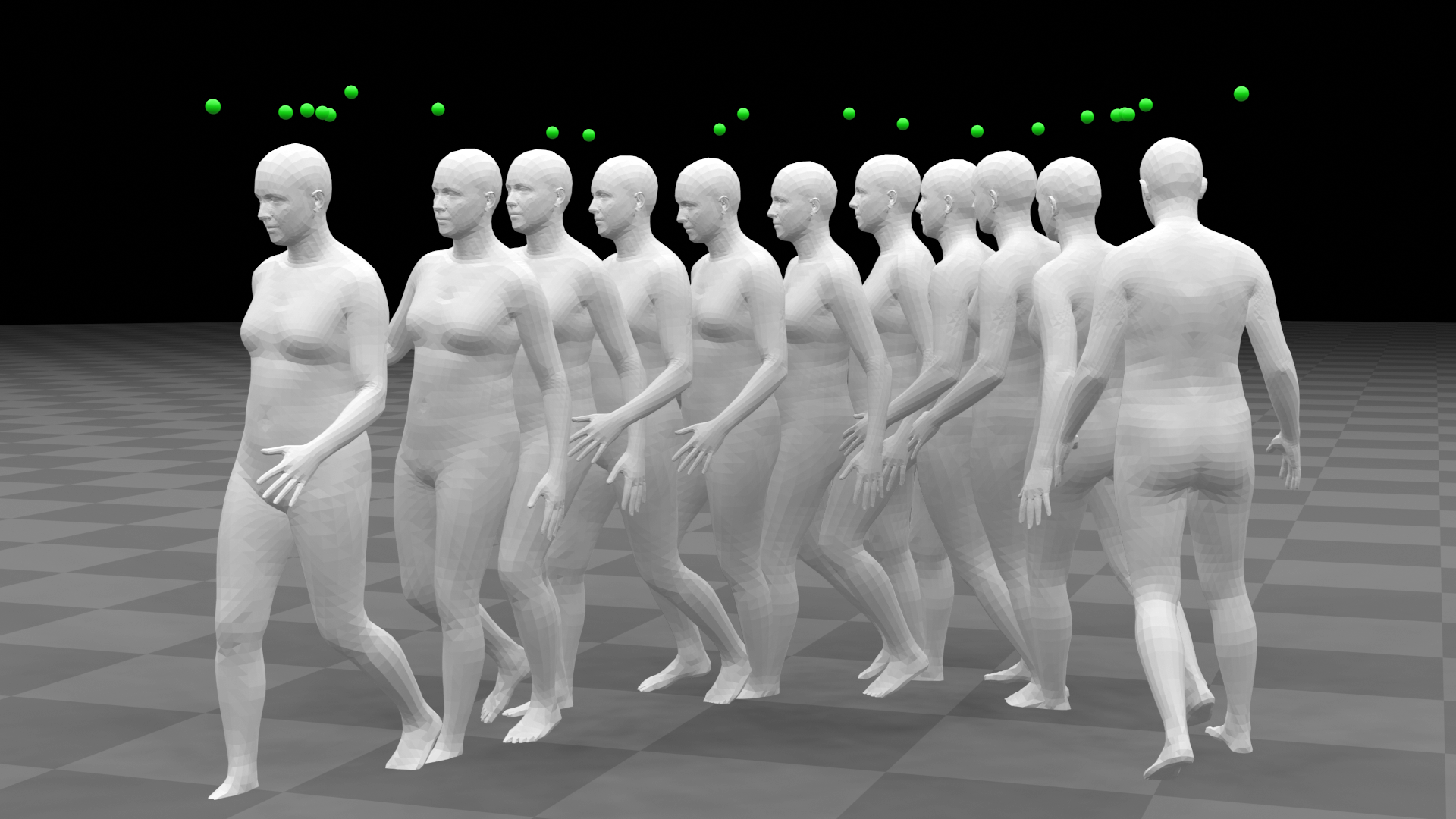}
        \end{subfigure}%
        \begin{subfigure}{0.5\linewidth}
        \includegraphics[width=0.99\linewidth]{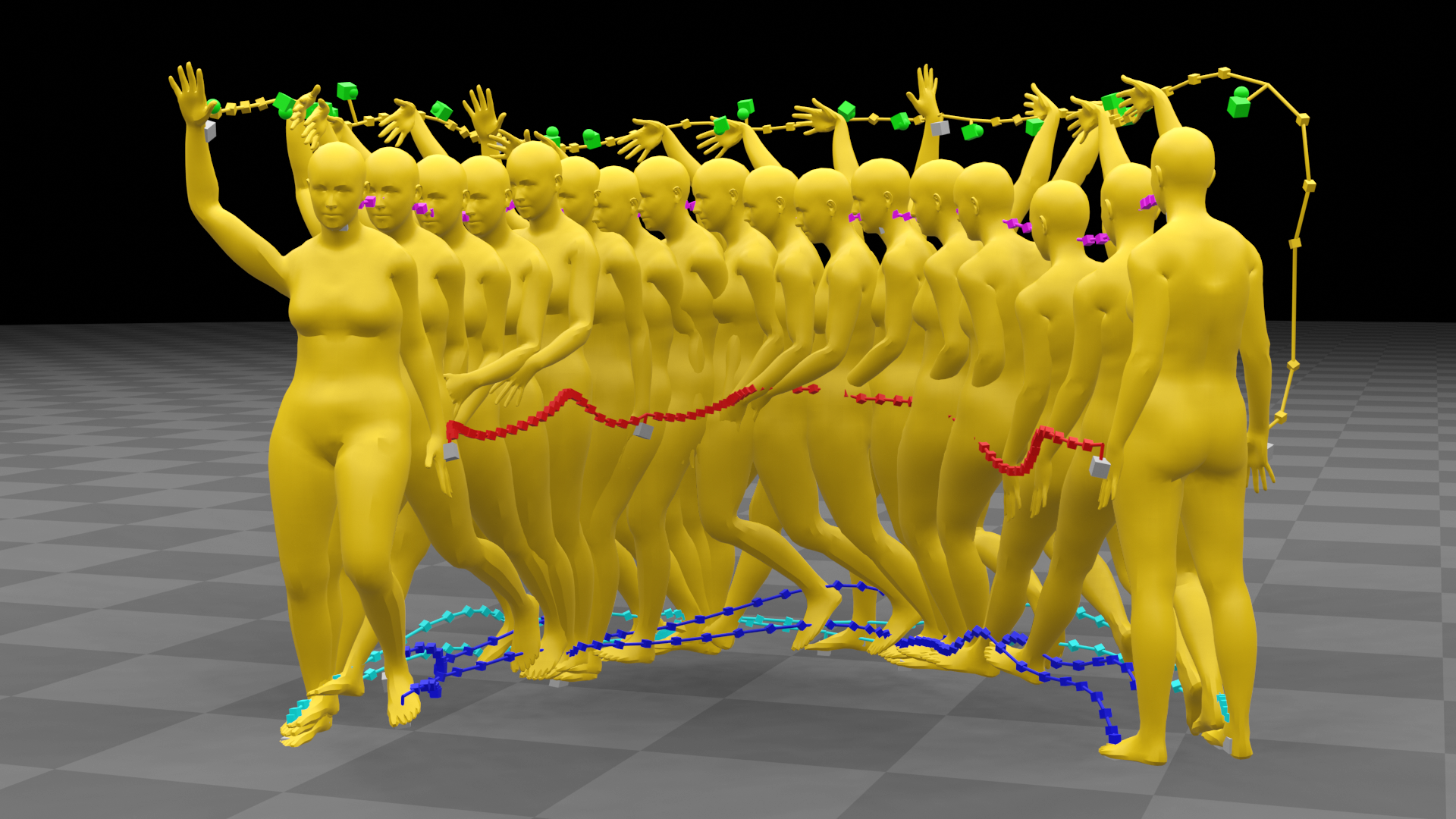}
        \end{subfigure}
        \caption{Editing to lift the right hand with sparse constraints}\label{fig:comparison_motion_lab_raise_hand}
    \end{subfigure}   
    \caption{ 
    Comparison of scheduled inpainting (right) to MotionLab~\cite{guo2025motionlab} (left). MotionLab cannot extend the walking distance (a) and ignores sparse constraints (green) (b), while our approach follows them accurately.}
    \label{fig:comparison_motion_lab}
\end{figure}

In \figref{fig:comparison_motion_lab}, we show a comparison with the MotionLab~\cite{guo2025motionlab} model. While MotionLab can match hip trajectories when inside the learned distribution, it does not generalize well when stretching outside. For example, taking a walking forward motion from the MotionFix validation set and moving the hip trajectory further back results in walking backward only halfway, as shown in \figref{fig:comparison_motion_lab_move_backwards}. This problem is even more prominent when using sparse constraints, instead of dense trajectories, as shown in green in \figref{fig:comparison_motion_lab_raise_hand}, where we edit the arm position. MotionLab completely ignores the constraints while the SF-control model using our method successfully generates the required edits. 

Lastly, in our video we show an example of direct manipulation with CondEditor. We can observe that it struggles to follow constraints and cannot perform structural changes as the deformation becomes larger. In contrast, our method maintains both constraint precision and natural results throughout the editing process.

\subsubsection{Inference-based baselines}

Scheduled inpainting, thanks to our alignment and normalization (see \secref{sec:inpainting_space}), allows users to make large edits in real-time, in cases where even offline noise inversion baselines fail. For example, consider changing a rolling forward motion to a rolling backward motion, as shown in \figref{fig:motion_roll_baseline_comparison} and our video. We can see that both DNO and noise-inversion result in many artifacts and unnatural motion. Since the target motion is quite different from the original, the recovered noise of the inversion is no longer compatible with the target motion, causing artifacts such as unnatural poses and interpenetration during takeoff and landing. In particular, we can see that the backward turn happens too quickly with the character hovering, resulting in self-penetration artifacts. In contrast, our method results in natural motion, including taking more realistic timing to turn.

\subsubsection{Quantitative Analysis}

\tabref{tab:randomized_final_frame_edit} reports the impact of randomly displacing the final frame of a sequence by up to one meter horizontally when using IBMM with and without our scheduled inpainting.
To evaluate the results, we measure foot-sliding and root-relative reconstruction error of non-root joints, reported as L2 position error (L2P) and L2 rotation error (L2R), to quantify the motion naturalness and similarity to base motion.
Without displacement, our method matches the foot-sliding performance of IBMM while significantly improving base motion preservation.
When the final frame is displaced, our method's foot-sliding increases marginally (by ${\sim}1$ mm/frame) because the constraint of preserving the original trajectory conflicts with the edit.
Despite this, our approach consistently outperforms IBMM in maintaining the integrity of the original motion.

These results demonstrate that our method effectively executes edits while simultaneously preserving the original motion’s characteristics and naturalness.

\begin{table}
    \caption{Foot sliding ($m/frame$) and root-relative reconstruction metrics when randomly displacing the final frame of the animation sequence on our internal dataset.}
    \centering
    \begin{tabular}{l >{\centering\arraybackslash}p{0.20\linewidth} ccc}
        \toprule
        Task & Edited Final Frame? & Foot-sliding $\downarrow$ & L2P $\downarrow$ & L2R $\downarrow$ \\
        \midrule
        GT   & -  & 0.0116 & - & - \\
        IBMM & No  & \textbf{0.0101} & 0.0122 & 0.1484 \\
        Ours & No  & 0.0105 & \textbf{0.0034} & \textbf{0.0886} \\
        \hdashline
        IBMM & Yes & 0.0114 & 0.0131 & 0.1566 \\
        Ours & Yes & 0.0125 & \textbf{0.0063} & \textbf{0.1051} \\
        \bottomrule
    \end{tabular}
    \label{tab:randomized_final_frame_edit}
\end{table}

\subsection{Ablations}

In this section, we ablate our design choices. We quantitatively evaluate the impact of the inpainting schedule on the reconstruction error. Additionally, we qualitatively assess how our motion alignment and influence radius affect the generated motion.

\subsubsection{Inpainting Schedule}
\label{sec:ablation_inpainting_schedule}

\tabref{tab:recon_error_inpainting_schedules_ilm} compares different inpainting schedules on reconstructing $\baseMotion$ on our internal and LaFan1~\cite{robustInbetweening} datasets.
Compared to no inpainting ($\sigma_s = \sigma_e = 1000$), applying light inpainting with $\sigma_s = 1000$ and $\sigma_e = 700$ significantly reduces both reconstruction errors.
This stems from the fact that inpainting at high noise levels dictates the general structure of the motion and accounts for the large differences between the synthesized and base motion.
For stronger inpainting schedules, the reconstruction error improves linearly with the strongest schedule of $\sigma_s = 300$ and $\sigma_e = 50$ achieving the best metrics.
Note that we always include some inference steps without inpainting to ensure that the motion model resolves any interpolation artifacts.
We find that the schedule with $\sigma_s = 500$ and $\sigma_e = 50$ best balances maintaining $\baseMotion$ and performing natural-looking edits.

Additionally, we see that noise-inversion with $25$ evaluation steps performs similarly to an inpainting schedule with $\sigma_s = 700$ and $\sigma_e = 500$, which we find is insufficient to maintain the details from $\baseMotion$. While increasing the number of evaluation steps improves reconstruction error, we list results with $25$ evaluation steps to be consistent with our scheduled inpainting results. We refer to Appendix D for analysis of evaluation steps versus reconstruction quality for noise-inversion.

Similarly, $20$ optimization steps, which need $10.3$ sec to compute, with DNO only match an inpainting schedule of $\sigma_s = 1000$ and $\sigma_e = 500$ ($0.192$ sec) and are insufficient to reconstruct details in $\baseMotion$.
While more optimization steps help improve reconstruction, note that each step requires evaluating the entire diffusion process.
Although this optimization is only required once for each $\baseMotion$, it hinders more experimental tasks such as searching for sub-clips within long motion capture sequences.
We provide an analysis of increasing gradient steps in Appendix E.

Lastly, CondEditor similarly exhibits large reconstruction errors while struggling to perform realistic edits, as described earlier.

\begin{table}
    \caption{Reconstruction error using different inpainting schedule with IBMM model for inbetweening motion clips on our internal and LaFan1~\cite{robustInbetweening} datasets. CondEditor, noise-inversion~\cite{song2022denoisingdiffusionimplicitmodels} and DNO~\cite{dno} all result in too high errors for interactive motion editing while our method allows user control over the level of preservation and editing.}
    \centering
    \begin{tabular}{ccccccc}
        \toprule
         \multirow{2}{*}{$\sigma_s$} & \multirow{2}{*}{$\sigma_e$} & \multicolumn{2}{c}{Internal} & \multicolumn{2}{c}{LaFan1} & \multirow{2}{*}{\shortstack{Time \\ (s)}}\\
         
         \cline{3-6} & & L2P $\downarrow$ & L2R $\downarrow$ & L2P $\downarrow$ & L2R $\downarrow$ & \\
         \midrule
         300    & 50    & \textbf{0.003}    & \textbf{0.047}  & \textbf{0.004}    & \textbf{0.044} & \multirow{6}{*}{0.192}\\
         500    & 50    & 0.004    & 0.051  & 0.005    & 0.049 & \\
         700    & 500   & 0.014    & 0.075  & 0.017    & 0.071 & \\
         1000   & 500   & 0.028    & 0.091  & 0.033    & 0.089 & \\
         1000   & 700   & 0.049    & 0.109  & 0.058    & 0.108 & \\
         1000   & 1000  & 0.173    & 0.157  & 0.178    & 0.141 & \\
         \hdashline
         \multicolumn{2}{c}{CondEditor} & 0.035 & 0.148 & - & - & 0.192\\
         \multicolumn{2}{c}{Noise-inversion} & 0.017 & 0.069 & 0.018 & 0.068 & 4.5\\
         \multicolumn{2}{c}{DNO} & 0.030 & 0.156 & 0.029 & 0.128 &  10.3\\
         \bottomrule
    \end{tabular}
    \label{tab:recon_error_inpainting_schedules_ilm}
\end{table}

\subsubsection{Motion Alignment} \label{sec:ablation_motion_alignment}
\begin{figure}
    \centering
    \begin{subfigure}{0.5\linewidth}
        \includegraphics[width=0.99\linewidth, trim=3cm 2.2cm 3cm 0cm, clip]{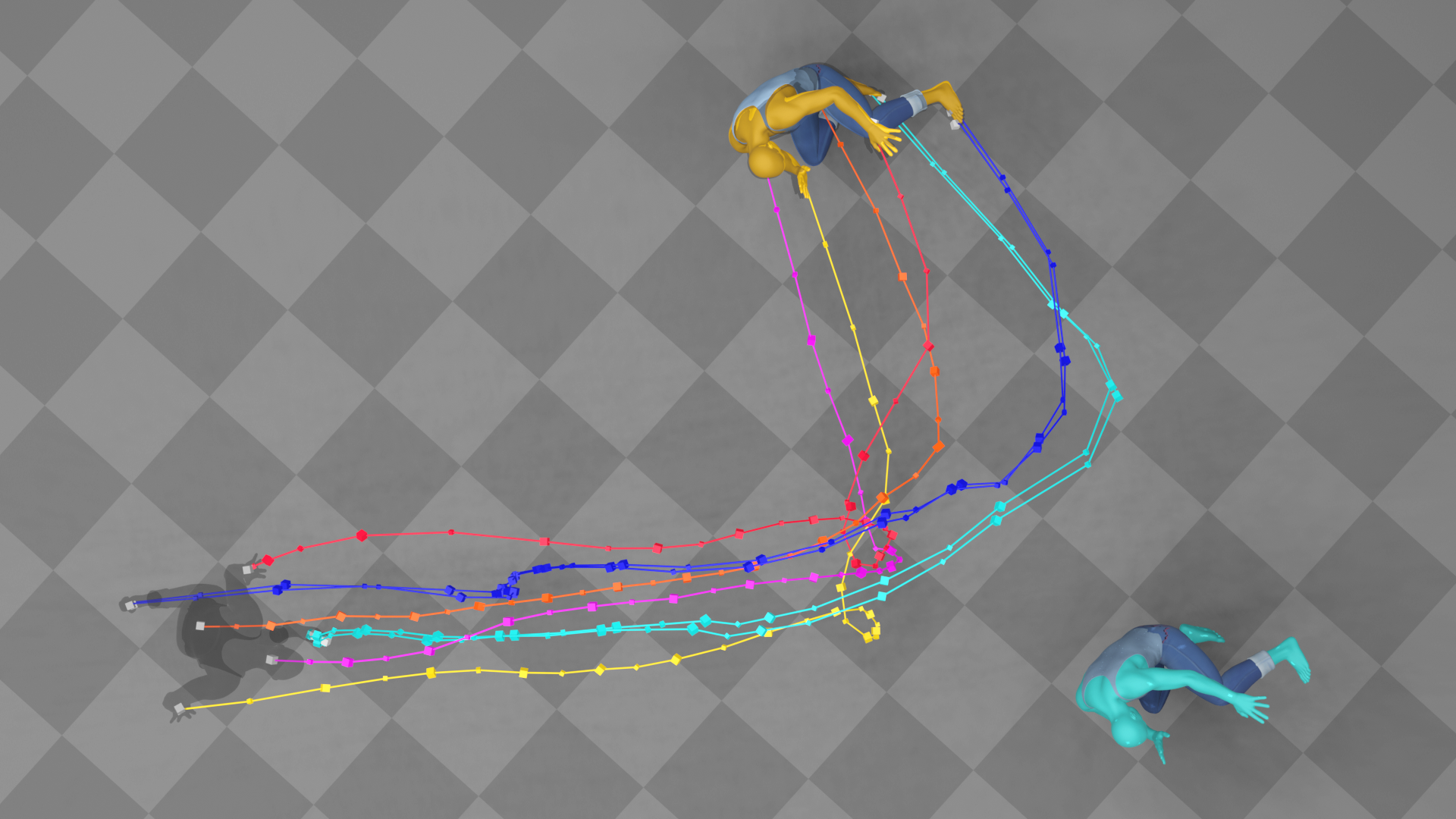}
        \subcaption{without alignment}
    \end{subfigure}%
    \begin{subfigure}{0.5\linewidth}
        \includegraphics[width=0.99\linewidth, trim=3cm 2.2cm 3cm 0cm, clip]{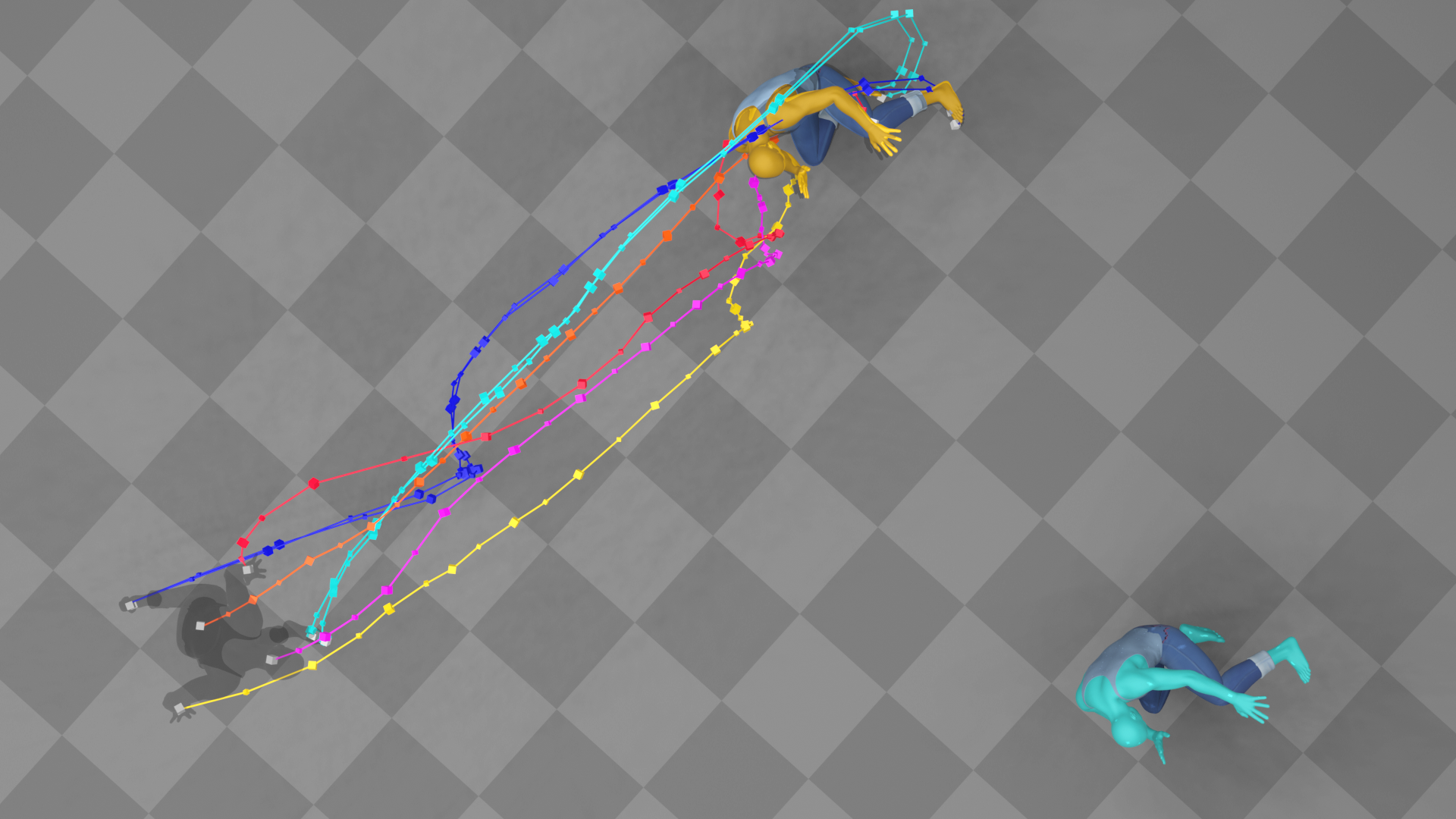}
        \subcaption{with alignment}
    \end{subfigure}
    \caption{Effect of aligning motions during inpainting. Without alignment, the character unnaturally moves towards the original keyframe (in cyan) before warping to the target keyframe. In contrast,  with alignment the motion remains natural and consistent.}
    \label{fig:ablation_alignment}
\end{figure}

In \figref{fig:ablation_alignment}, we edit a forward roll's end frame to demonstrate the importance of aligning the base motion and generated motions before inpainting.
When inpainting without aligning (a), the resulting motion drifts towards the original end frame before warping unnaturally.
Aligning forward directions (b), the motion moves naturally and consistently towards the new constraint.

\subsubsection{Influence Radius $\mu_{c_i}$}
\begin{figure}[ht]
    \centering
    \begin{subfigure}{0.5\linewidth}
        \centering
        \includegraphics[width=0.99\linewidth]{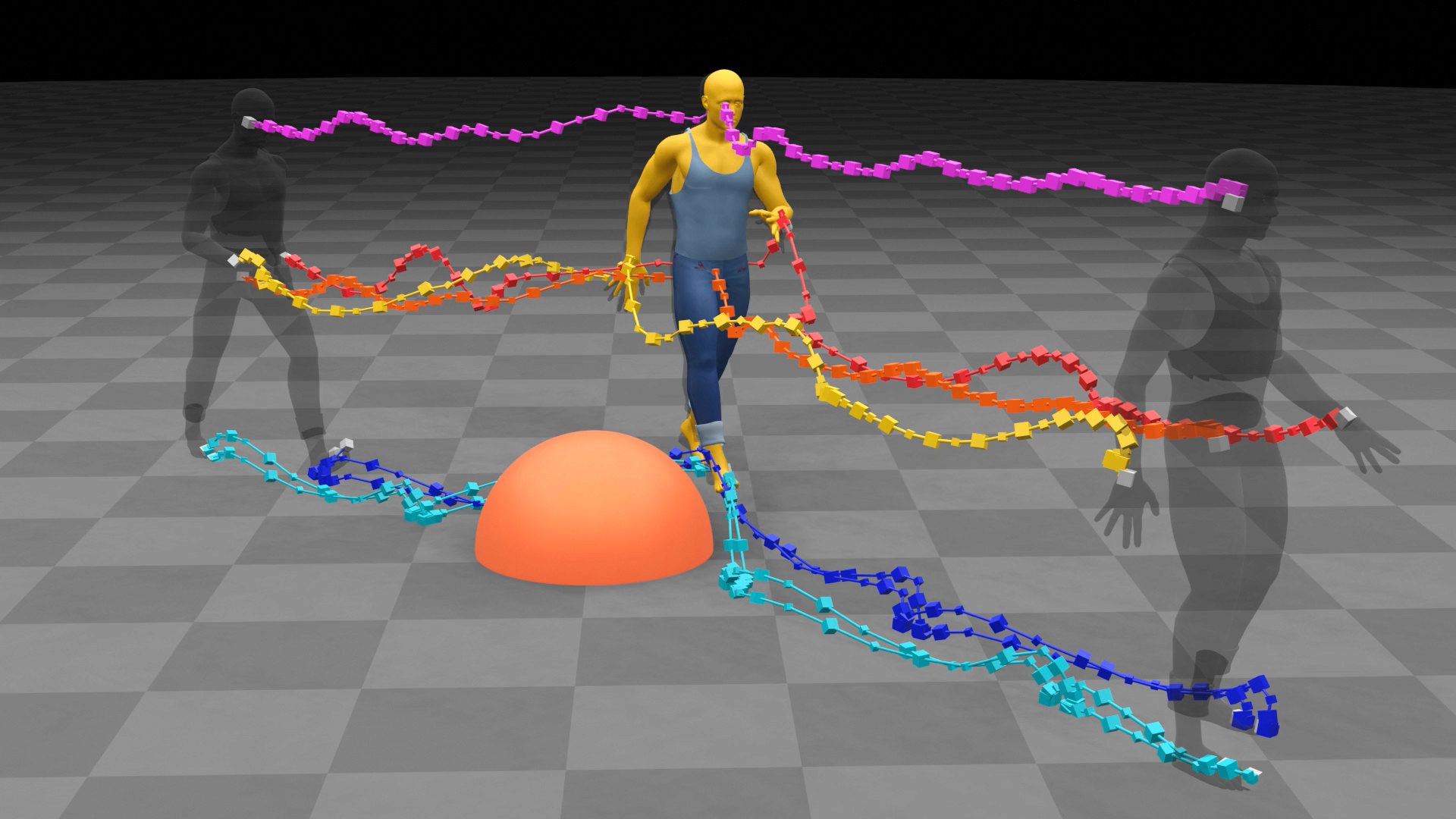}
        \subcaption{$\mu_{c_i} = 5$}
    \end{subfigure}%
    \begin{subfigure}{0.5\linewidth}
        \centering
        \includegraphics[width=0.99\linewidth]{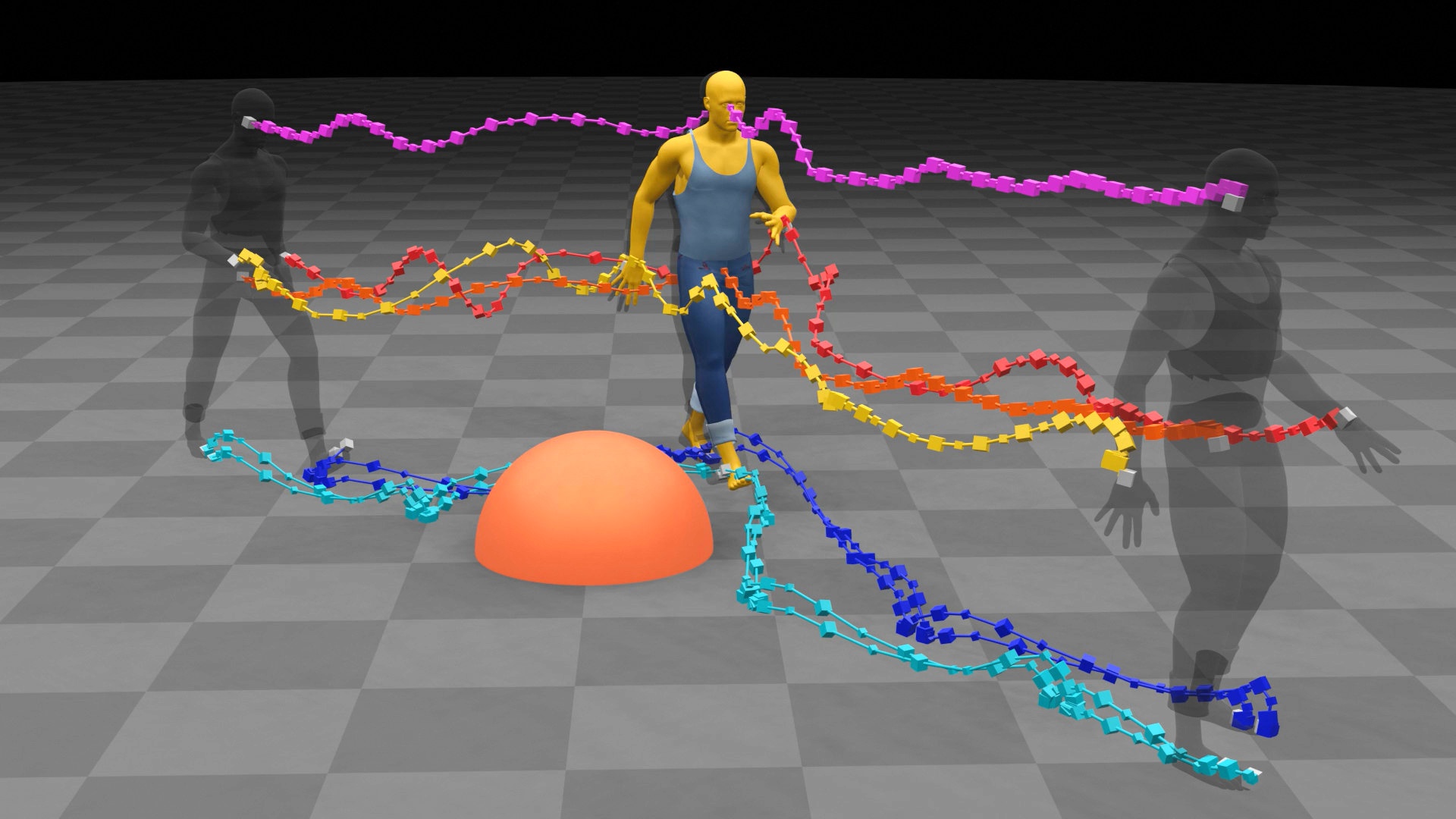}
        \subcaption{$\mu_{c_i} = 10$}
    \end{subfigure}
    \begin{subfigure}{0.5\linewidth}
        \centering
        \includegraphics[width=0.99\linewidth]{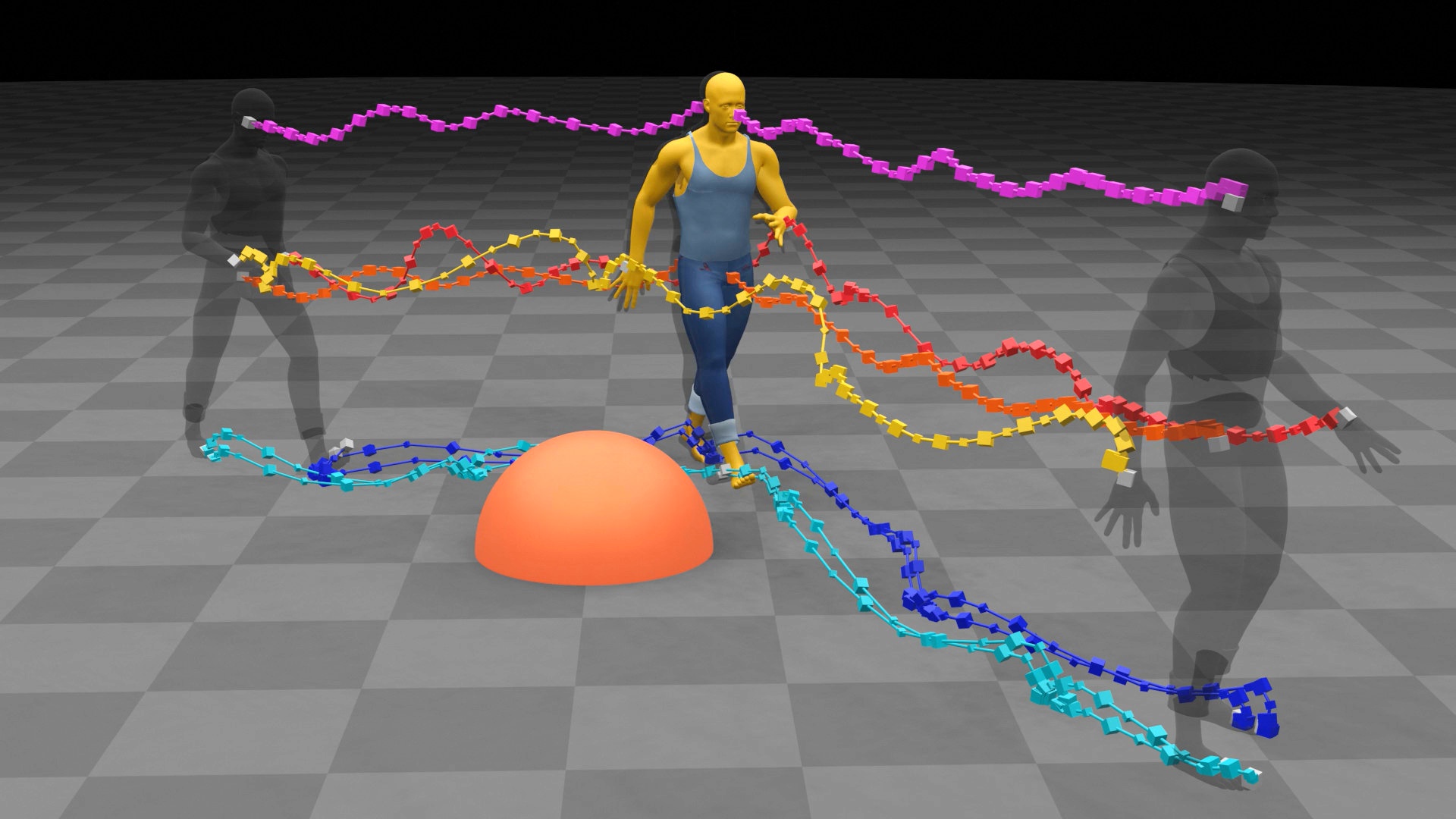}
        \subcaption{$\mu_{c_i} = 20$}
    \end{subfigure}%
    \begin{subfigure}{0.5\linewidth}
        \centering
        \includegraphics[width=0.99\linewidth]{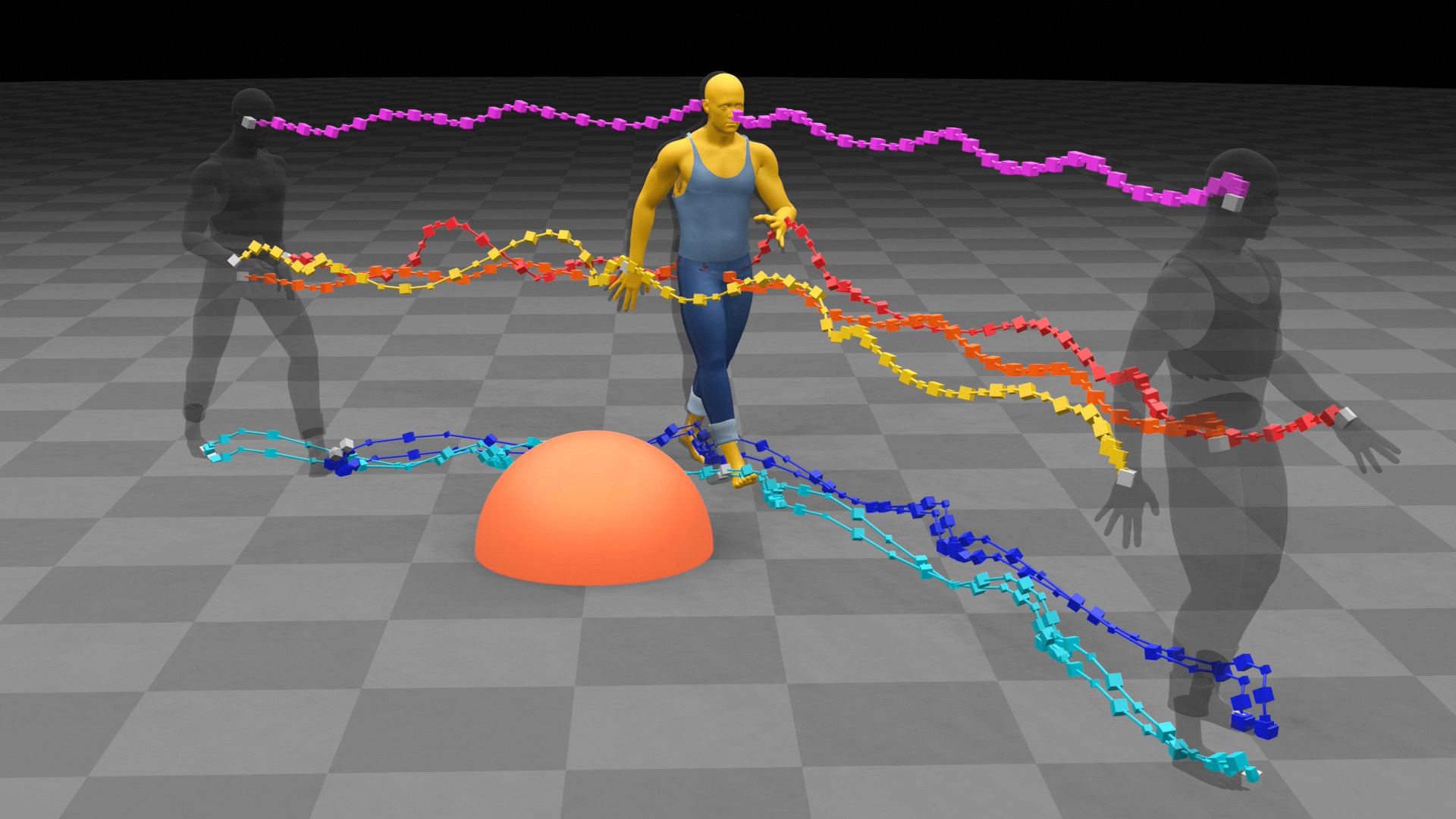}
        \subcaption{$\mu_{c_i} = 40$}
    \end{subfigure}
    \caption{We can see the effect of the \remove{parameter}\add{influence radius} $\mu_{c_i}$ when editing to avoid an obstacle. Notice how more frames are adapted to avoid the obstacle, as the radius increases.}
    \label{fig:ablation_influence_radius}
\end{figure}

\figref{fig:ablation_influence_radius} edits a 90-frame `walk forward' motion to avoid the obstacle using a fixed schedule ($\sigma_s=500$ and $\sigma_e=50$).
Increasing the radius of influence $\mu_{c_i}$ from $5$ to $40$ smoothly expands the number of adjusted frames, as seen by the motion curves.
Even for $\mu_{c_i} = 40$, the Gaussian kernel allows sufficient inpainting to preserve the base motion's style and cadence.

\subsubsection{Kernel Shape} \label{sec:ablation_kernel_shape}
We ablate the shape of the kernel $G(j, t; c_i)$ defined in \myeqref{eq:appDirectManip}, used to construct the spatiotemporal mask for direct manipulation. We require that the partial inpainting at the kernel's edges sufficiently anchors the generated motion to the base motion, so the edit follows a trajectory close to the original rather than diverging arbitrarily. Additionally, the resulting motion should not have abrupt changes in velocity or cadence at the kernel boundaries.

\begin{figure}
    \centering
    \begin{subfigure}{0.5\linewidth}
        \includegraphics[width=0.99\linewidth]{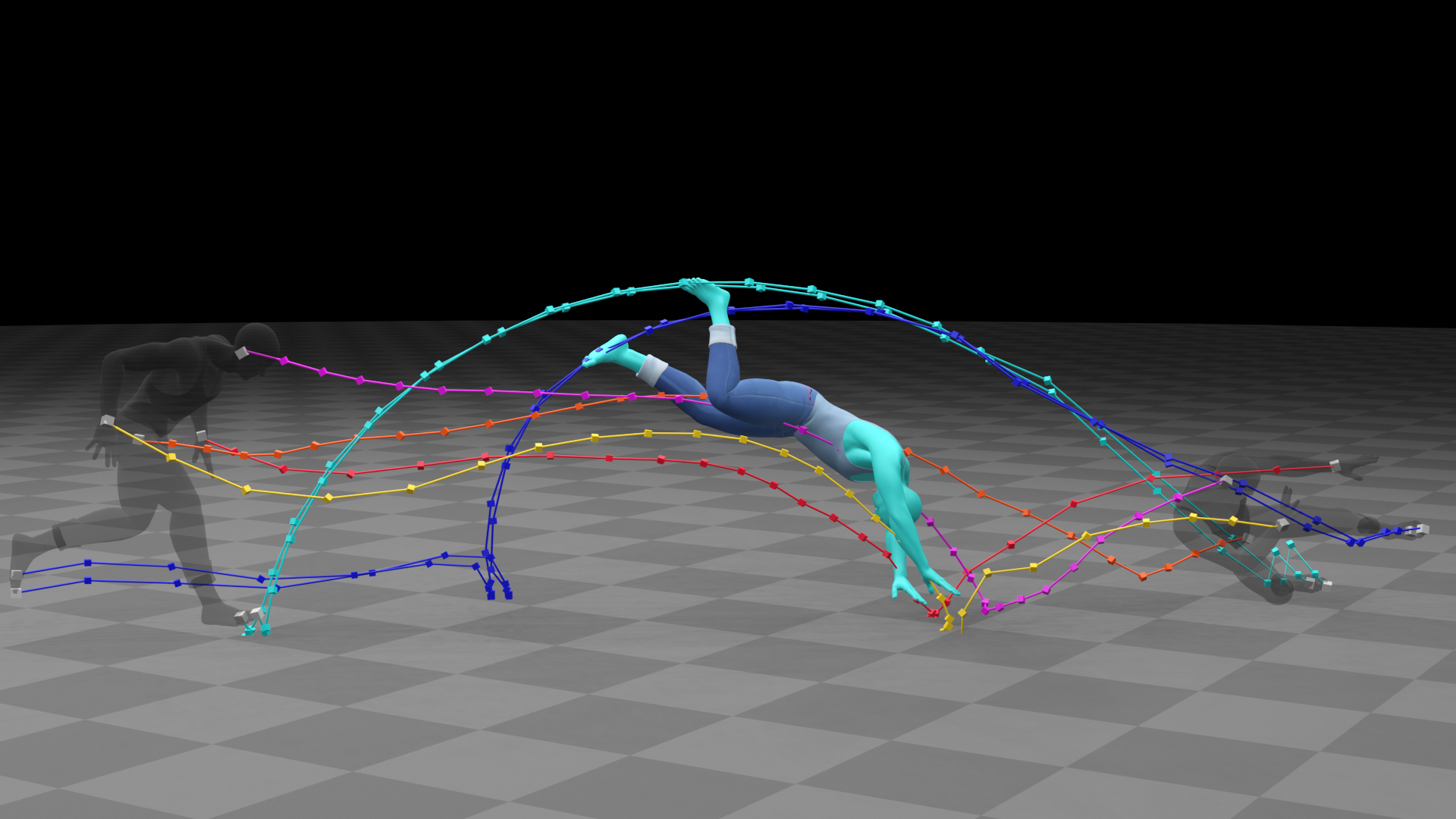}
        \subcaption{\add{Base motion}}
    \end{subfigure}%
    \begin{subfigure}{0.5\linewidth}
        \includegraphics[width=0.99\linewidth]{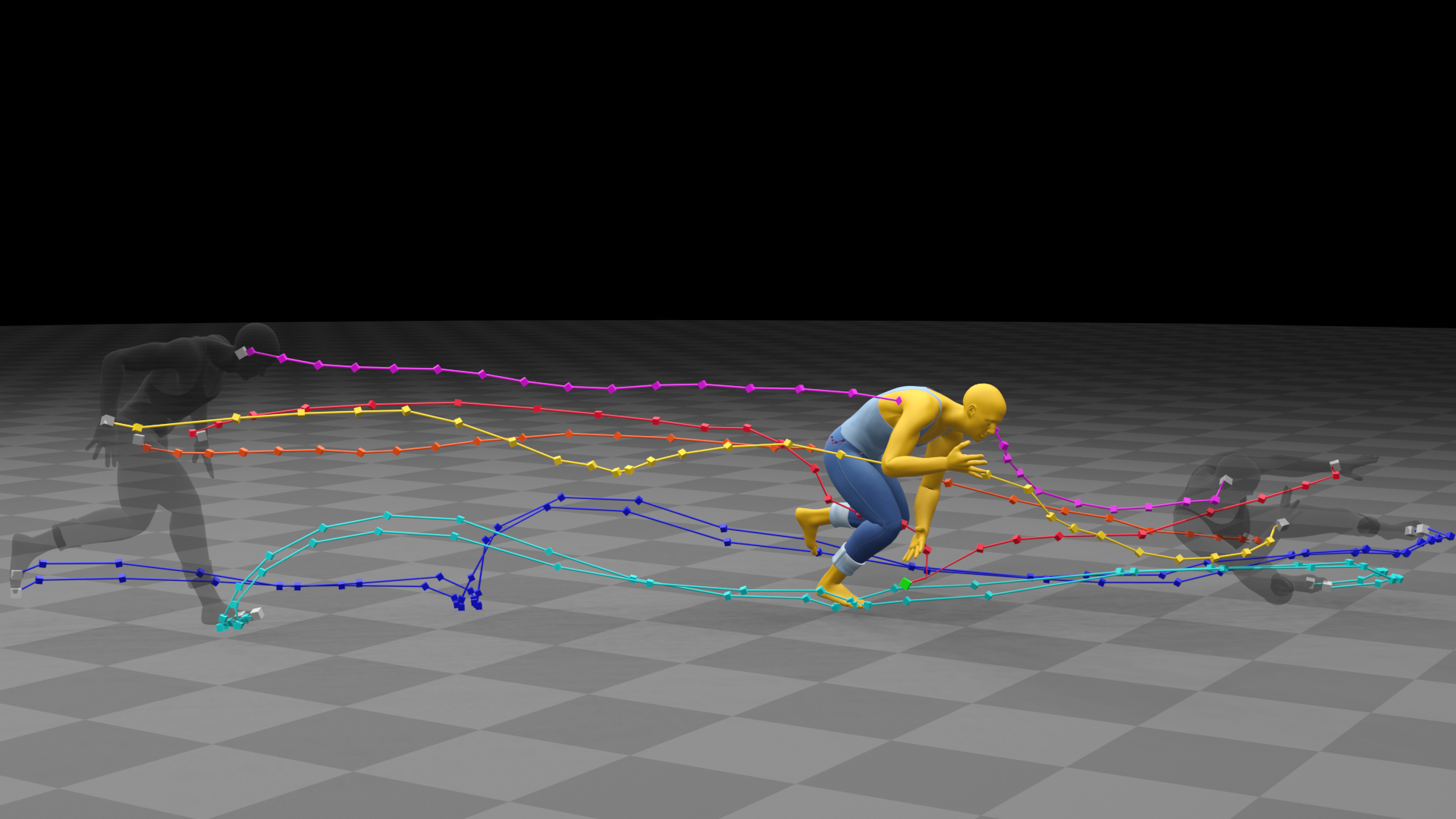}
        \subcaption{\add{Square kernel}}\label{fig:motion_roll_kernel_comparison_square}
    \end{subfigure}
    \begin{subfigure}{0.5\linewidth}
        \includegraphics[width=0.99\linewidth]{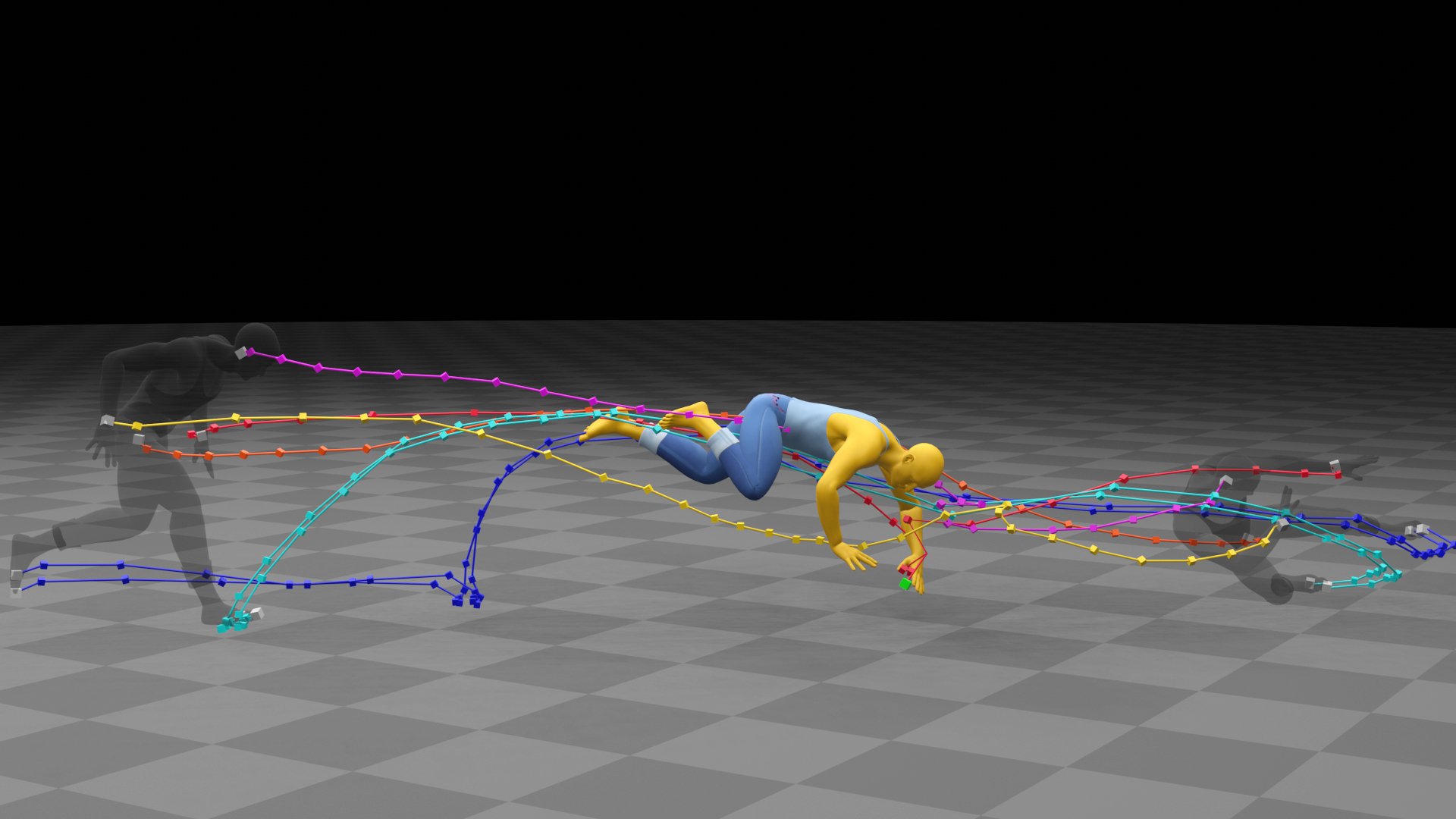}
        \subcaption{\add{Triangular kernel}}
    \end{subfigure}%
    \begin{subfigure}{0.5\linewidth}
        \includegraphics[width=0.99\linewidth]{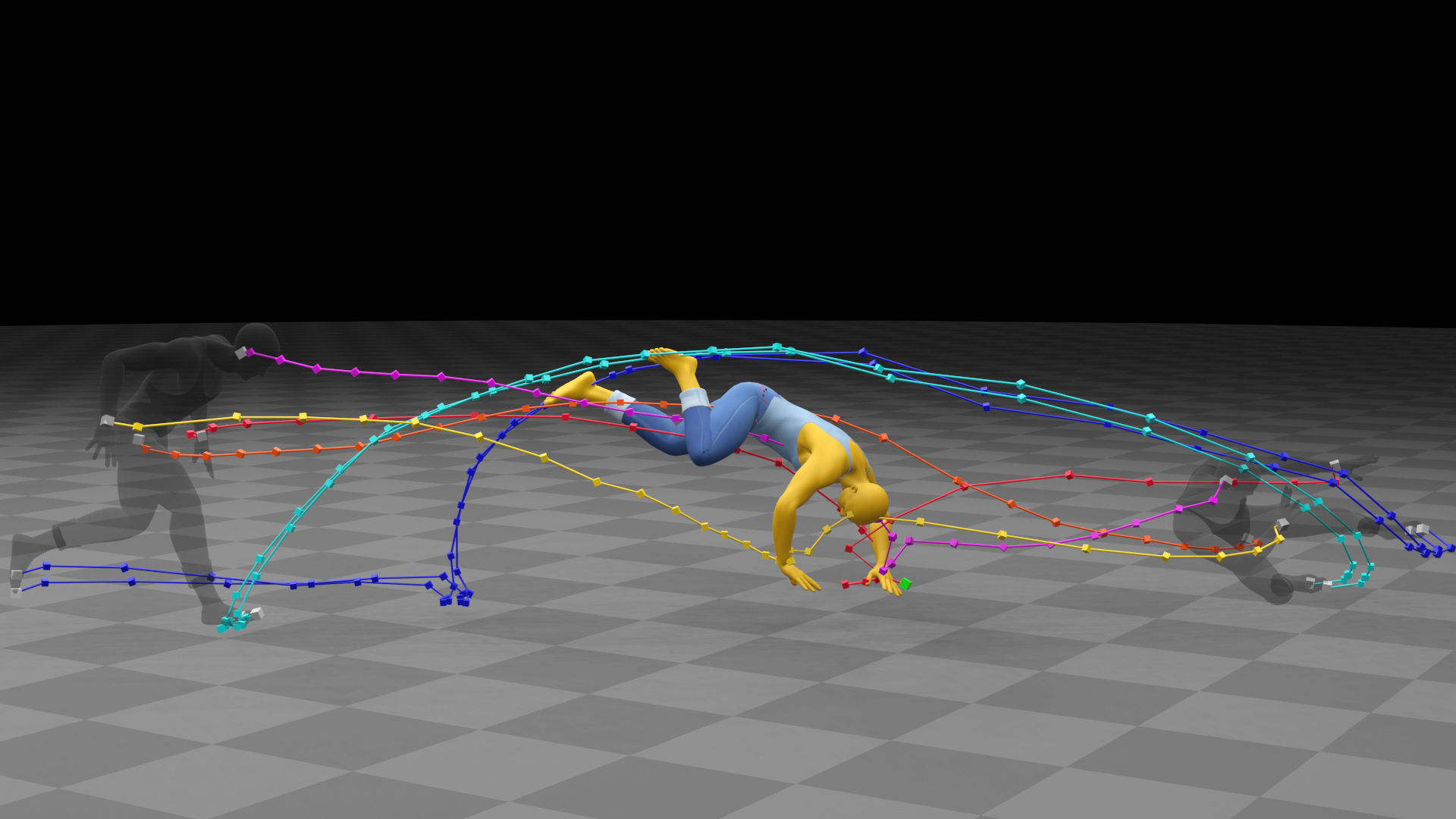}
        \subcaption{\add{Gaussian kernel}}
    \end{subfigure}
    \caption{\add{Ablating the weight decay kernel function for direct manipulation. While the piecewise linear and gaussian kernels can recreate the roll, the square wave results in a ducking motion.}}
    \label{fig:motion_roll_kernel_comparison}
\end{figure}

In \figref{fig:motion_roll_kernel_comparison}, we edit the forward roll replacing the Gaussian with a width-modulated square wave and a triangular kernel. With each kernel's minimum value set to zero, the square wave inpaints no fraction of the roll pose, and the edit collapses into a ducking motion, as seen in \figref{fig:motion_roll_kernel_comparison_square}. Raising the square wave's minimum weight can recover the roll, but at the cost of manipulability, as the motion is then interpolated away from the user's constraints.
In contrast, both the Gaussian and triangular kernels retain the roll but using the gaussian kernel more closely matches the base motion.

\subsection{Usability Test}

\begin{figure}
    \centering
    \begin{subfigure}{\linewidth}
        \begin{subfigure}{0.5\linewidth}
            \includegraphics[width=0.99\linewidth]{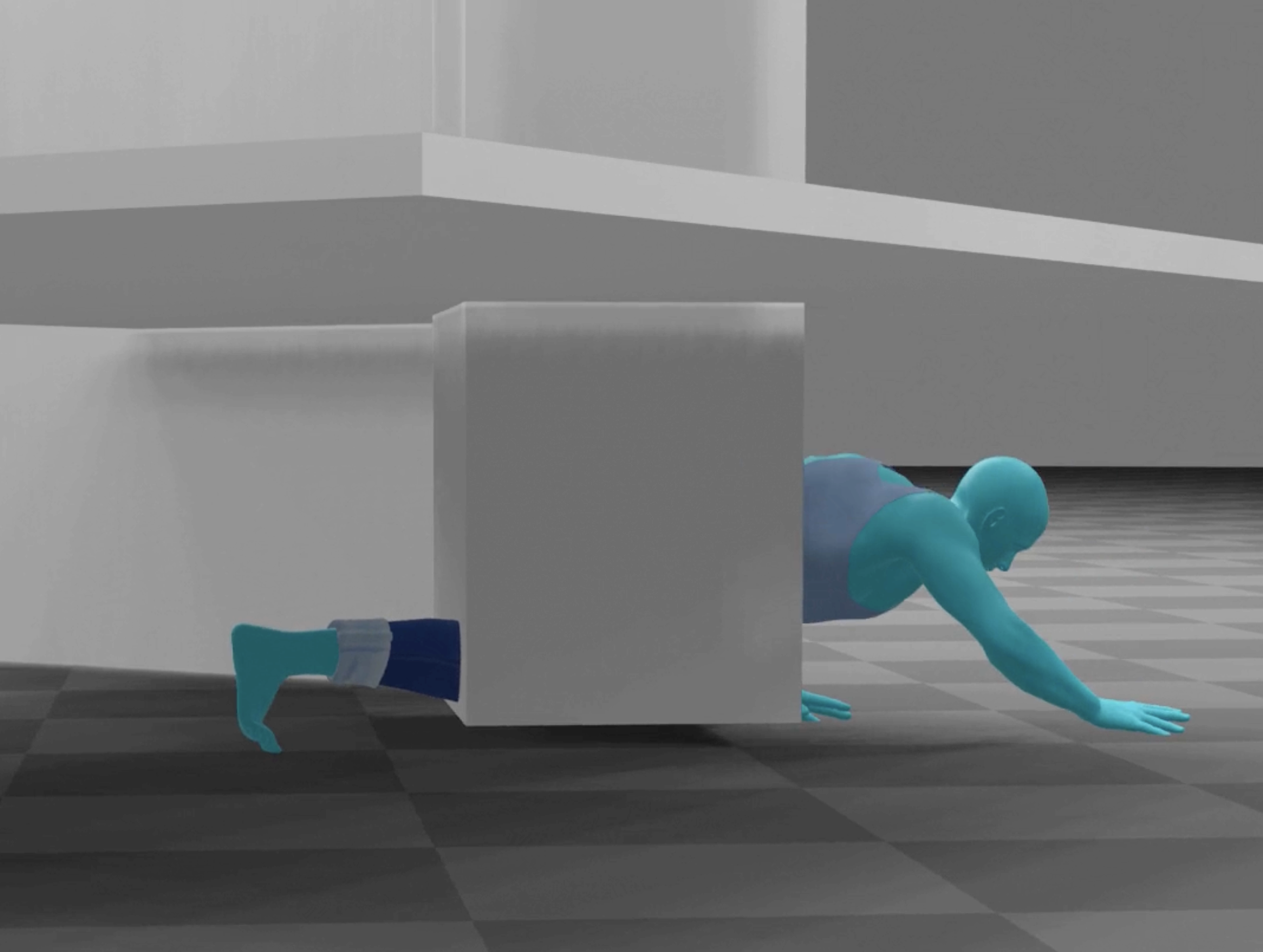}
        \end{subfigure}%
        \begin{subfigure}{0.5\linewidth}
            \includegraphics[width=0.99\linewidth]{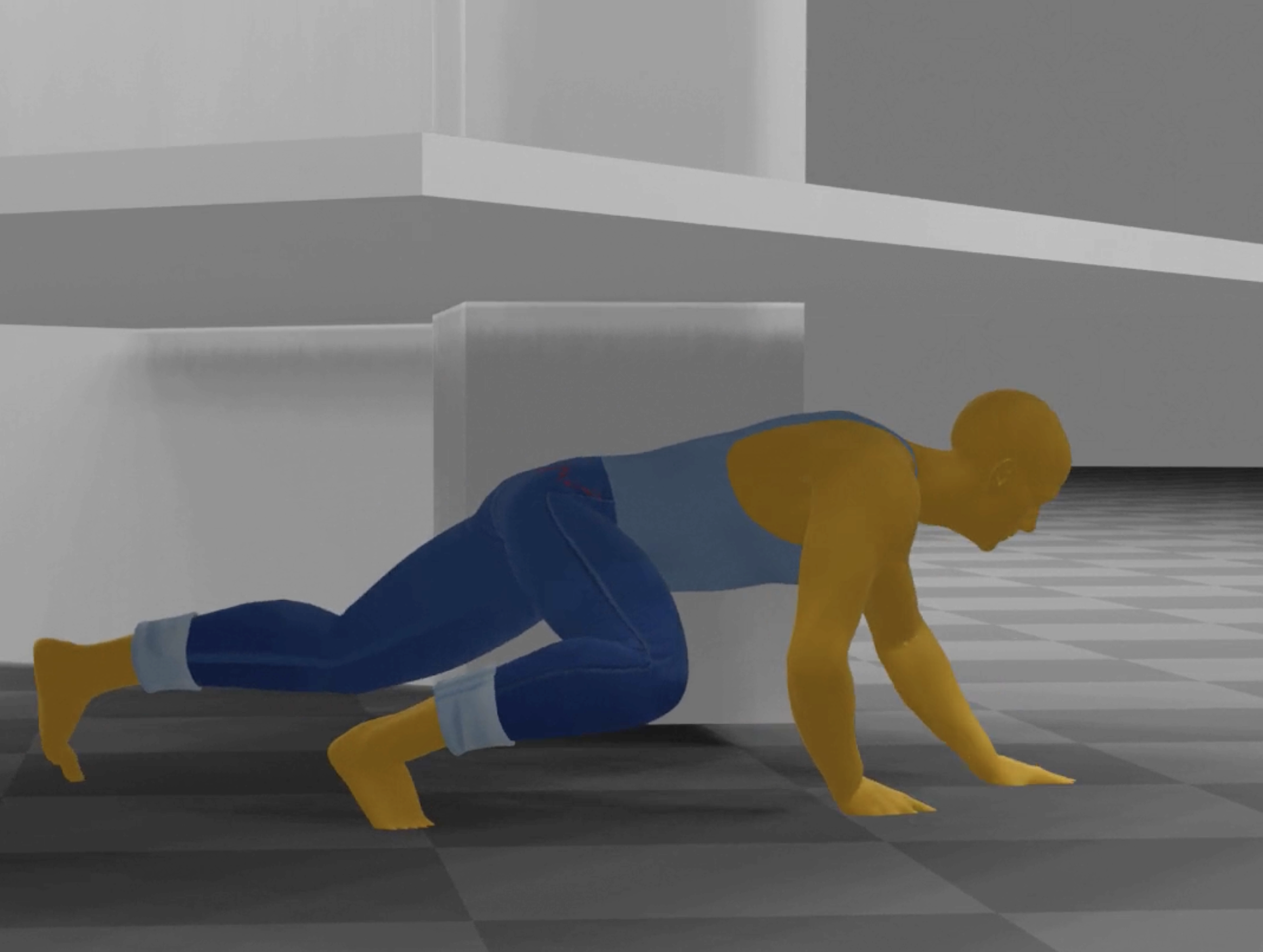}
        \end{subfigure}
        \subcaption{Fixing scene penetration}
    \end{subfigure}
    \begin{subfigure}{\linewidth}
        \begin{subfigure}{0.5\linewidth}
            \includegraphics[width=0.99\linewidth]{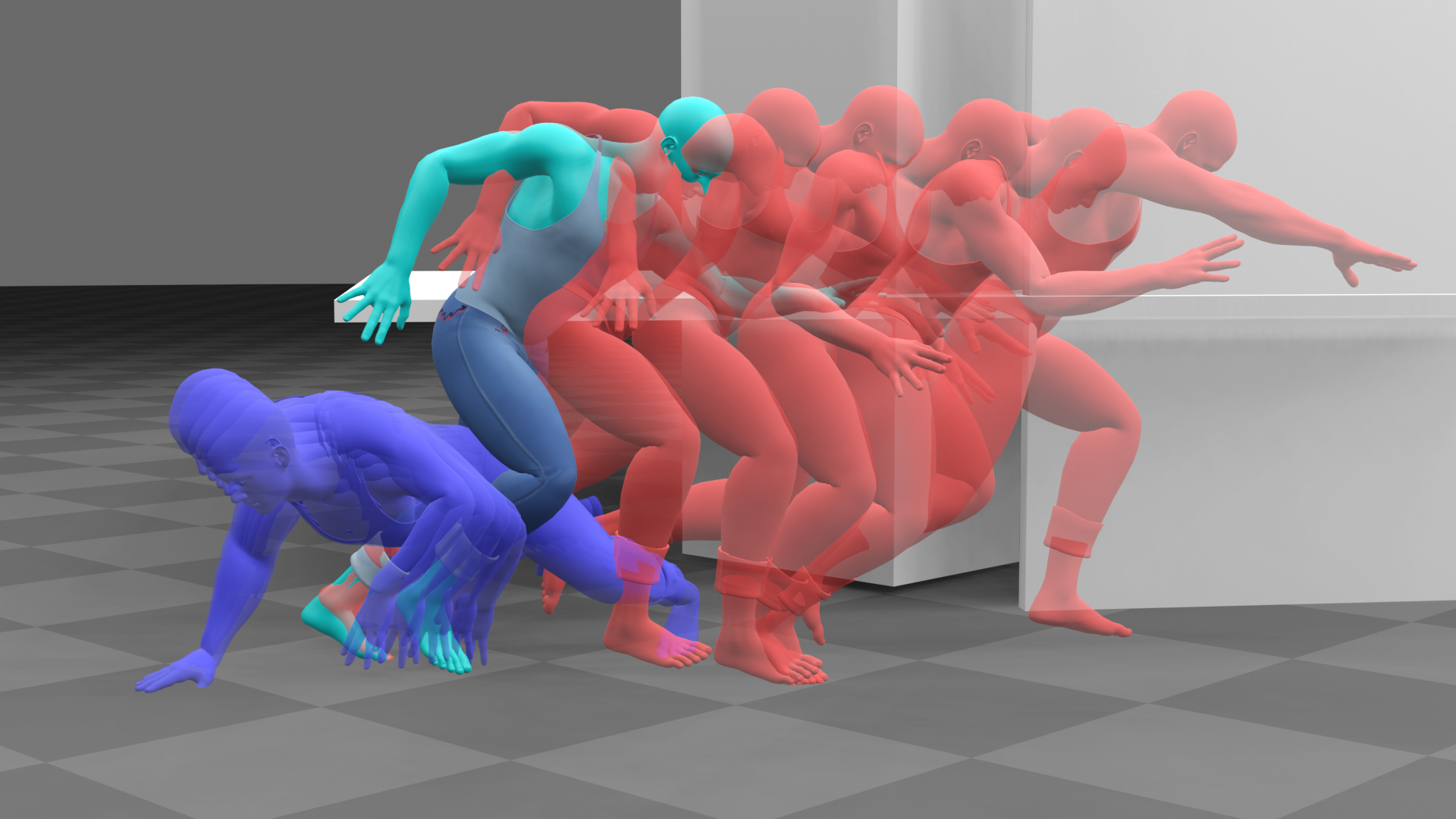}
        \end{subfigure}%
        \begin{subfigure}{0.5\linewidth}
            \includegraphics[width=0.99\linewidth]{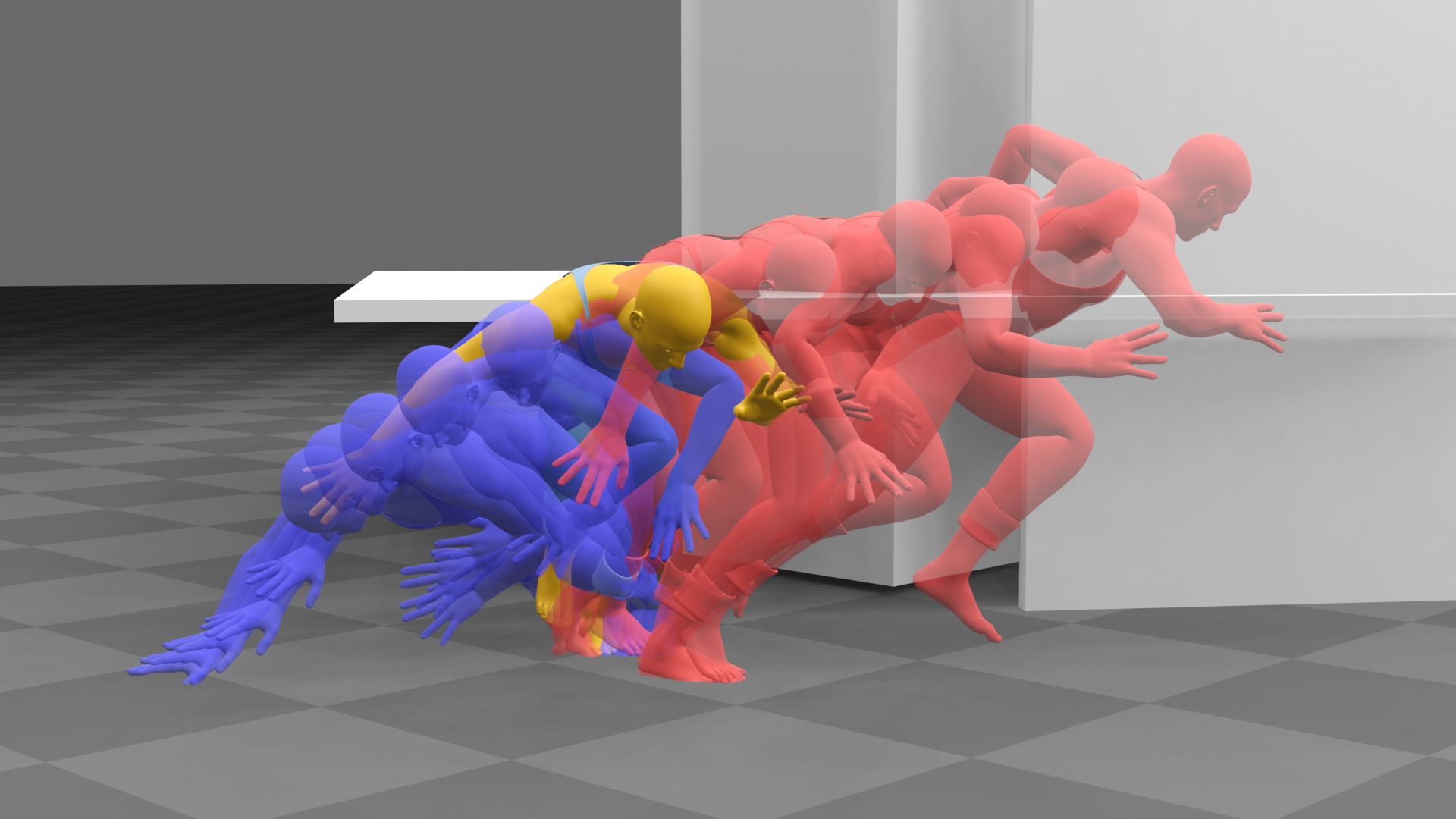}
        \end{subfigure}
        \subcaption{Stitching}
    \end{subfigure}
    \begin{subfigure}{\linewidth}
        \begin{subfigure}{0.5\linewidth}
            \includegraphics[width=0.99\linewidth, trim={5cm 0cm 5cm 0cm}, clip]{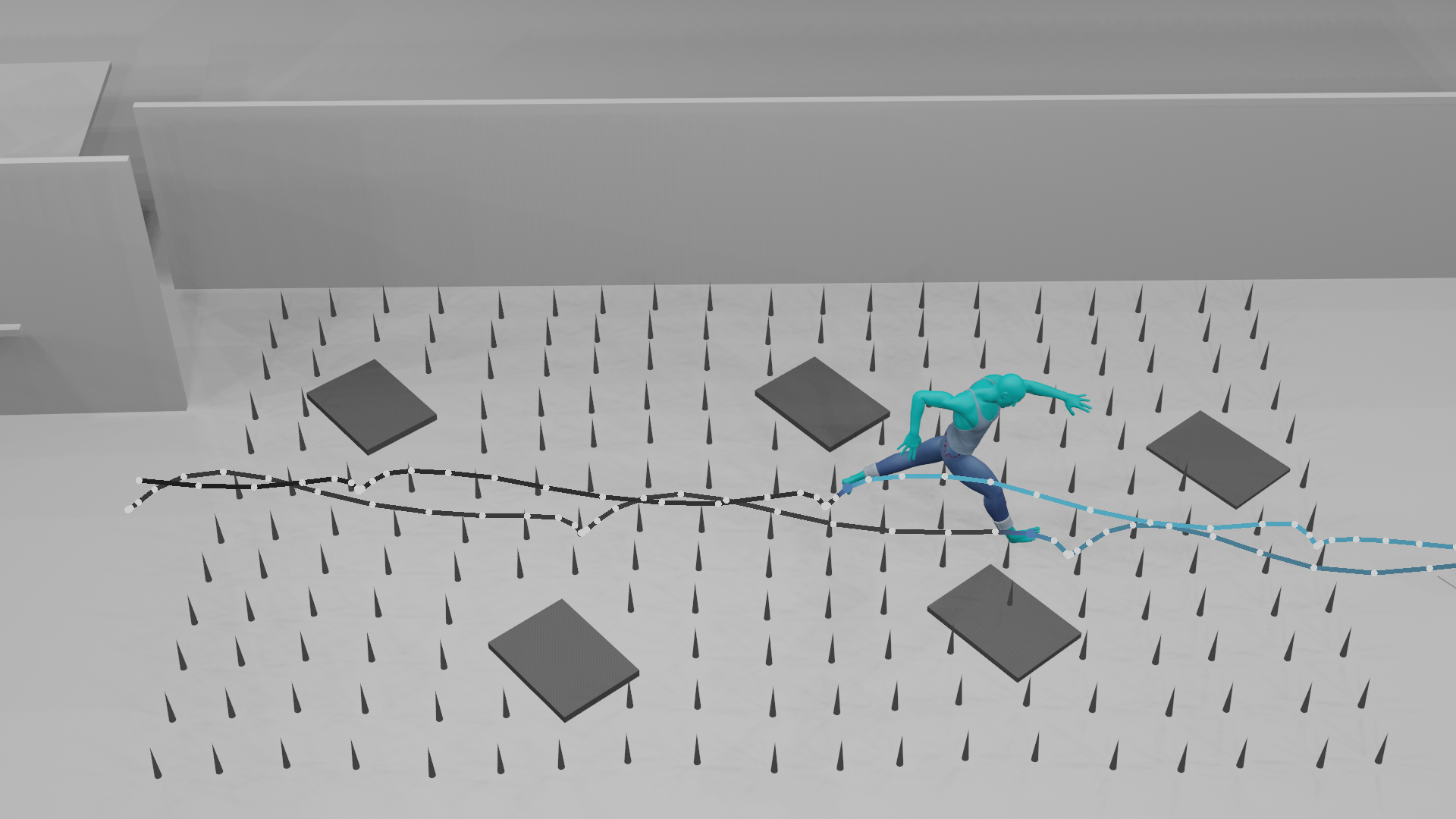}
        \end{subfigure}%
        \begin{subfigure}{0.5\linewidth}
            \includegraphics[width=0.99\linewidth, trim={5cm 0cm 5cm 0cm}, clip]{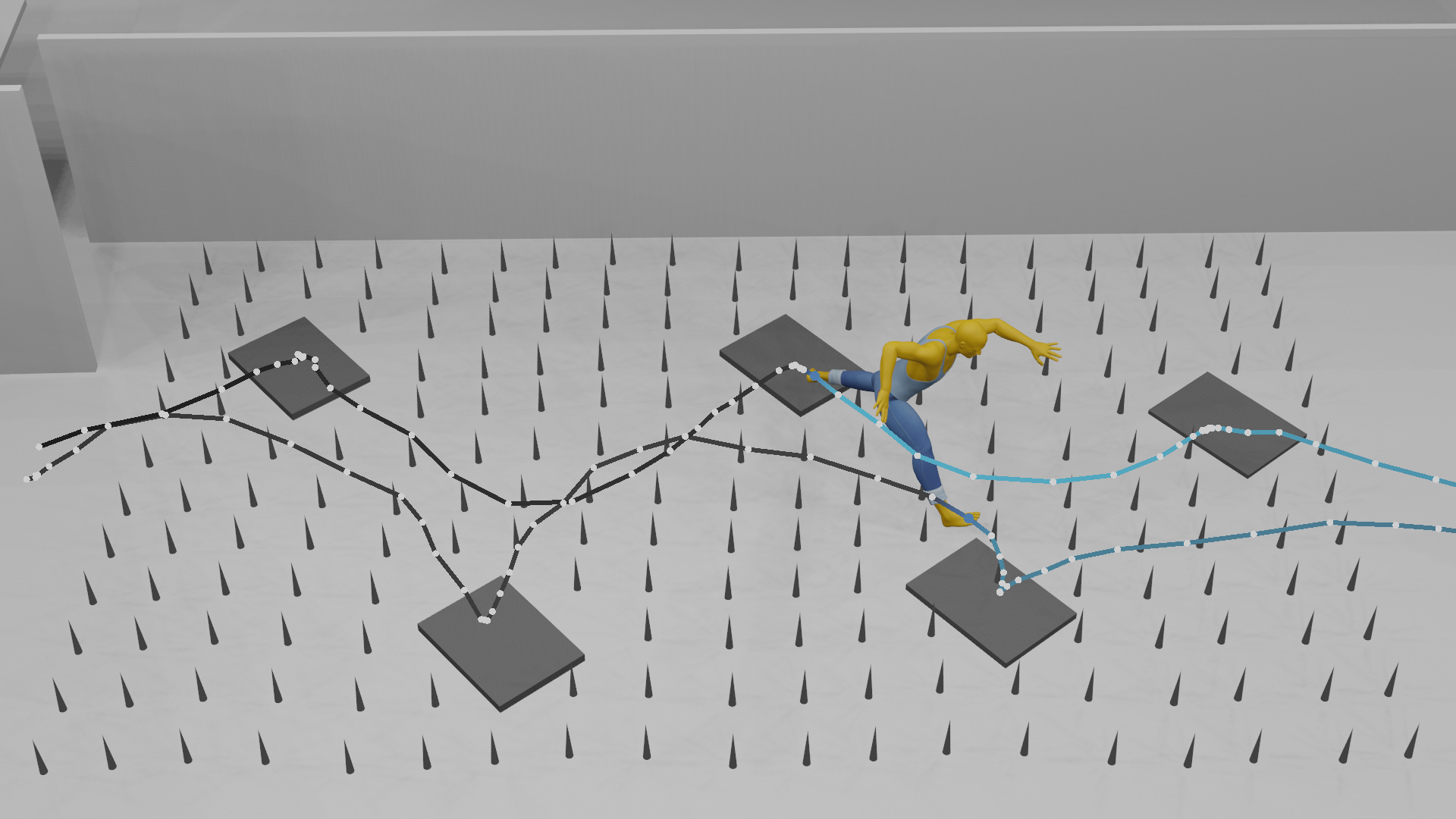}
        \end{subfigure}
        \subcaption{Foot trajectory}
    \end{subfigure}
    \caption{Before (left) and after (right) fixing scene penetration and stitching artifacts in the artist test scene.}
    \label{fig:user_test_result}
\end{figure}

To evaluate the usability of our method, we conducted a test with two professional digital artists tasked with animating a character through a parkour course.
This task required running, crawling, and jumping, using a library of two running clips and one crawling motion clip.
As shown in \figref{fig:user_test_result}, the provided motion capture contained artifacts such as environmental penetration (e.g., the character crawling through a pillar) and discontinuities at the boundary between naively stitched motions.

The artists were given one hour to fix these artifacts. Both participants reported that this time limit would be infeasible with traditional motion editing tools, but they successfully completed the task using our framework, as shown in our supporting video. During the process, both artists leveraged the capability to preserve specific motion segments while sampling variations for the rest. One of the artists noted that the transition from crawling to running would traditionally require significant manual effort, while it was automatic using our method. Additionally, they reported that the editing process being non-destructive was an advantage.

The test also identified areas for future refinement.
While neural motion curves~\cite{agrawal2024skel} were intuitive, a better interface was needed for interacting with foot contact points.
Furthermore, due to the computational overhead of the full diffusion process, the artists had to adapt to a lower frame rate compared to traditional interpolation. However, it is worth noting that slower frame rates are standard in professional animation workflows utilizing complex non-linear rigs.
Finally, we observed that when generating sequences significantly longer than those in the training set, the method's ability to closely preserve the base motion begins to degrade.

%% file: content/05_Discussion.tex
In the context of image editing, interactive generative editing capabilities, such as generative fill, image extension, object removal, image blending, and stylization, have found great success in professional software.
While interactive manipulation and motion editing have each been explored individually within generative motion research, combining these applications holds vast potential for creating similar transformative workflows in professional animation, increasing reusability of existing data and reducing artist effort, but remains underexplored.
To this end, our intuitive method addresses this gap by implementing important applications in animation, including direct manipulation, stitching, and compositing, within a single framework. This allows users to seamlessly switch between applications or even combine multiple together, as shown in our supporting video, where we combine direct manipulation with extension and stitching.

Inference-time techniques for pretrained models have been explored in previous works, such as inpainting in MDM~\cite{hmdm} and blending in PriorMDM~\cite{shafir2024human}. However, the binary inpainting mask used in MDM renders the inpainted section completely unmodifiable. Similarly, PriorMDM leverages the blending of boundaries between different samples to generate longer sequences, but crucially without any manipulation control. Consequently, both methods restrict an artist's ability to fine-tune the generated motion.
In contrast, our method extends these techniques to expose granular control over both the inpainting schedule and the spatial regions to the user. This maximizes user agency over the preservation, blending, and manipulation of the motion within a single, unified framework.

As previously discussed, DDIM inversion requires many evaluation steps to generate high quality samples. To accelerate this process, prior work \cite{shi2024dragdiffusion} has explored partially inverting to timestep $t \ll T$. Similarly to our method, this results in a mixture of base motion and noise at timestep $t$. However, unlike these methods, which lack fine-grained spatial control over preservation and generation, our method provides intuitive spacetime control, allowing more creative freedom.

To enable direct manipulation, we focused on models such as IBMM~\cite{vogeli2025implicit} and SF-control~\cite{hwang2025motion}, which provide more granular interaction than models relying solely on full poses and trajectory constraints such as CondMDI~\cite{cohan2024flexible}.
Nonetheless, as we build upon pretrained diffusion models, we inherit their limitations as well. Most notably, reconstructing movements with a generative model results in a small loss of high-frequency details, which by extension affects the edited movements we generate with our method.
Furthermore, preserving a base motion and satisfying new constraints are inherently adversarial tasks, which can cause some generative models to begin ignoring constraints, as the level of preservation increases.

Generative models are inherently limited by the motion distribution on which they were trained. Therefore, when a user edit goes out of the training distribution, generative motion authoring can produce unexpected results, causing users to spend time trying to create a motion infeasible for the model. In the future, it would be valuable to explore visualizing feasibility for a given model.

Finally, we have shown that simply overwriting different clips for stitching and compositing already results in natural motion with our method. However, by integrating traditional authoring tools one can leverage layer-based editing to create base motion with fewer starting artifacts, further improving the generative results and potentially unlocking new workflows.

%% file: content/06_Conclusion.tex
In this work, we introduced a new motion editing task that expands motion editing with generative capabilities while simultaneously allowing for interactive manipulation. To achieve this, we introduce scheduled inpainting, an inference framework that supports many real-time editing applications, from stitching and extending to compositing diverse movements and generating stylized motion, while consistently resulting in natural motion. Our inference-based approach is training-free and compatible with any pretrained interactive motion diffusion model and dataset.

Animating and capturing movements is famously challenging and time-consuming. Generative motion authoring significantly lowered the entry barrier. However, it lacked the ability to preserve movements. By unifying preservation and generation into a shared workflow, we bring generative motion authoring closer to solving real-world production tasks.
We hope that our work will inspire and serve as a foundation for new workflows leveraging such generative models for real world tasks in VFX such as motion editing, enabling creatives with different backgrounds and skill sets to achieve their full creative visions more directly and intuitively.

%% file: content/07_Appendix.tex
\section{Scheduled Inpainting for Constraints}

\begin{figure}[h]
    \centering
    \begin{subfigure}{0.75\linewidth}
        \includegraphics[width=\linewidth]{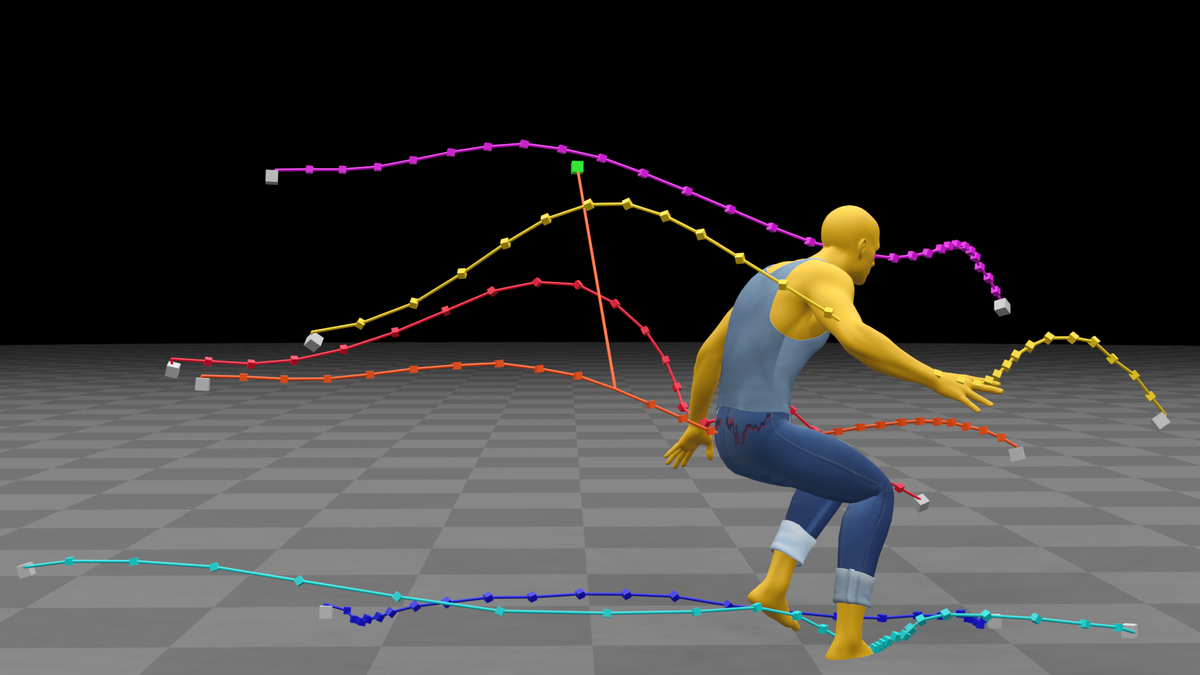}
        \subcaption{Base CondMDI}\label{fig:constraints_inpainting_base_condmdi}
    \end{subfigure}
    \begin{subfigure}{0.75\linewidth}
        \includegraphics[width=\linewidth]{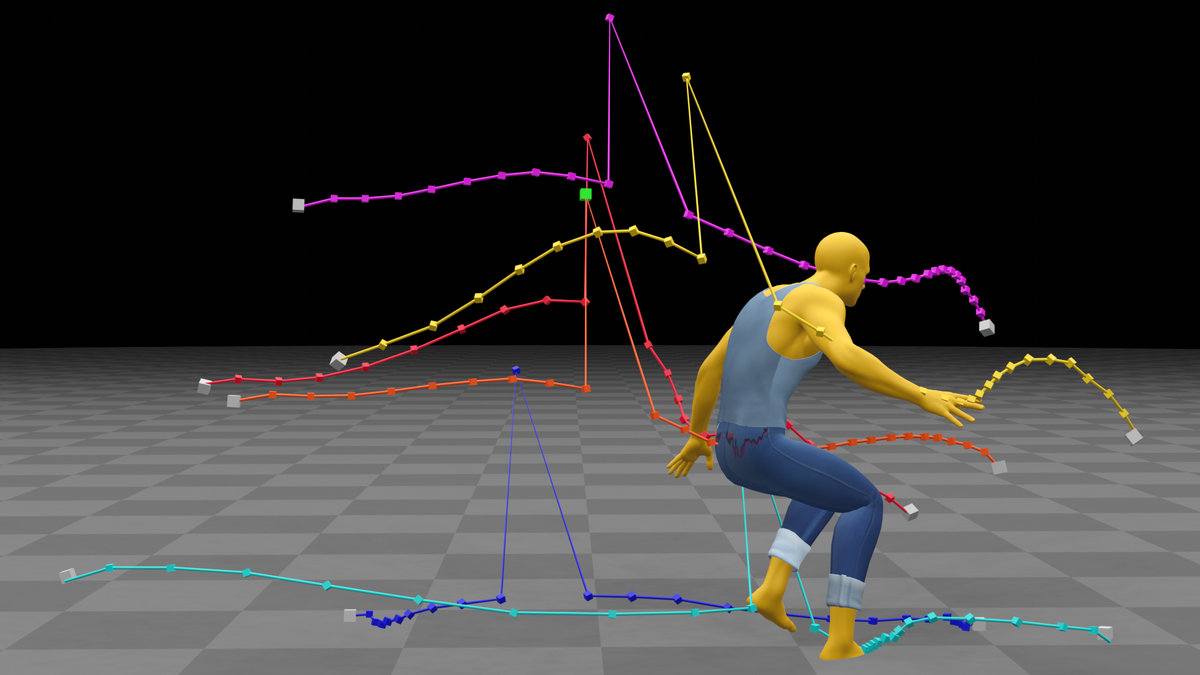}
        \subcaption{Standard Inpainting}\label{fig:constraints_inpainting_inpainting}
    \end{subfigure}
    \begin{subfigure}{0.75\linewidth}
        \includegraphics[width=\linewidth]{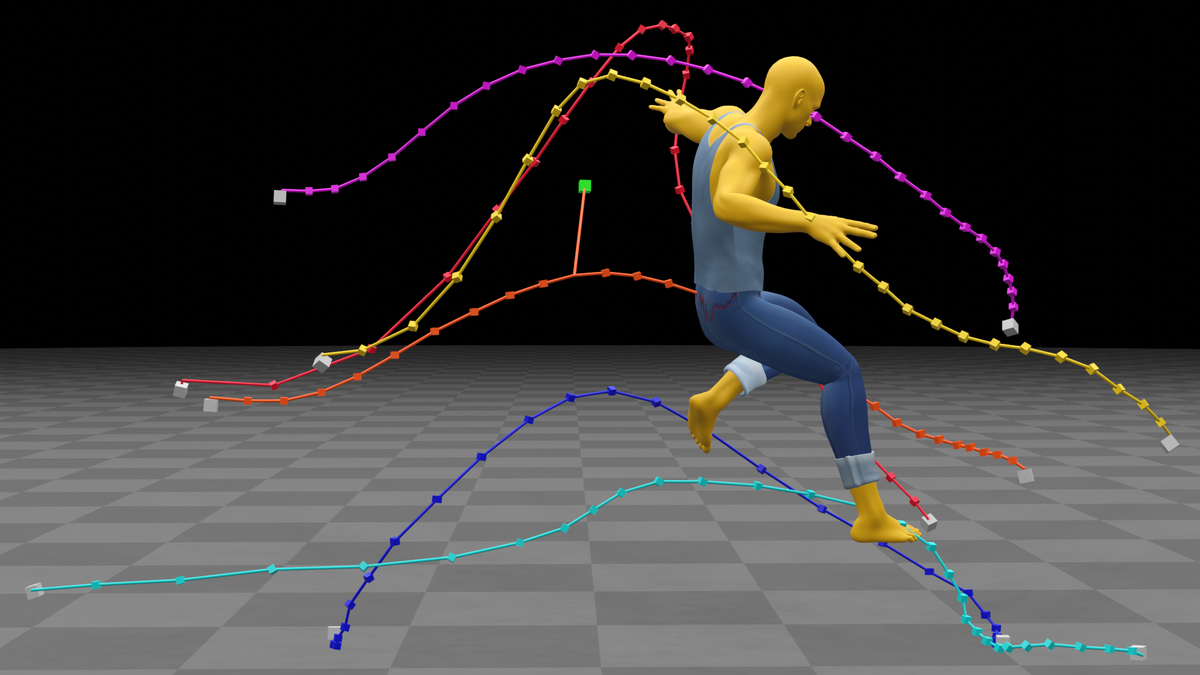}
        \subcaption{Scheduled Inpainting (Ours)}\label{fig:constraints_inpainting_SI}
    \end{subfigure}
    \caption{Comparison of using sparse constraints with CondMDI~\cite{cohan2024flexible}. By default, CondMDI completely ignores the sparse hip constraint (in green), while using standard inpainting results in large spikes. Using scheduled inpainting better follows the constraint while creating smooth motion.}
    \label{fig:constraints_inpainting}
\end{figure}

Dense motion representation models such as CondMDI~\cite{cohan2024flexible} struggle at click-and-drag interactions due to the sparsity of point constraints in the input \cite{studer2024factorized}.
As shown in \figref{fig:constraints_inpainting_base_condmdi}, a single hip constraint fails to trigger a jump, resulting in large errors and a running sequence.
Using simple inpainting results in harsh peaks in the generated motion (see \figref{fig:constraints_inpainting_inpainting}) although the constrained joint now satisfies the condition, it does not address the sparsity of information that causes the conditions to be ignored.
Consequently, the neighboring frames are not updated correctly.

In contrast, we propagate constraints to neighboring frames using a Gaussian kernel.
We schedule the kernel width to shrink with the diffusion noise such that broad kernels guide global trajectory at high noise, while narrow kernels enforce local constraints at low noise.
This results in the neighboring frames being updated according to the point constraints and producing a natural jumping motion, as seen in \figref{fig:constraints_inpainting_SI}.

\section{Ablating Normalization}\label{app:ablation_normalization}

\begin{figure}
    \centering
    \begin{subfigure}{0.75\linewidth}
        \centering
        \includegraphics[width=\linewidth]{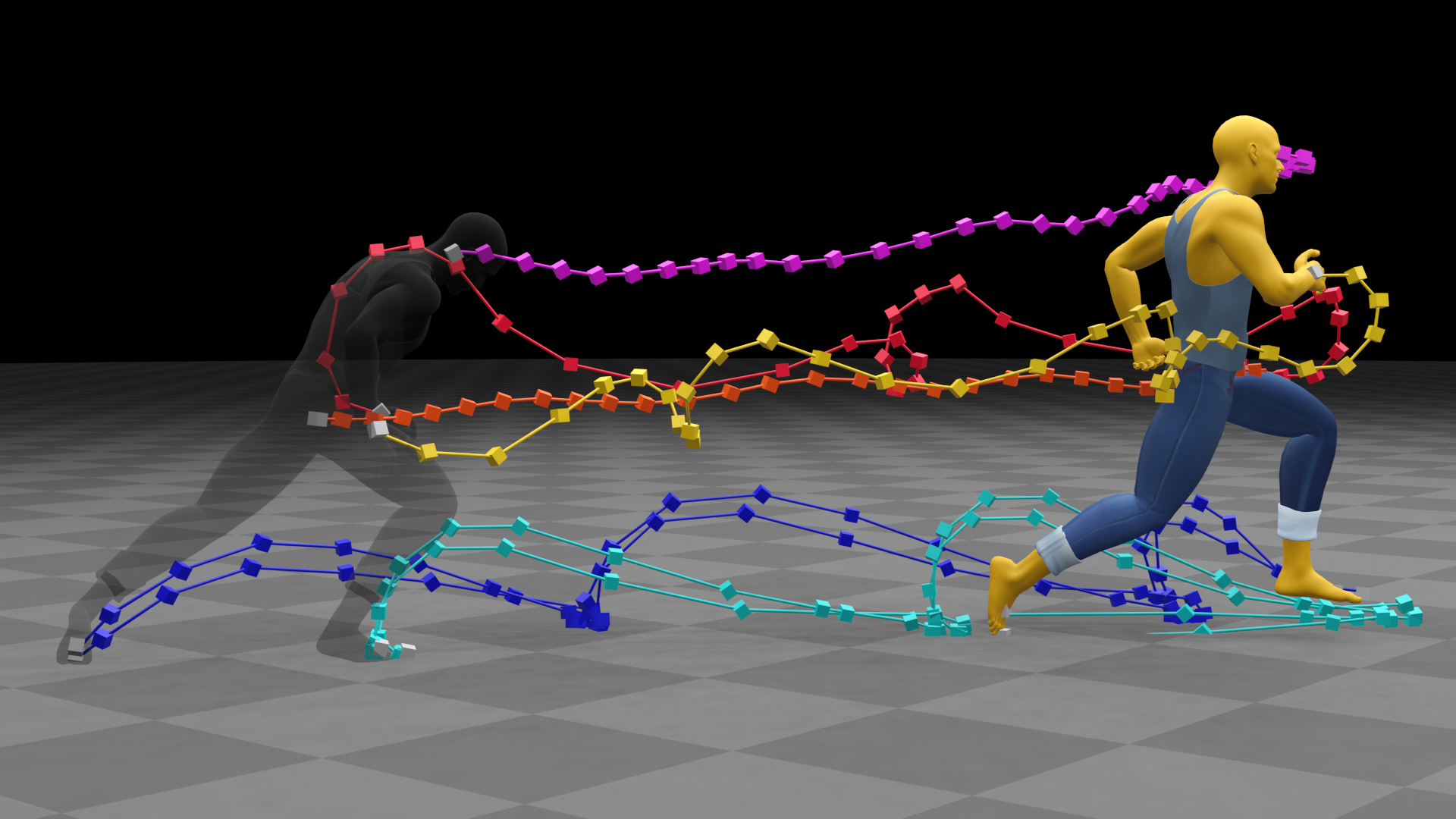}
        \subcaption{Without normalization}
    \end{subfigure}
    \begin{subfigure}{0.75\linewidth}
        \centering
        \includegraphics[width=\linewidth]{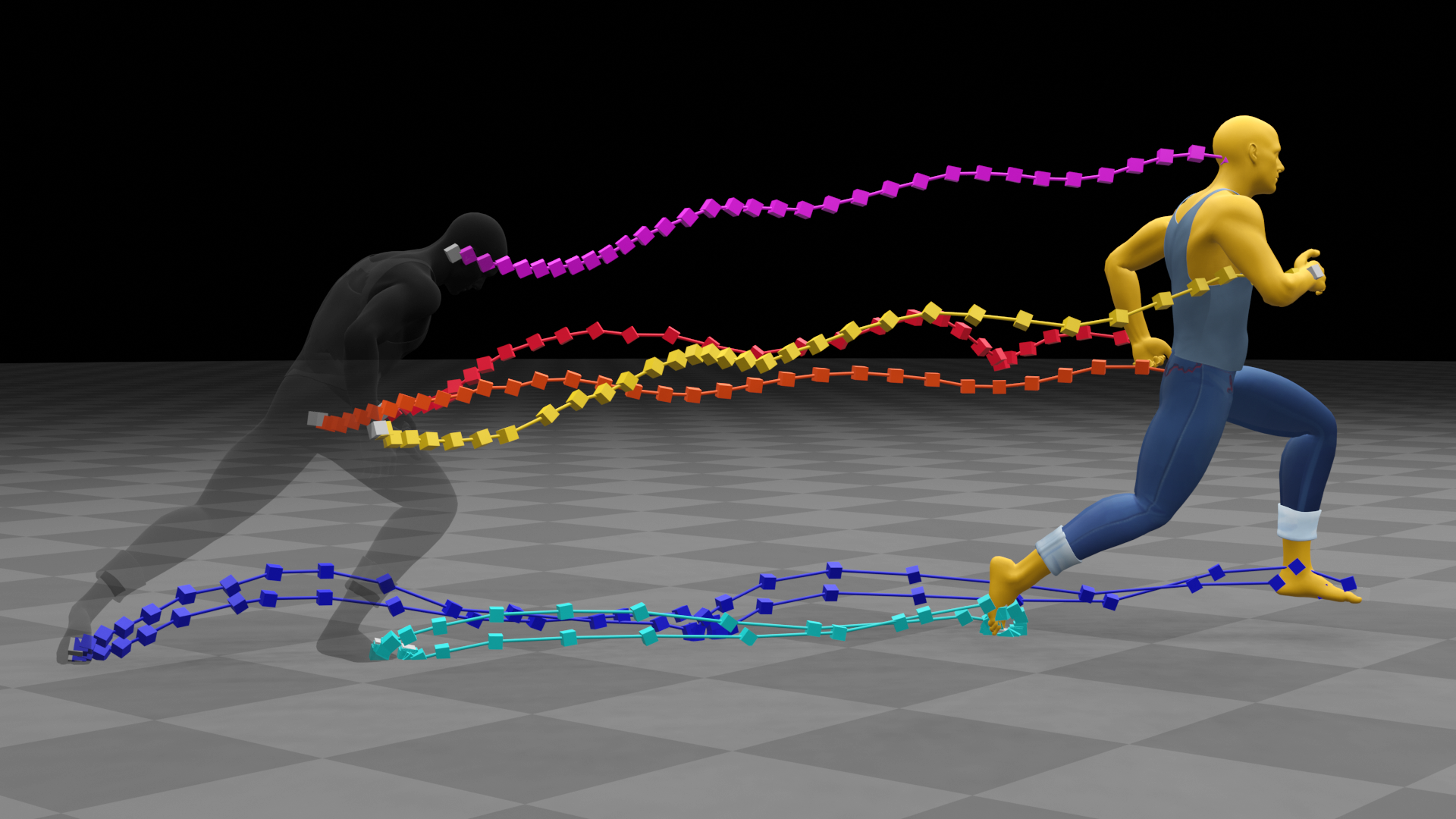}
        \subcaption{With normalization}
    \end{subfigure}
    \caption{We edit a running clip to run half the distance of the original sequence. Without normalization, the base motion results in an unnatural warp to the final frame with foot sliding. In contrast, with normalization, the resulting motion does not have such artifacts.}
    \label{fig:ablation_normalization}
\end{figure}

When moving the boundary frames to stretch or squash an original animation, the overall scale of the animation changes for inpainting.
If this is ignored, it can cause artifacts in the generated motion.
In \figref{fig:ablation_normalization}, we squash a running forward sequence to run only half of the original distance.
Without normalization, this edit creates artifacts where the motion goes past the final frame and then slides backwards to the target frame.
In contrast, with normalization the motion correctly adjusts the scale and generates a slower running motion.

\section{Editing Outside the Convex Hull}
\begin{figure}
    \centering
    \includegraphics[width=\linewidth, trim={0 5cm 0 5cm}, clip]{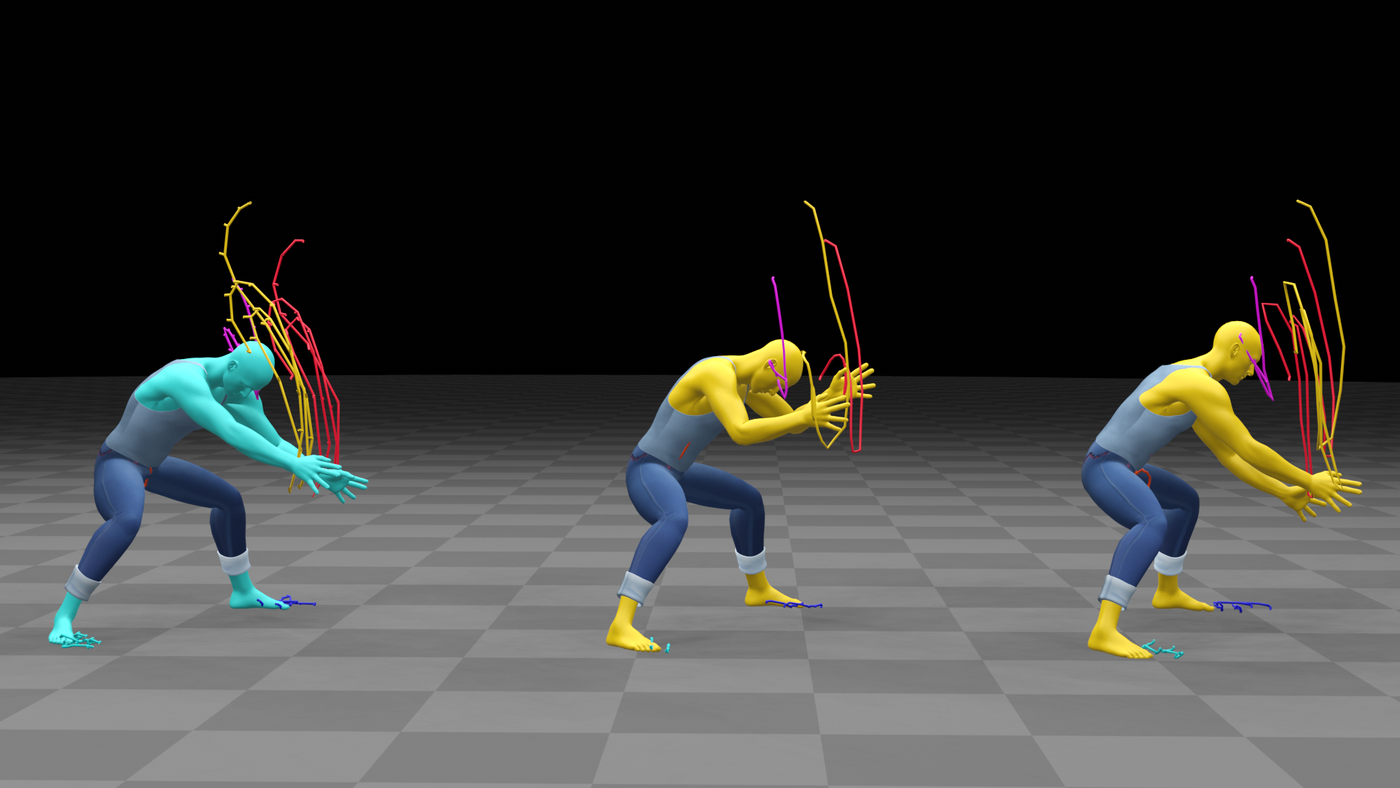}
    \caption{From an exaggerated base motion (left), we can see how automatically adding some sparse constraints with our method (shown on the right), better re-constructs the motion than without any constraint (shown in the middle).}
    \label{fig:exiting_the_convex_hull}
\end{figure}

We find that current generative motion models struggle to sample highly exaggerated or periodic motions that are far away from the input conditions, such as the `smashing an object' action shown by the cyan character in \figref{fig:exiting_the_convex_hull}.
We refer to this loss of variability as the \textit{convex hull problem for motion generation}, as the generated motion is mostly bound within the convex hull of the input conditions.
Consequently, when reconstructing such motion using our inpainting method, the regenerated motion loses the periodic dynamics, as seen in the middle of \figref{fig:exiting_the_convex_hull}.

To better reconstruct such motions, we automatically detect and add the extrema points on the base motion as additional constraints for the motion model.
Given these constraints, the base motion is now within the convex hull of the conditions, and hence the motion model can reconstruct it more closely.
In \figref{fig:exiting_the_convex_hull}, this results in the motion curves of the right character matching more closely those of the cyan character than those of the middle character.
While those additional constraints help to better match the original base motion, they also come at the cost of managing them when one wants to edit on top. We hope that future research can better mitigate this convex hull problem to avoid this additional use of added constraints. 

\section{Number of Diffusion Steps for Noise-inversion}\label{app:eval_steps_for_noise_inversion}

\begin{figure*}
    \centering
    \begin{subfigure}{0.25\linewidth}
        \centering
        \includegraphics[width=\linewidth]{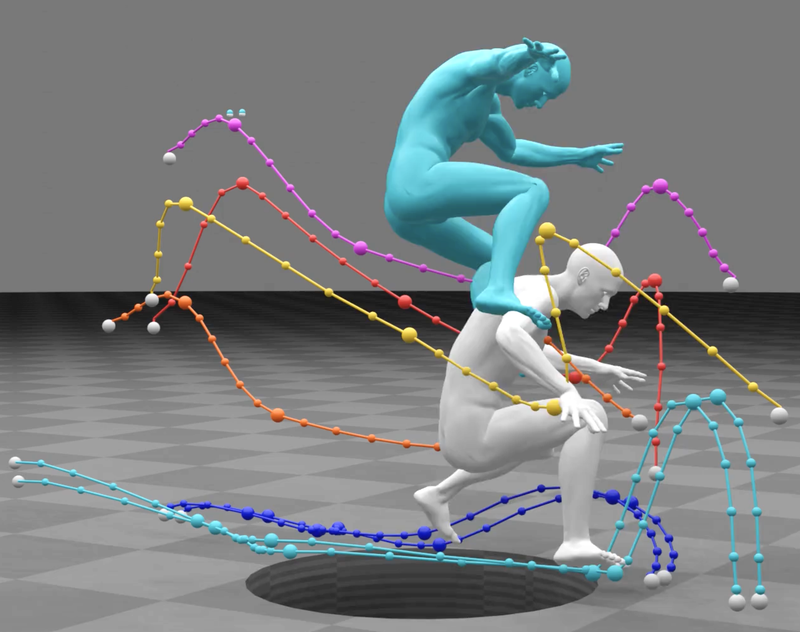}
        \subcaption{25 steps}
    \end{subfigure}%
    \begin{subfigure}{0.25\linewidth}
        \centering
        \includegraphics[width=\linewidth]{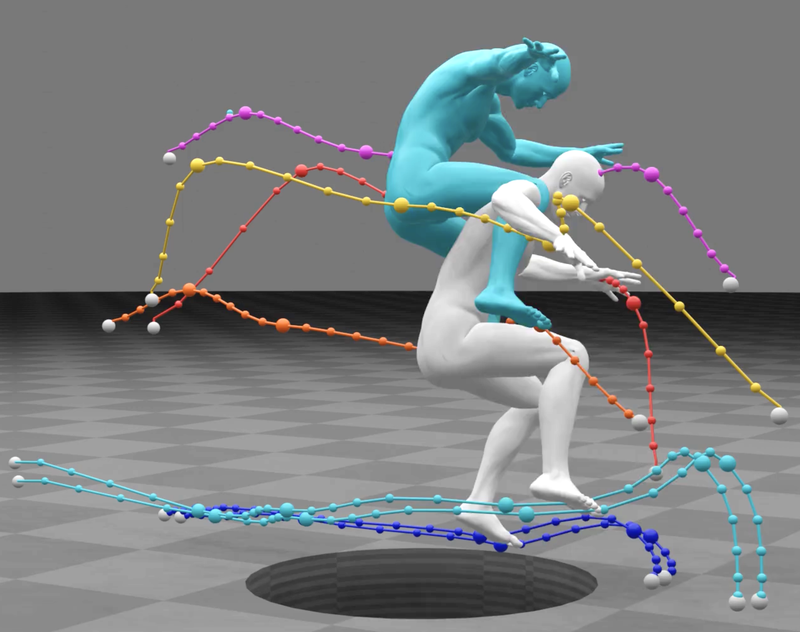}
        \subcaption{50 steps}
    \end{subfigure}%
    \begin{subfigure}{0.25\linewidth}
        \centering
        \includegraphics[width=\linewidth]{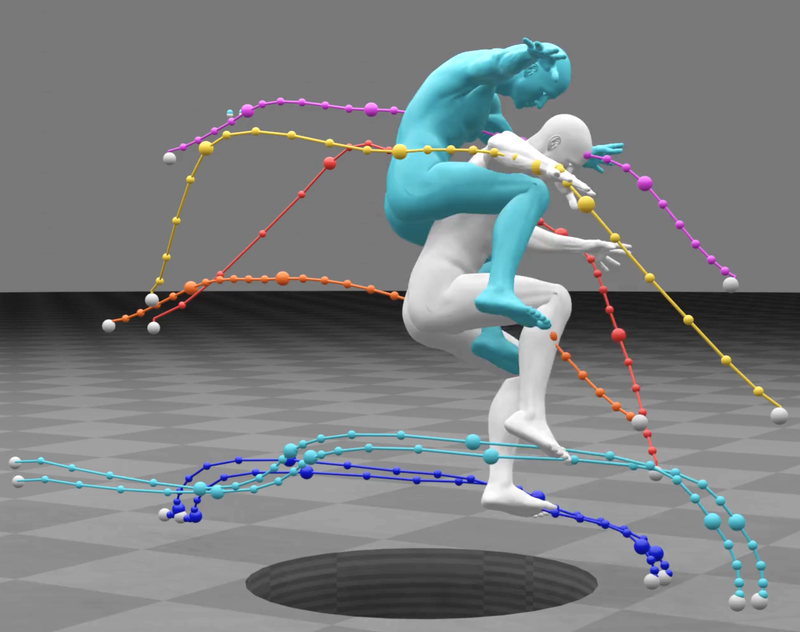}
        \subcaption{100 steps}
    \end{subfigure}%
    \begin{subfigure}{0.25\linewidth}
        \centering
        \includegraphics[width=\linewidth]{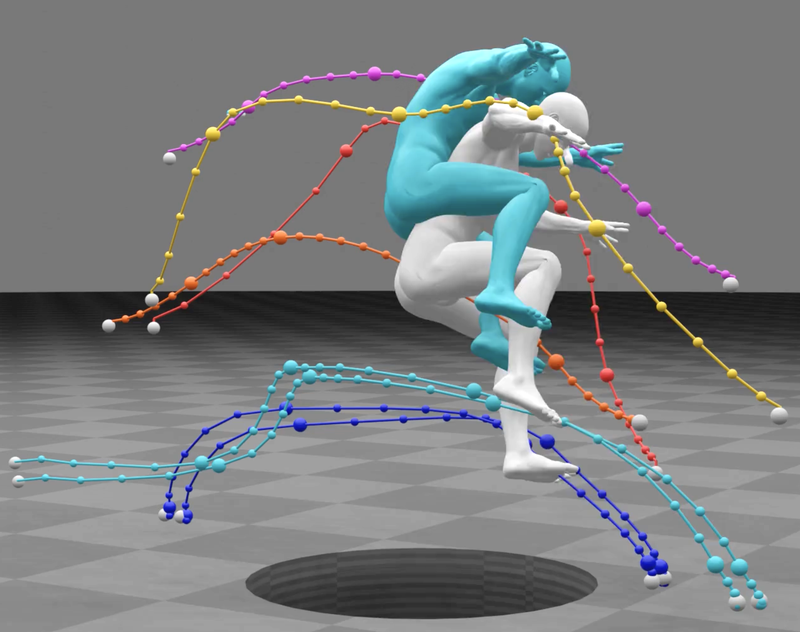}
        \subcaption{200 steps}
    \end{subfigure}
    \begin{subfigure}{0.25\linewidth}
        \centering
        \includegraphics[width=\linewidth]{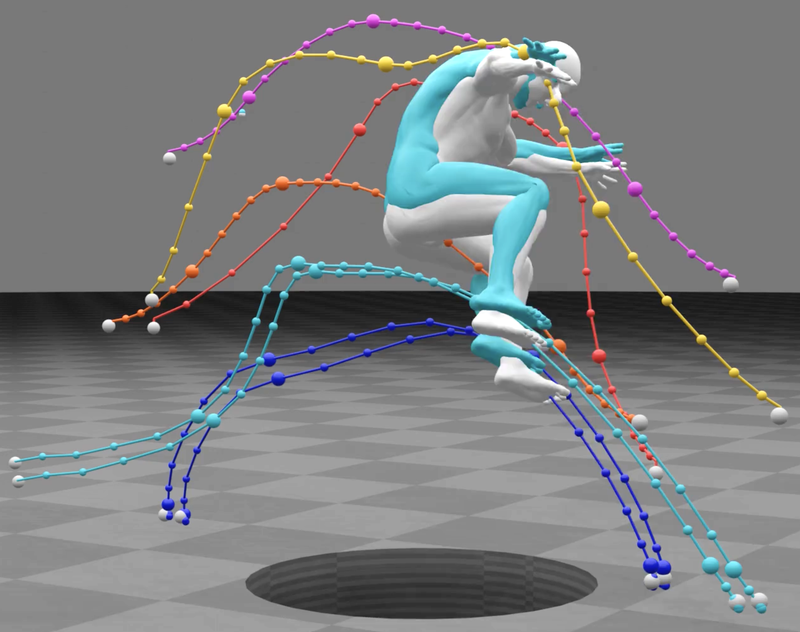}
        \subcaption{400 steps}
    \end{subfigure}%
    \begin{subfigure}{0.25\linewidth}
        \centering
        \includegraphics[width=\linewidth]{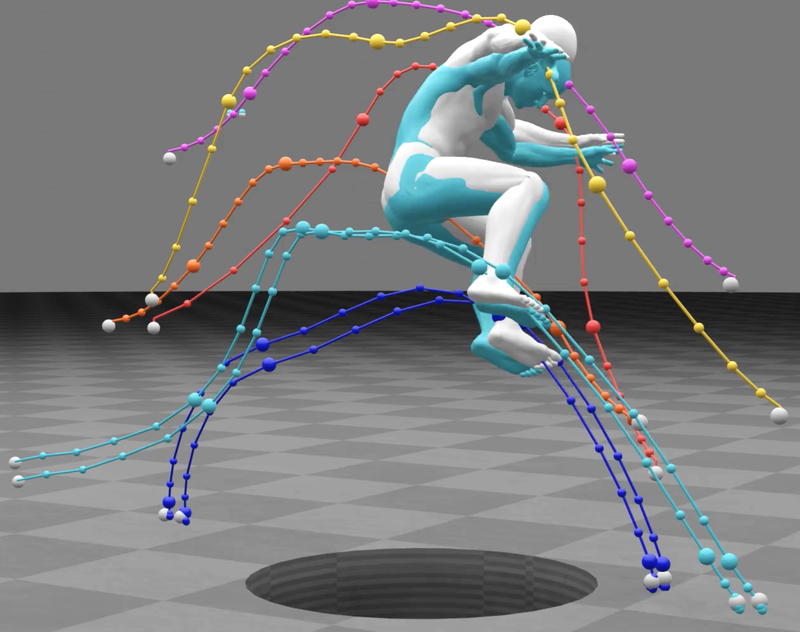}
        \subcaption{600 steps}
    \end{subfigure}%
    \begin{subfigure}{0.25\linewidth}
        \centering
        \includegraphics[width=\linewidth]{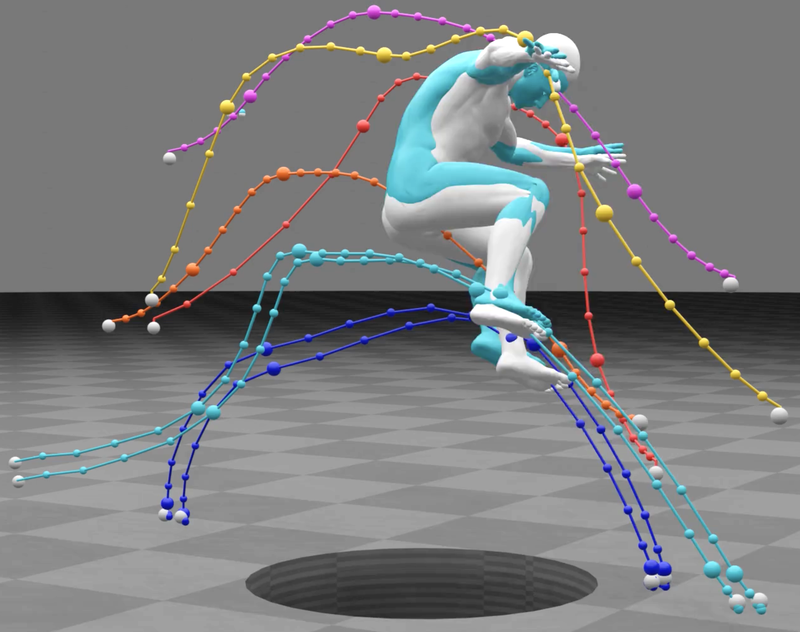}
        \subcaption{800 steps}
    \end{subfigure}%
    \begin{subfigure}{0.25\linewidth}
        \centering
        \includegraphics[width=\linewidth]{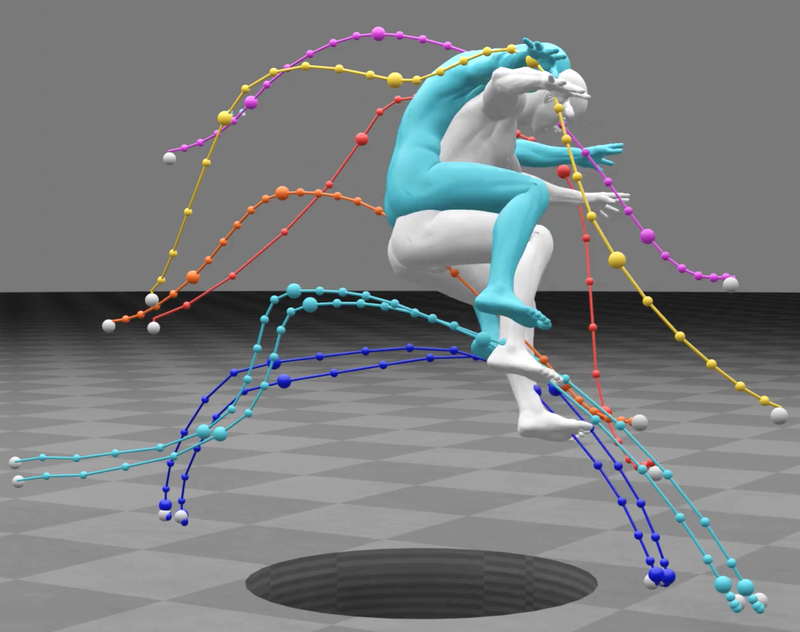}
        \subcaption{1000 steps}
    \end{subfigure}
    \caption{Number of steps required for noise-inversion. Evaluating too few steps results in the inverted motion (white) not matching the ground truth (cyan) in height. At higher number of evaluation steps, the method is not real-time as shown in our video.}
    \label{fig:noise_inversion_num_steps}
\end{figure*}

In \figref{fig:noise_inversion_num_steps}, we use DDIM noise inversion to invert a jumping clip using different number of evaluation steps.
We use the same number of steps for the inversion and the denoising processes, since we observed that using different numbers of steps resulted in artifacts in our experiments.
As seen in the figure, using fewer than 400 steps results in large offsets between the ground truth in cyan and the generated motion in white.
Interestingly, for this example, 600 evaluation steps results in the best performance with more steps also hurting reconstruction.

DDIM-Inversion~\cite{song2022denoisingdiffusionimplicitmodels} assumes that the noise added in time steps $t$ and $t+1$ is almost identical.
Therefore, it uses the noise removed by the motion model $\motionModel$ at $t$ as the noise to be added to create the noised motion at timestep $t+1$.
However, this assumption does not hold when using much fewer diffusion steps.
Consequently, the inverted motion does not closely match the base motion.

\section{Optimization Steps for DNO}\label{app:grad_steps_for_DNO}
In \tabref{tab:num_optim_steps_DNO}, we measure the number of optimization steps required for DNO \cite{dno} to match the reconstruction performance of our recommended inpainting schedule ($\sigma_s = 500$ \& $\sigma_e=50$) on our internal dataset. We see that even after $100$ optimization steps, which need $\sim50$ seconds to compute a batch of $256$ samples, DNO performs marginally worse than our method.
However, the computation time makes DNO too slow for more experimental workflows, where artists try out different edits on different clips to see which best fits their vision.
In contrast, we can better reconstruct base motion in just one inference process, which requires $0.192$ seconds for the same batch size, and additionally allow the user to control the level of preservation desired.

\begin{table}
    \centering
    \caption{Measuring the reconstruction error on our internal dataset for different amounts of optimization steps using DNO \cite{dno}. Even with 100 optimization steps, our scheduled inpainting gives better reconstruction with much faster compute time.}
    \begin{tabular}{cccc}
         \toprule
         $\#$of optimization steps & L2P $\downarrow$ & L2R $\downarrow$ & Time (s) $\downarrow$\\
        \midrule
         20  & 0.0303 & 0.1424 & 10.306\\
         40  & 0.0162 & 0.1234 & 20.032\\
         60  & 0.0104 & 0.1064 & 29.553\\
         80  & 0.0077 & 0.0921 & 39.179\\
         100 & 0.0060 & 0.0812 & 49.129\\
        \hdashline
        Ours ($\sigma_s = 500$ \& $\sigma_e=50$) & \textbf{0.0044} & \textbf{0.0516} & \textbf{0.192} \\
        \bottomrule
    \end{tabular}
    \label{tab:num_optim_steps_DNO}
\end{table}


\section{Connections to Classifier Free Guidance}
Classifier-Free Guidance (CFG) \cite{classifier_free_diffusion_guidance}, moves the conditioned generated motion away from an unconditionally generated motion to better satisfy the given conditions.
Applying CFG with a static weight results in over-saturation in the generated samples.
Consequently, previous works have explored applying CFG on different interpolation schedules.
Wang et al.~\cite{wang2024analysis} provide an analysis of these different schedules.
However, these schedules are quite simple and fixed, i.e.\ not interactive for users.
For motion generation, in place of evaluating the model unconditionally, one could consider $-\baseMotion$ as a fixed unconditional sample that the motion generation should move away from.
This results in a formulation similar to our scheduled inpainting.
But the models need to be specifically trained for CFG while our method can be applied to any motion generation model.
Additionally, by formulating the task as inpainting, we allow the user to manipulate the schedule for closer control.

When applying CFG, a common exploit is to set the interpolation coefficient to greater than 1.
However, we find that for scheduled inpainting, setting $\alpha^t > 1$ in Eq.~(1) results in artifacts where the generated motion goes beyond the convex hull of the base motion, similar to over-saturation in images.
Sadat et al.~\cite{sadat2024eliminating} show that adding only the orthogonal component for CFG, which they call APG, eliminates over-saturation in images.
We leave exploring parallels of APG for scheduled inpainting for future work.